\documentclass[fleqn,usenatbib]{mnras}

\usepackage{newtxtext,newtxmath}
\usepackage[T1]{fontenc}

\DeclareRobustCommand{\VAN}[3]{#2}
\let\VANthebibliography\thebibliography
\def\thebibliography{\DeclareRobustCommand{\VAN}[3]{##3}\VANthebibliography}

\usepackage{graphicx}	% Including figure files
\usepackage{amsmath}	% Advanced maths commands

\newcommand{\target}[0]{Phoebe}

\title[]{AMPM II. --- A Lunar-Mass Primordial Black Hole Microlensing Candidate in the Milky Way Halo}

\author[R. Key et al.]{Renee Key,$^{1,2}$\thanks{E-mail: rkey@swin.edu.au}
Edward N. Taylor,$^{1}$
Ken C. Freeman,$^{3}$
Jeremy Mould,$^{1,2}$
Abhijit Saha,$^{4}$
Anais M{\"o}ller,$^{1}$
\newauthor
Timothy M. C. Abbott,$^{5}$
and Alan R. Duffy$^{1,2}$
\\
$^{1}$Centre for Astrophysics and Supercomputing, Swinburne University of Technology, Melbourne, 3122, VIC, Australia\\
$^{2}$ARC Centre of Excellence for Dark Matter Particle Physics\\
$^{3}$Research School of Astronomy and Astrophysics, Australian National University, Canberra, 2611, ACT, Australia\\
$^{4}$NSF NOIRLab National Optical-Infrared Astronomy Research Laboratory, 
950 North Cherry Ave., Tucson, 85719, AZ, USA\\
$^{5}$Cerro Tololo Inter-American Observatory, Casilla 603, La Serena, Chile}
\date{Accepted XXX. Received YYY; in original form ZZZ}

\pubyear{\the\year{}}

\begin{document}
\label{firstpage}
\pagerange{\pageref{firstpage}--\pageref{lastpage}}
\maketitle

\begin{abstract}
Primordial Black Holes (PBH) are hypothesised to form during inflation \cite{Bean_2002, Clesse_2015} and have long been considered a candidate for compact dark matter \cite{Carr_Hawking_1974, Green_2024, GarciaBellido_2019}. Gravitational microlensing is known as a productive method for exoplanet discovery and characterisation, but also provides an experimental avenue to constraining the PBH abundance in the mass regime 
from $\sim 10^{-11}\ \mathrm{M}_\odot$ to $\sim 10^5\ \mathrm{M}_\odot$ \citep[][and references therein]{Carr_2024}. We performed a high-cadence, optical microlensing survey with DECam over five nights towards the Large Magellanic Cloud, sensitive to microlensing timescales from minutes to days. Here, we report the discovery of a one hour-long microlensing event. An optical depth probabilistic analysis indicates that the lensing object, which we refer to as Phoebe, is 5 orders of magnitude more likely to be part of the Milky Way's dark matter halo than part of the stellar content of the Milky Way and Large Magellanic Cloud. No matter the location of Phoebe, it is among the fastest and lowest mass microlensing signals ever detected, with a Einstein timescale of approximately 60 minutes. Using Bayesian modelling, we interpret Phoebe as a PBH with mass $0.032^{+0.227}_{-0.027} M_{\oplus}$, or approximately 3 lunar masses. Phoebe suggests a population of compact, lunar-mass objects associated with the dark matter distribution of the Milky Way, and potentially opens a new window to the physics of inflation.
\end{abstract}

\begin{keywords}
gravitational lensing: micro -- galaxies: Magellanic Clouds
\end{keywords}

%%%%%%%%%%%%%%%%%%%%%%%%%%%%%%%%%%%%%%%%%%%%%%%%%

%%%%%%%%%%%%%%%%% INTRODUCTION %%%%%%%%%%%%%%%%%%

\section{Introduction}
Microlensing and detection of gravitational radiation are among the only tools available for the detection of distant, cold and compact objects \citep{Paczyski_1996}. Together with gravitational wave technology,
it is the only way to detect macroscopic dark matter substructures (such as Primordial Black Holes; PBHs) if it is present in galaxy halos \citep{Carr_2025}.
The generic challenge to contend with for microlensing detections is that the lensing signal depends only on the geometry, relative motion and mass of the microlens, and is insensitive to the physical nature of the lens itself \citep{WittMao_1994}. As it stands, any information gleaned about the microlens is model dependent, but the galactic models of microlensing densities are precise and well understood \citep{Li_2025, DeRocco_2023}, and so allow for likelihood analysis on the microlensing event. Microlensing has been used in the detection of exoplanets and Free-Floating Planets (FFPs) towards the galactic bulge \citep{Sumi_2023} with several candidates at terrestial masses \citep{Mroz_FFP12020, Koshimoto_2023}. In the former case, both the host star and the exoplanet are lensed. A review is given by \cite{Tsapras_2018}.
Astronomers have observed towards extragalactic fields for black holes and PBHs, see a recent review by \cite{Green26}, with recently \citet{Sugiyama_2026} present 12 new PBH candidates at lunar masses towards M31 using Subaru-HSC data.

We have used DECam on the Cerro Tololo Blanco 4m telescope for a fast cadence experiment towards the Large Magellanic Cloud (LMC) as described in the companion paper to this publication (Key et al., 2026a). Our search revealed a single candidate event which is the topic of this paper, and our goal is to infer the properties of this microlensing event. 

In Section \ref{sec:maths}, we introduce the necessary microlensing theory for the detection of low-mass PBHs, and discuss second order considerations in microlensing such as the finite source effect.  Section \ref{sec:obs} briefly describes our observations. We reconsider the photometry of the candidate, and examine the detection under many potential contaminating origins, such as atmospheric noise, blending and confusion with neigbouring stars and regular stellar variability in Section \ref{sec:phoebe} and its possible origins. In Section \ref{sec:parameters}, we introduce the model of the microlensing system, quantifying a distribution for the source star and describing the various necessary models of galactic dark matter and stellar density required to model the candidate event. The nature of the event and to what component of the Galaxy the lens might belong is discussed in Section \ref{sec:Origins}, and a discussion on the impact of such PBH detections is provided. 

\section{Gravitational Microlensing}\label{sec:maths}
Gravitational microlensing occurs when an intervening massive, compact object moves between an observer on Earth and a distant source star, which produces two images of the source star. As the microlens is small compared to the source star, the images are separated by microarcseconds \citep{Schneider_1992, WittMao_1994}. These images are projected radially from the source along the Einstein ring radius \citep{NO_1984}.
\begin{equation}
    r_{E} = \sqrt{\frac{4 \, G \, M \, D_{L} (D_{S} - D_{L})}{c^{2} \, D_{S}}}\, ,
\label{eq:rE}
\end{equation}
Here, $M$ is the mass of the microlens, $D_{L}$ is the distance to the lens from the observer, $D_{L}$ is the distance to the source star from the observer, and $c$ and $G$ are the speed of light and gravitational constant.

Although the microlensing images are unresolved from the source flux, the two images combine and magnify the source’s brightness. The compact microlens moves with some trajectory in time, $u(t)$, making the microlensing amplification a transient event. Assuming that both the source star and microlens are point-like objects, the amplification is \citep{VO_1983, Paczynski_1986},
\begin{equation}
    A(u)_{PS} = \frac{u^{2} + 2}{u \sqrt{u^{2} + 4}} \, ,
\label{eq:Aps}
\end{equation}
where $u(t)= \sqrt{u_{0}^{2} + (\frac{t - t_{0}}{t_{E}})^{2}} $, $t_{0}$ is the time at peak amplification, and $u_{0}$ is the impact parameter. The timescale of the event, $t_{E}$, is the time taken for a lens to travel the Einstein radius ($r_{E}$) with a transverse velocity ($v_{\perp}$).
\begin{equation}
    t_{E} = \frac{r_{E}}{v_{\perp}}\, ,
\label{eq:tE}
\end{equation}
here, $c$ and $G$ are the speed of light and gravitational constant. As the Einstein timescale depends on the microlens mass ($M$) from Equation \ref{eq:rE}, the lower the lens mass, the faster the timescale of the microlensing event.

The assumption that a source star is an infinitesimal point of light breaks down when the angular Einstein radius of the microlens ($\theta_E$) is comparable to the star's angular radius ($\theta_{S}$) \citep{Smyth_2020, Griest_2011}.  The parameter $\rho$ is the ratio of the lens and source angular radii, and describes the impact of the ‘finite’ size of the source \citep{WittMao_1994, Griest_1991}:
\begin{equation}
\label{eq:Rho}
    \rho = \frac{\theta_{S}}{\theta_{E}} = \frac{R_{S}\,D_{L}}{R_{E}\,D_{S}} \, ,
\end{equation}
where $R_{S}$ refers to the physical radius of the source. 
Finite source effects appear in microlensing as $\rho \leq 1$, which arises from a combination of very low-mass lenses, giant source stars or nearby sources. The microlensing amplification in Equation \ref{eq:Aps} is integrated over the source's 2D disc \citep{Lee_2009} to incorporate finite-source effects. 
\begin{equation}
\label{eq:AFS}
A_{FS}(u, \rho) = \frac{2}{\pi\,\rho}\int_{0}^{\pi}\int_{u_{1}}^{u_{2}} A_{PS}(u')\, u' \, du' \, d\theta\, ,
\end{equation}
The maximum amplification of the FS$-$PL event as $u_{0}$ tends towards zero is bounded as \citep{Paczyski_1996, Cieplak_2013},
\begin{equation}
    A_{max} = \frac{\sqrt{4 + \rho^{2}}}{\rho}\, .
\label{eq:amax}
\end{equation}
For a given microlensing system and fixed $A_{max}$ threshold, a distance limit ($D_{max}$) defines the maximum lens distance from the observer that is capable of producing such amplification, 
\begin{equation}
    D_{max} = \frac{2\,R_{E}\,D_{S}}{R_{S}\sqrt{1 + A_{max}^2}} \, .
\label{eq:dmax}
\end{equation}
The optical depth measures the likelihood that a star will intersect the lensing cross-section of a microlens population with a given density distribution $\varrho(D_L)$ along the line of sight between the observer and the source. To account for finite source effects, the optical depth calculation is restricted by the maximum line-of-sight distance $D_{max}$ to the source star \citep{Griest_1991}.
\begin{equation}
    \tau_{FS} =\frac{4\pi G u_{T}^{2}}{c^{2}}\int^{D_{max}}_{0}\varrho({D_{L}})\frac{D_{L}(D_{S} - D_{L})}{D_{S}}dD_{L}
\label{eq:tau}
\end{equation}
Similar to the the $D_{max}$ threshold, the $u_{T}$ factor represents the value for $u$ defined by a set amplification threshold $A_{T}$. For the basic point-source assumption, the optical depth equation is integrated along the complete line-of-sight to the source, $D_{max} = D_{S}$.

In this paper, we distinguish between the point-source microlensing treatment and the finite-source extension as PS$-$PL and FS$-$PL.
%%%%%%%%%%%%%%%%%%%%%%%%%%%%%%%%%%%%%%%%%%%%%%%%%

%%%%%%%%%%%%%%%%% OBSERVATIONS %%%%%%%%%%%%%%%%%%

\begin{figure*}
\centering
\includegraphics[scale = 0.7]{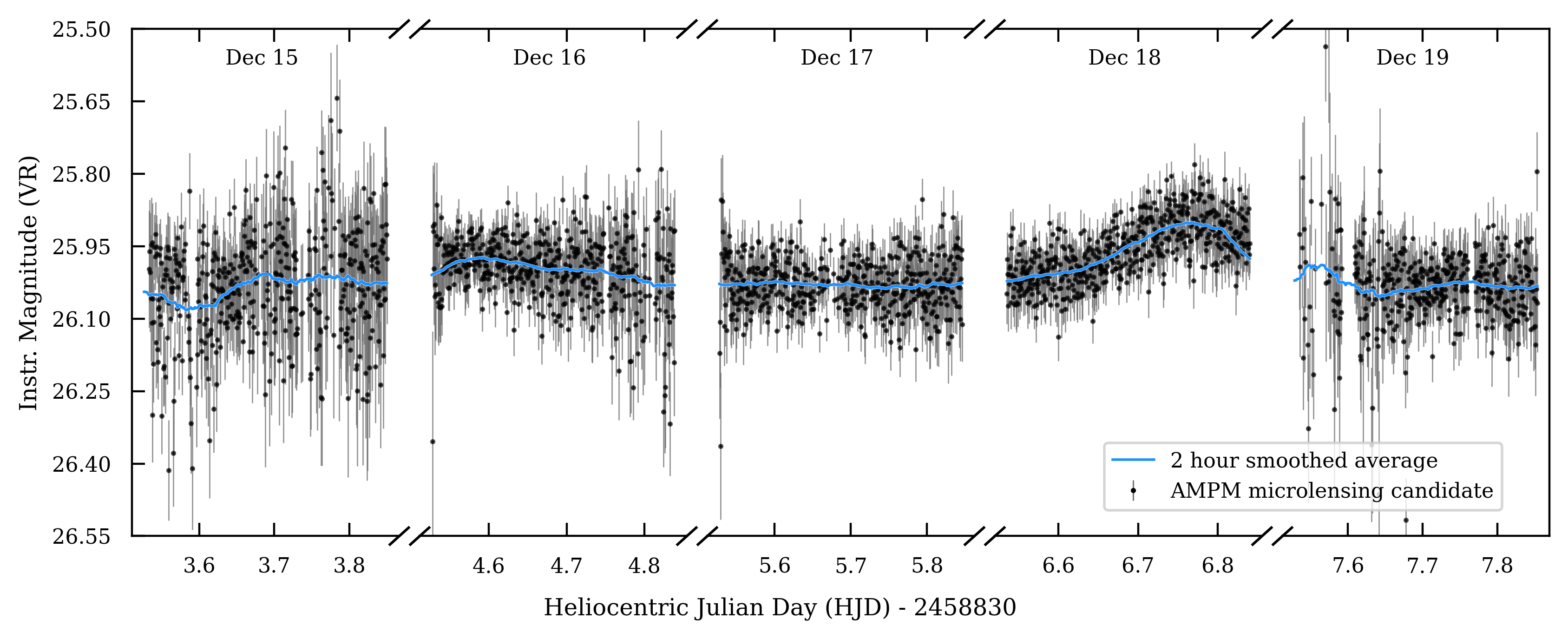}
\caption{The detection light curve for \target\ from December $15^{\rm th} - 19^{\rm th}$, with the magnitude measurements of the source star to \target\ from \texttt{DoPHOT} photometry. Poor observing conditions on December $15^{\rm th}$ produce strong scatter in stellar magnitudes that significantly affect all stars in the field. Similar localised sections of atmospheric scatter are evident in the light curve of \target\ towards the end of December $16^{\rm th}$ and the beginning of the night on December $19^{\rm th}$. The smoothed signal in blue is a convolution of the light curve with a 2-hour sliding box kernel.}
\label{figfulllightcurve}
\end{figure*}

\section{Observations and Photometry and Microlensing Detection} \label{sec:obs}
In this section, the observation sequence, photometric reduction and microlensing detection pipeline aspects of the AMPM survey are briefly summarised. A comprehensive analysis of each component of AMPM, including the motivation and output from each processing stage, is provided in the first paper in this series (see Key et al. 2026a).
\newline\newline
AMPM ran for five consecutive nights in December 2019, using the Dark Energy Camera (DECam) to target a single LMC field in the broadband optical VR filter. The 3 $\rm deg^{2}$ field of interest coincides with Field 51 in the SMASH LMC survey \citep{Nidever_2019, Nidever_2021}, centered at coordinates $\alpha =$ 05:59:51.799, $\delta =$ -70:12:19.001 (J2000). Aligning the AMPM field with a SMASH field additionally provided photometry in \textit{u, g, r, i} and \textit{z} to complement the VR observations. The AMPM survey had a continuous sequence of 20-second integration of Field 51 in VR, which produced an average cadence of 50 seconds when combined with the detector overhead and readout time. No colour measurements were taken during the 2019 observations so as to avoid disrupting the cadence rate and completeness of the light curves. The exposures from each of the five nights of observation were processed using \texttt{DoPHOT} \citep{Dophot_1993}. The sources on each image were collated by coordinate position into five catalogues comprising several million light curves for each night.
\newline\newline
We first imposed a quality control on the full catalogue of sources in order to select the well-sampled, reliable stars for the microlensing search. The quality control first removed any light curve not identified by \texttt{DoPHOT} as a single, well-fitted star on at least $80 \%$ of the exposures. Additionally, any remaining star that had more than 5 consecutive missing measurements within the night was removed during this stage of processing, with this particular subset of stars often lying very close to the edges of the DECam CCDs. The quality control removed contaminated, saturated or faint objects with low signal-to-noise measurements from the microlensing search. After quality control, of order $O(10^{5})$ stars remain in each of the five catalogues, where the variation in source counts is a consequence of the change in observing conditions between each night of the survey.

We detect microlensing signals using an optimised set of statistics tailored to discover light curves with a lensing-like brightening. The detection pipeline functions by combining peak detection with an optimal smoothing analysis, the Von Neumann parameter \citep{VNstat1941, Kim2014}, atmospheric blending levels \citep{Irwin_2007}, variability period analysis, cross-matching with Gaia DR3 variability \citep{GAIAVari} and source catalogues \citep{GAIASource} for each star \citep{GAIA1, GAIA2, GaiaVarInfo}, and a microlensing $\chi^{2}$ model fit comparison. The catalogues are processed through the microlensing pipeline individually, with the detection threshold for microlensing events informed by the statistical behaviour of a suite of microlensing injection simulations. Finally, of order $O(10^2)$ light curves remain and are manually inspected for microlensing. The majority of candidates are RR-lyrae stars with relatively long brightening periods, and fast flare stars with $10$ minute outbursts.  

\section{Phoebe, a Short Duration Microlens} \label{sec:phoebe}
\begin{figure}
\centering
\includegraphics[width = \columnwidth, ]{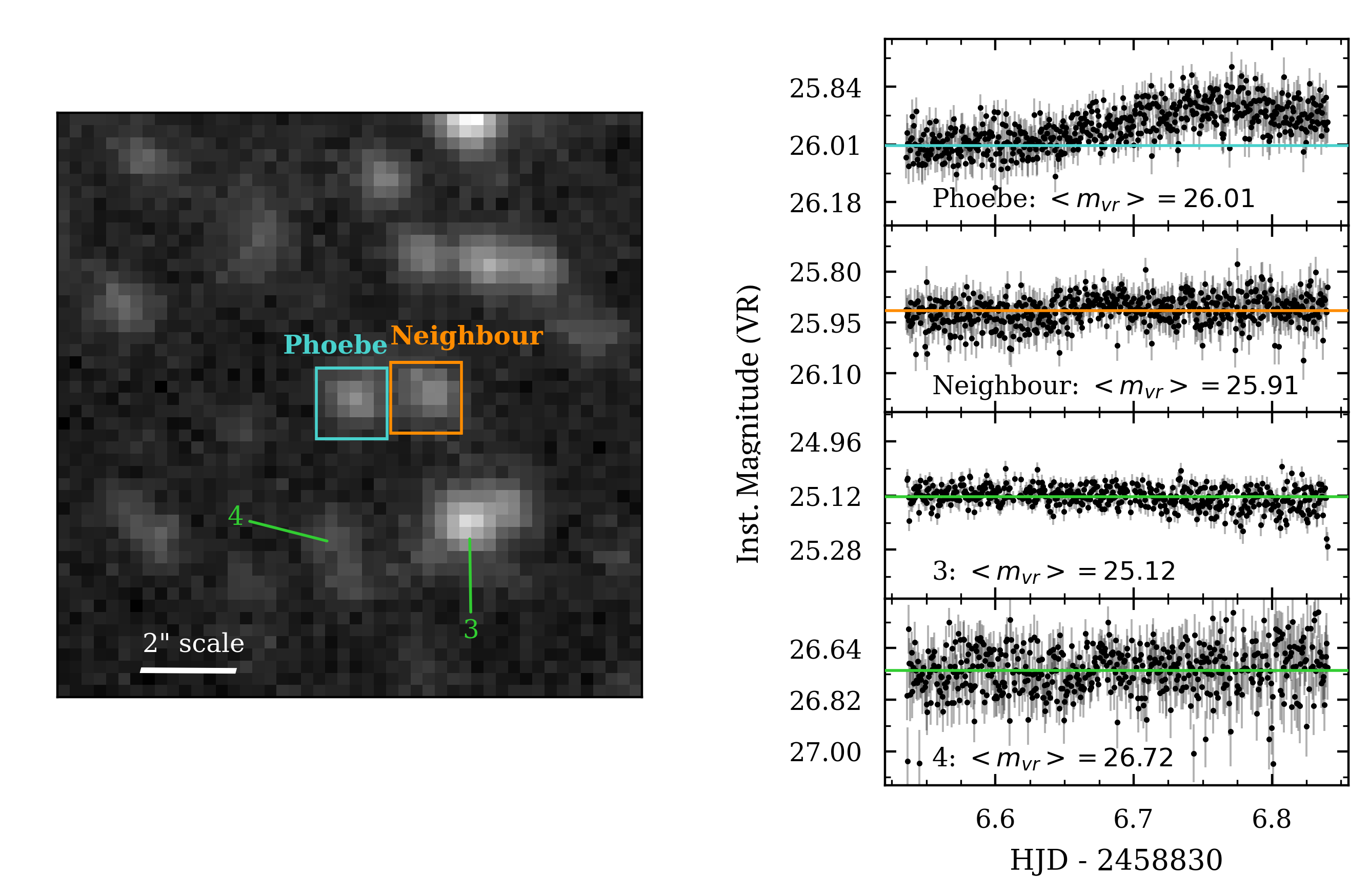}
\caption{The on-sky neighbourhood of \target\, along with several example lightcurves of neighbouring stars. The sky image is a 10-arcsecond region around the source star, taken at HJD 2458836.761019, UTC 06:14:37, during the peak amplification of \target. The source star (bounded by a blue box) sits near a comparably bright neighbouring star (indicated by a pink box). Two other nearby stars are marked by the numbers 3 and 4. The December $18^{\rm th}$ light curves for the four stars are shown below, with the mean instrumental magnitude for each star presented as a horizontal line and in the legends.}
\label{figNeighbourhood}
\end{figure}

One microlensing event was detected from the AMPM pipeline, with the lensing signal occurring during the fourth night of observations (18th of December). The light curve of the event exhibits the typical traits of a microlensing amplification; it is symmetric and smooth on December 18, and the remaining adjacent nights of the AMPM survey are free from repeated variability. We dub the event `\target' as a phonetic nod to the FFP and PBH, which both produce isolated microlensing events at comparable FWHM timescales. The source star with \target\ in its light curve (which we will continue to refer to as the \target\ Star) has J2000 coordinates of $\alpha$ = 05:52:25.6, $\delta$ = -70:44:21.8. Figure \ref{figfulllightcurve} shows he full light curve for the candidate and Figure \ref{figNeighbourhood} shows an image region identifying the source \target\ Star and the nearest neighbouring star. The accompanying light curves for the local region are also shown, where the stable light curves of three surrounding stars demonstrate our data quality and stability, supporting that the signal of Phoebe is unique and not a simultaneous effect across nearby stars.

\subsection{Microlensing Signal Validation}
Given that microlensing is such a rare phenomenon, it is important to consider whether \target\ was, in fact, produced by an alternative process. Complications with the crowded field photometry and other misattributed astrophysical variability, as in the case for true asteroids in the Kepler dataset \citep{Griest_2014} and the MACHO Blue Bumper variables\footnote{Although these particular variable stars are more relevant to microlensing events with timescales spanning some days.} \citep{Alcock_2000} can mimic microlensing events. So, we give substantial consideration to the issue of microlensing confusion in the following section. 

\subsubsection{The Local Neighbourhood around \target}
\begin{figure}
\centering
\includegraphics[width = \columnwidth]{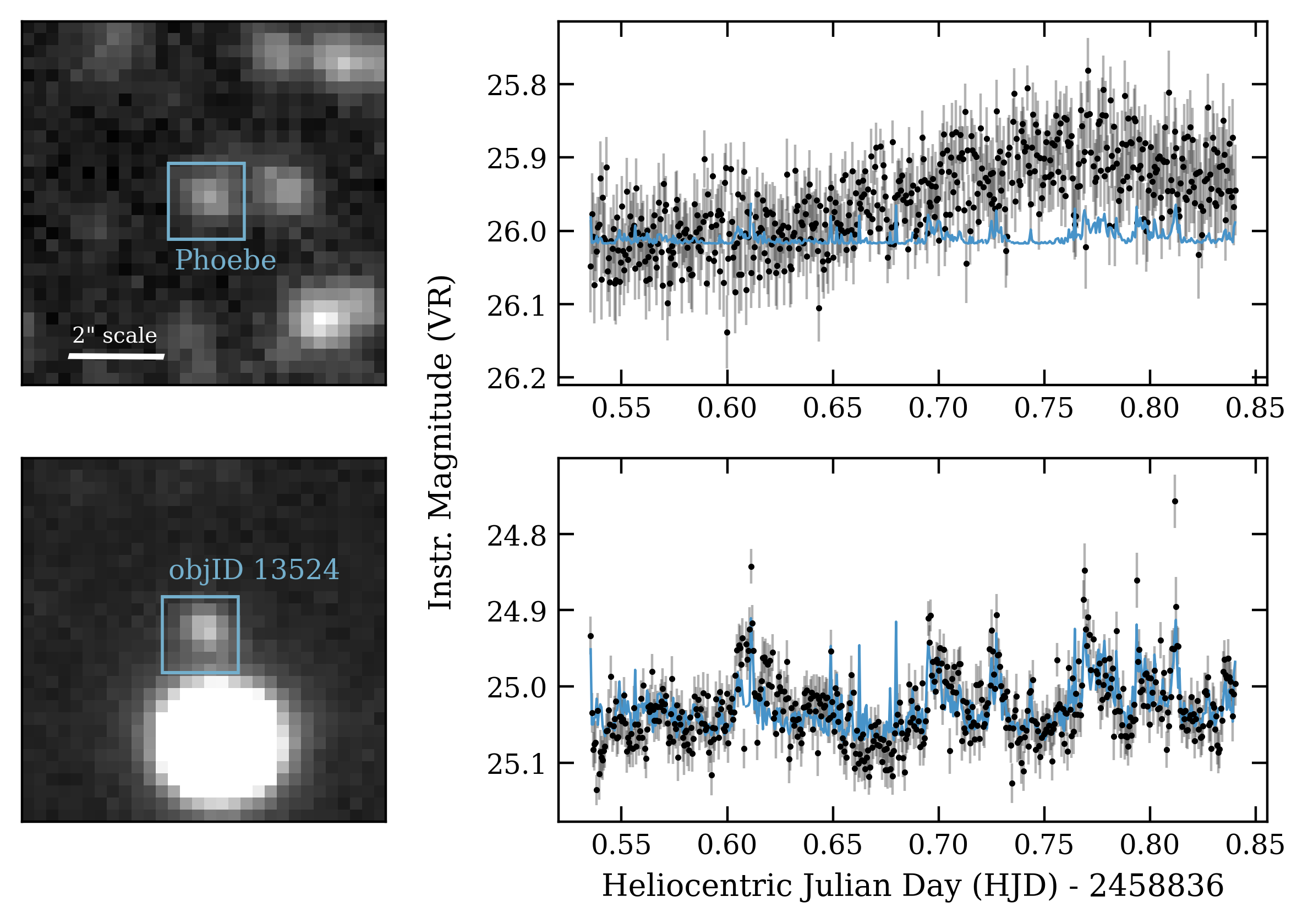}
\caption{A comparison between the signal of \target\ and typical seeing-contaminated fluctuations in AMPM light curves. The uppermost row shows the $30\times30$ pixel cut-out region around the \target\ star indicated by blue bounding box, with the \target\ detection light curve shown to the right. Plotted on the light curve in blue is the best-fit atmospheric blending model for \target\, as determined from correlations between the global image PSF full width-half maximum and magnitude offset. The bottom row presents the sky cut-out and light curve $+$ blending model for a star (recorded as object ID
13424) located within 3 pixels of a saturated source. It is clear that the light curve of objID 13524 is severely contaminated with PSF induced blending from the dominant, saturated star.}\label{figDETREND}
\end{figure}

As an initial verification of \target\ as a microlensing event, we confirmed that similar episodes of brightening were not recorded for any star in the raw catalogues in a 40-arcsecond region around the \target\ Star located on CCD N20, which is on the South-West corner of the DECam array. While some instances of stellar variability are present in the local region, it mostly takes the form of high levels of atmospheric seeing contamination. A single eclipsing binary star located 33 arcseconds away from the \target\ Star at the J2000 coordinates of $\alpha$ = 05:52:32.4, $\delta$ = -70:44:22.1. 
\newline\newline
The AMPM pipeline includes a summary statistic referred to as the $B$ parameter \citep{Irwin_2007} that quantifies the amount of atmospheric contamination in the light curve compared to the change in the PSF full width-half maximum over the night. The $B$ parameter is a value that represents the normalised comparison between the $\chi^{2}$ of a constant magnitude model, to the $\chi^{2}$ of the best-fit PSF contaminated model. 
\begin{equation}
    B = \frac{\chi^{2}_{\mbox{const}} - \chi^{2}_{\mbox{fit}}}{\chi^{2}_{\mbox{const}}}\, .
\end{equation}
The average contamination for all stars across the field-of-view on December 18th is $B_{avg} = 0.03$, indicating that the typical atmospheric disturbance for that night is low compared to the nights with worse atmospheric conditions like December 15th with $B_{avg} = 0.13$. Across December 18th, \target's $B$ parameter is very low at 0.002, which fundamentally suggests that a constant magnitude model is a better fit to the light curve than the blended model. Figure \ref{figDETREND} shows a comparison between the light curve of \target\ and that of a highly contaminated star, with the best-fit atmospheric blending model superimposed on the magnitude light curve. The example of a contaminated star shows the blending conditions that produce poor PSF photometry; a bright, saturated neighbour is situated close to the target star. While \target\ sits near to a neighbouring star, both stars are of similar magnitude and therefore do not bleed excessive flux across PSF radii.

\subsubsection{An Independent Reanalysis of Phoebe's Photometry}
\begin{figure}
\centering
\includegraphics[width = \columnwidth]{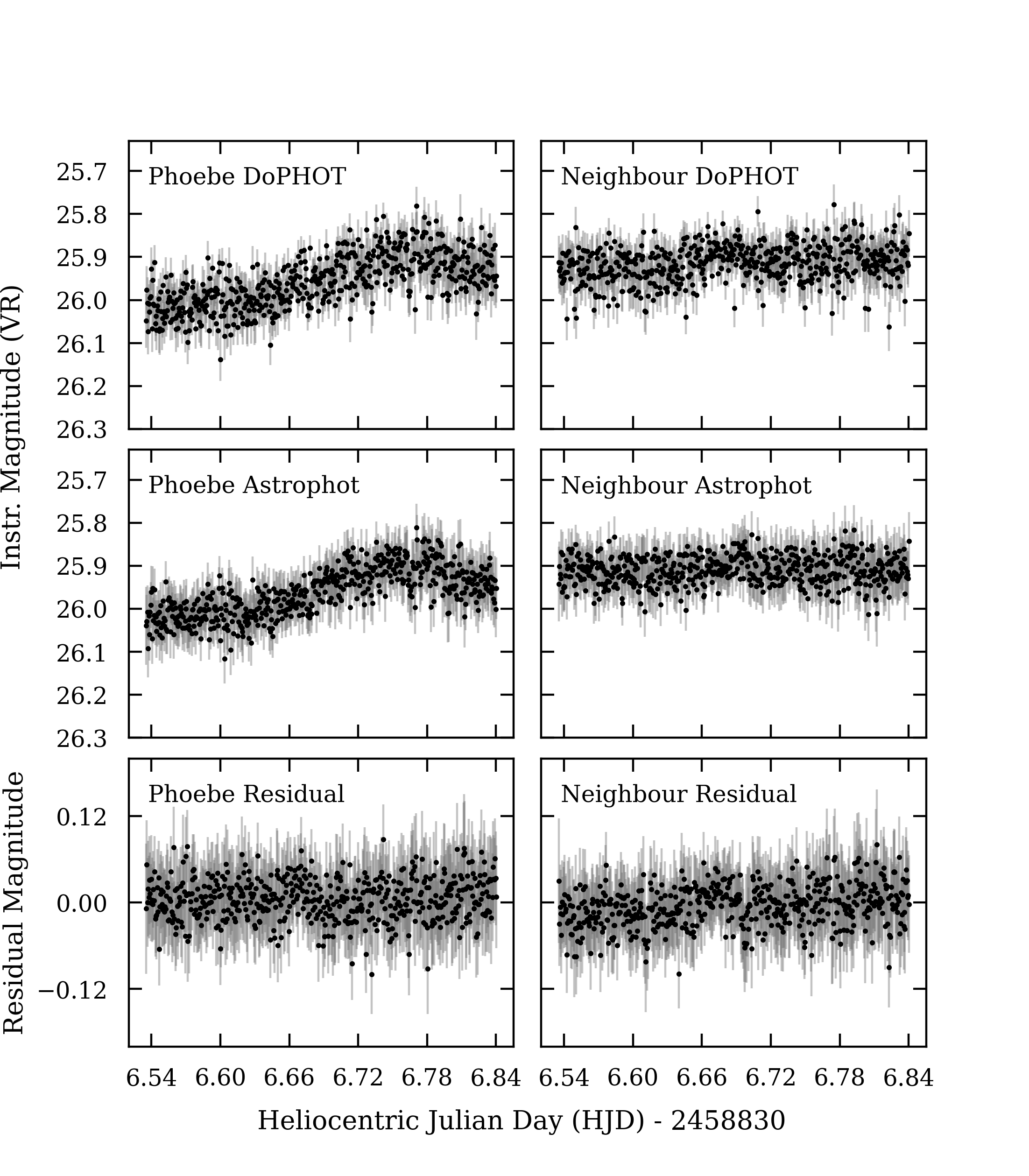}
\caption{The light curves for \target's source and the neighbouring star from the \texttt{DoPHOT} and \texttt{Astrophot} photometry routines. The uppermost panels are the \texttt{DoPHOT} photometry from which \target\ was detected with the AMPM pipeline. The middle panels present the \texttt{Astrophot} light curves from the detrending stage. For clarity, the \texttt{Astrophot} light curves have been converted from flux to the magnitude space, and a constant zero-point has been applied to shift the light curves to the same instrumental magnitude scale as the \texttt{DoPHOT} measurements. The final row of panels show the residual curves produced by subtracting the \texttt{Astrophot} light curves from the \texttt{DoPHOT} light curves. The residual light curves for both the source and neighbour are stable and flat, showing that \texttt{Astrophot} recovers the signal of \target\ in the \texttt{DoPHOT} photometry. }
\label{figCOMP}
\end{figure}

We considered whether the microlensing signal could be a systematic artifact of a misaligned stellar point spread function in the photometry routine. The \target\ Star is located close to a neighbouring star of comparable magnitude, with a on-sky separation of 1.623'' between the pair. The two stars are sufficiently close that the wings of their PSF distributions overlap. While it is clear that the light curve of \target\ is not highly contaminated by the neighbouring star, smaller levels of flux from either star may leak into the PSF of the other and produce a false microlensing amplification. In a similar manner, a faint, typically undetectable source behind the \target\ Star may have variability that is misattributed to \target. We check the SMASH DR2 catalogues (with a $5\sigma$ r-band depth of 24.5, which is $\sim 3$ magnitude deeper than \target) for background stars, and find no additional records of stars in a 1'' box centered around \target\ \citep{Nidever_2021}. To verify that the proximity of neighbouring stars has no impact on the microlensing status of \target, we perform an independent reanalysis of the full AMPM photometry using \texttt{AstroPhot} \citep{Astrophot}. \texttt{Astrophot} is a forward modelling tool that fits parametric models to pixel data and uses automatic differentiation utilities with a Levenberg-Marquardt (LM) minimiser to robustly find the maximum likelihood solution for the model parameters.
\newline\newline
The astrometric solution and precise centroid coordinates for each exposure are determined using a set of 200 stars from CCD N20, with reference to Gaia DR3 \citep{GAIA1, GAIA2, GAIASource}. The PSF model is a circular Moffat profile, with the average size and shape parameters of each image defined by fitting a subset of 75 bright, isolated and non-varying stars with \texttt{Astrophot}. We later use the 75 reference stars to standardize all images to a common zero point, by modelling the inter-image variation as the mean of all reference stars. \target\ and its nearest neighbours are modelled within a 25 $\times$ 25 pixel cutout from each image. We use robust sigma-clipped Gaussian statistics to estimate the local pixel-to-pixel variance. The flat sky background is set to the NOIRlab community pipeline instant calibration value computed for each exposure \citep{NoirlabCP}. In our \texttt{Astrophot} routine, the central intensity of the Moffat profile is the only free parameter, with size, shape, position, and background previously and independently determined. The output from this process is the total flux and flux uncertainty of both the neighbour and source stars. The joint modelling of the two stars produces a deblended light curve of \target\ that is free from neighbourhood contamination. A comparison between the \texttt{DoPHOT} and \texttt{Astrophot} photometry is shown in \ref{figCOMP}. The two photometry routines agree; the amplification, duration and symmetry of \target\ persist in both the \texttt{DoPHOT} and the detrended \texttt{Astrophot} light curves. Likewise, the neighbouring star is equally stable in the \texttt{DoPHOT} and \texttt{Astrophot} light curves. We conclude that the microlensing signal in \target\ is not impacted by systematic effects arising from PSF variations or blending. We highlight that the point-to-point variation is less in the \texttt{Astrophot} photometry and yields the optimal Signal-to-Noise measurements. For this reason, the microlensing modelling and analysis is performed on the \texttt{Astrophot} photometry.

\subsubsection{Typical Variable Sources}
Visual inspection of the light curve of \target\ rules out strong periodic variability that could mimic a microlensing signal. The absence of consistent variability in the nights before and after observation suggests that \target\ is unlikely to be a pulsating star or binary system. Additionally, the smooth and low-magnitude deviation of \target\ is unlike typical chaotic variability from accretion. Comparing \target\ to the accreting stars monitored by TESS \citep{Robinson_2022} with similar minute-wise cadence, shows that accretion will produce light curve deviations of several magnitudes on hourly to daily timescales. Such accretion onto circumstellar disks around young-stellar objects (YSO) \citep{Labdon_2021}, and binary companion accretion events \citep{Izzard_2017} are characterised by jagged outbursts superimposed on a continuous rise in luminosity \citep{Hillenbrand_2019, Hakala_2019}. Even intrinsic variability of the source star cannot replicate the clear microlensing signal of \target, as flickering from star spots is far smaller than the typical signal-to-noise background of ground-based observations \citep{Cranmer_2015}. Therefore, if the microlensing event of \target\ were produced by stellar variability, it would be a dampened and solitary event produced by a novel variability mechanism, and unlike any other present in the AMPM data set.

\subsubsection{A Discussion of Flare Stars}

A stellar flare is an energetic explosion from a star, typically resulting from magnetic reconnection in its convective envelope, leading to high-energy emissions across a broad spectrum \citep{Webb_2021}. The lifespan of a flare event is set by the combination of magnetic field strength, convective winds and external mass-transfer mechanisms between binary systems of stars \citep{Jackman_2021}. The flare profile as exhibited in a light curve can be broken into two stages centered around the peak brightness: the duration of the fast outburst ($t_{rise}$) and the duration of exponential decay into quiescence ($t_{decay}$) \citep{Kowalski_2024, Yan_2021}. As flare stars have a typical light curve signal with long periods of inactivity followed by sudden, often solitary brightening events, they might replicate microlensing events. The flare misattribution issue has been discussed for past microlensing events, namely the TESS microlensing candidate TIC 107150013 \citep{Kunimoto_2024, Mroz_TESS2024, Yang_2024}, and whether the TESS event could better be explained through the flare star mechanism. In the AMPM data, we do observe a significant number of flare events, with a total of 8 stars detected exhibiting 9 flare events in the quality data catalogue, with one star exhibiting a double flare. On December 18th alone, the AMPM detection pipeline identified 4 flare events. 

However, \target\ is strongly dissimilar to the classic flare shape, with a long, gradual rise in flux leading up to the event peak. To illustrate the difference between the microlensing and flare events, we imagine \target\ is a flare and compare the shape of \target\ to a variety of known flare events. We source flare stars from various catalogues through \texttt{LightKurve} \citep{lightkurve}. From the TESS flare catalogue \citep{Yang_2023}, we select nine stars with radii close to $2 R_{\odot}$ as analogues to the \target\ Star (see \ref{ssec: rstar} for the stellar radius determination). The set of TESS flares is directly comparable to the AMPM light curves as both surveys have similar minute-wise cadences. We additionally query ten Kepler flare events from the Kepler Flare Catalogue \citep{Davenport_2016} as examples of longer duration flare events, although these are measured with a slower 30-minute cadence. Finally, we collect all flare events detected from the AMPM survey. In total, 28 flare events are used to calculate exemplar values for $t_{rise}$ and $t_{decay}$, all of which are compared to the entire TESS catalogue of 60,810 events. 

We define a fitting function to measure the rising and decaying times of flares across different optical surveys. All light curves are initially converted to normalised flux measurements using the methods described in \citep{Yang_2023}, 
\begin{gather}
F_{ave} = \frac{(F_{max} + F_{min})}{2}\\
F_{norm} = \frac{(\mbox{Flux} + F_{ave})}{F_{ave}}\, .
\end{gather}
The time of peak flux ($t_{peak}$) is detected using a simple maxima query, and the light curves are trimmed around the peak time to 1 day for TESS light curves, 3 days for the Kepler stars and 1 night for AMPM events. To evaluate the quiescent magnitude of the star, we mask 6 before and 18 data points after the peak time, calculating the median magnitude from the remaining light curve. In the case of \target, the `flare' signal consumes most of the light curve, and so the median magnitude is evaluated from the first 50 data points.

\begin{figure}
\centering
\includegraphics[width = \columnwidth]{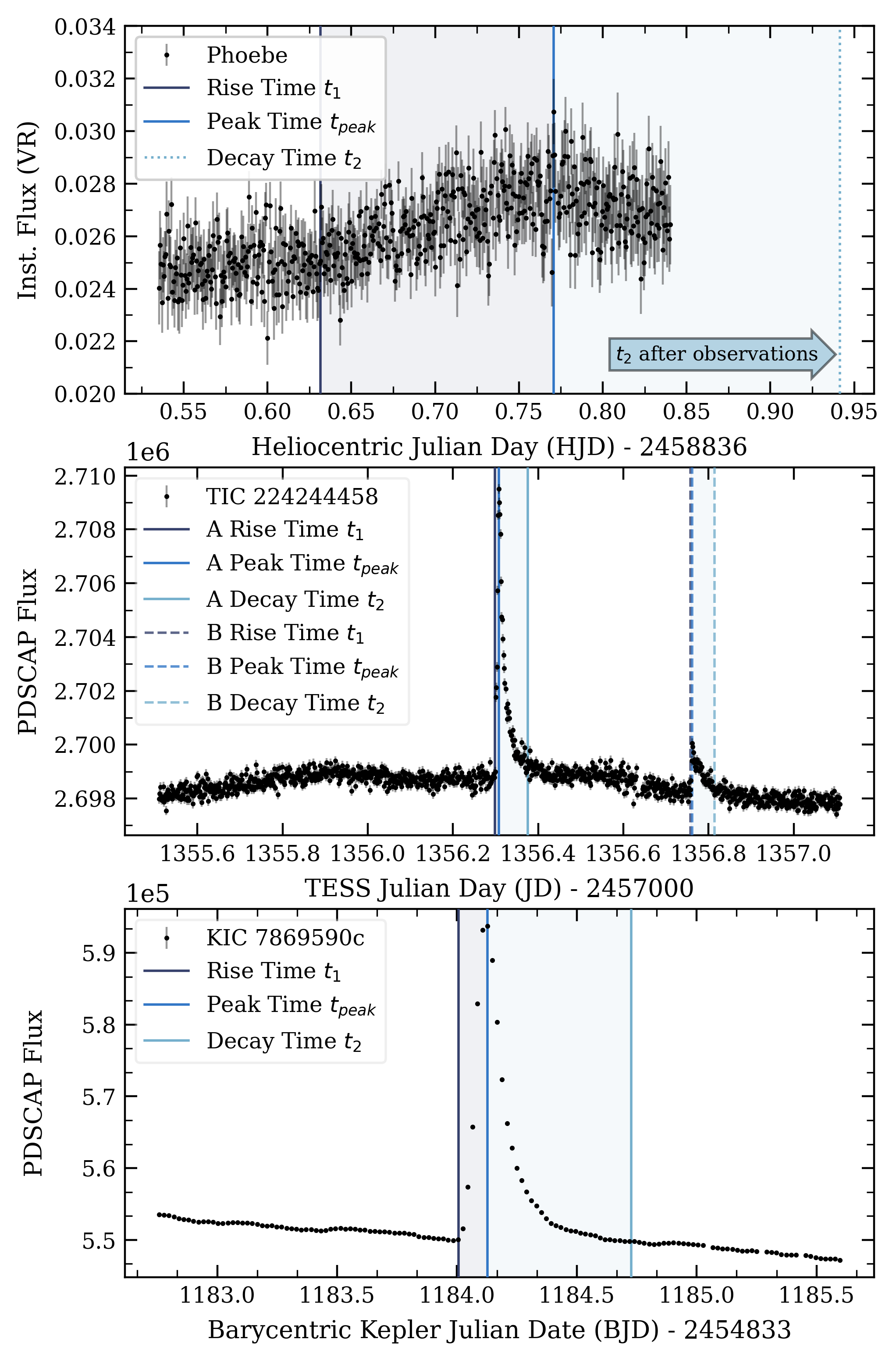}
\caption{Three examples of flare events with rise and decay times evaluated from the Flare Fitting Routine. Flare events are shown from AMPM 1-minute, TESS 2-minute and Kepler 30-minute cadence light curves.}\label{figflareexample}
\end{figure}

\begin{figure}
\centering
\includegraphics[width = \columnwidth]{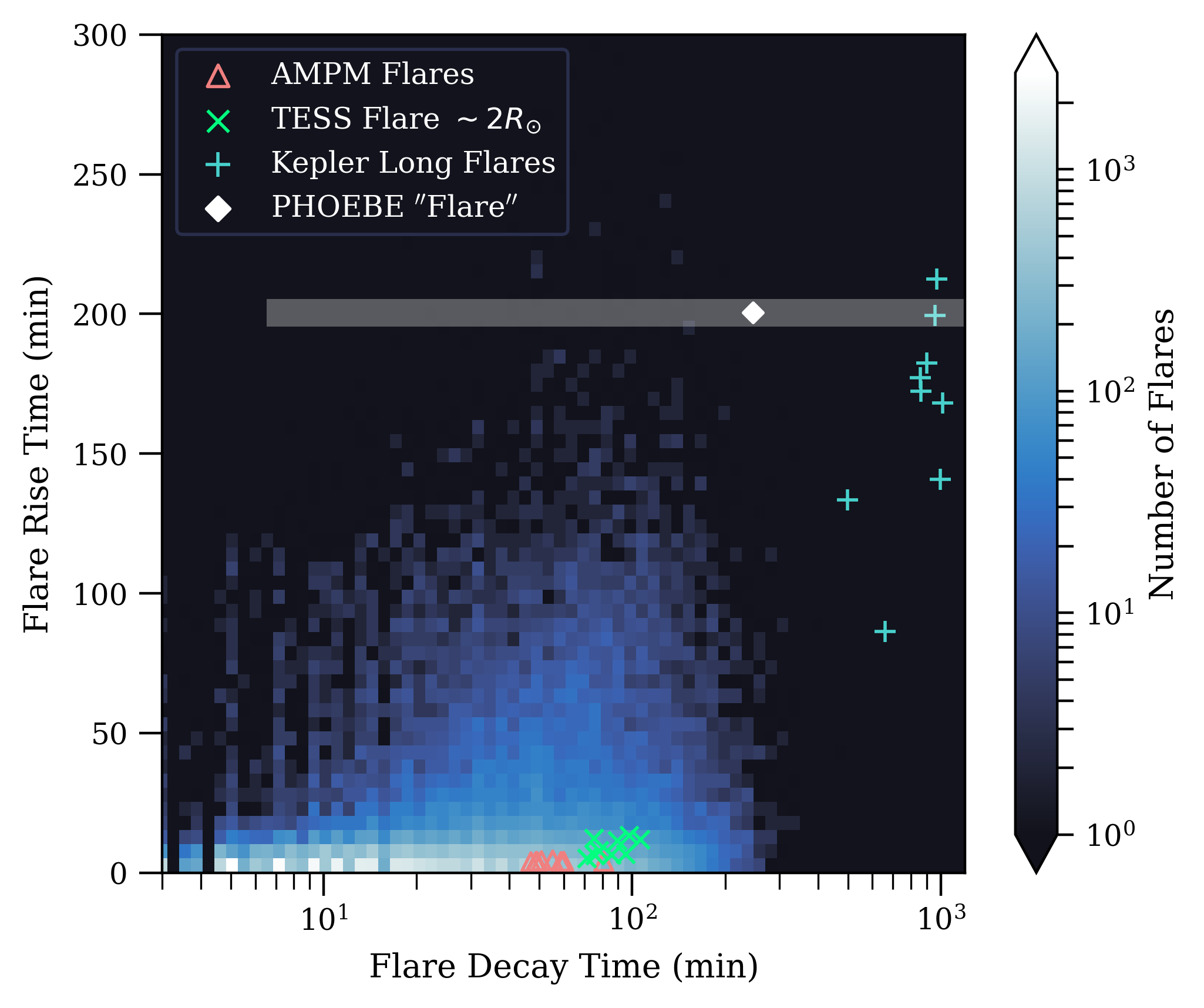}
\caption{Two dimensional distribution of flare rising and decaying times for the set of events in TESS 2-minute cadence data \citep{Yang_2023}. A broad peak in the number density of flare events spans across rising times of $1-10$ minutes, with decay times of 5-10 minutes. Shown separately (green crosses) to the TESS dataset are 8 examples of TESS flares from sub-giant stars with radii $\sim 2R_{\odot}$, similar to the \target\ Star. Kepler flare events in 30 minute cadence data are plotted as blue crosses, with comparatively long rising and decay times. The fast flare events from the AMPM survey are shown as pink triangles, and have a similar clustering in the time space to the TESS events. The discrepancy between the Kepler and TESS/AMPM flares is predominantly from the difference cadence of the observations. The \target\ microlensing candidate is shown as the white diamond at flare rise and decay times dissimilar to the average distribution of flares in fast cadence data. The uncertainty in $t_{decay}$ spans upwards from 90 minutes, and results from the scheduling constraints on the detection night. }\label{figflareloc}
\end{figure}

\begin{figure}
\centering
\includegraphics[width = \columnwidth]{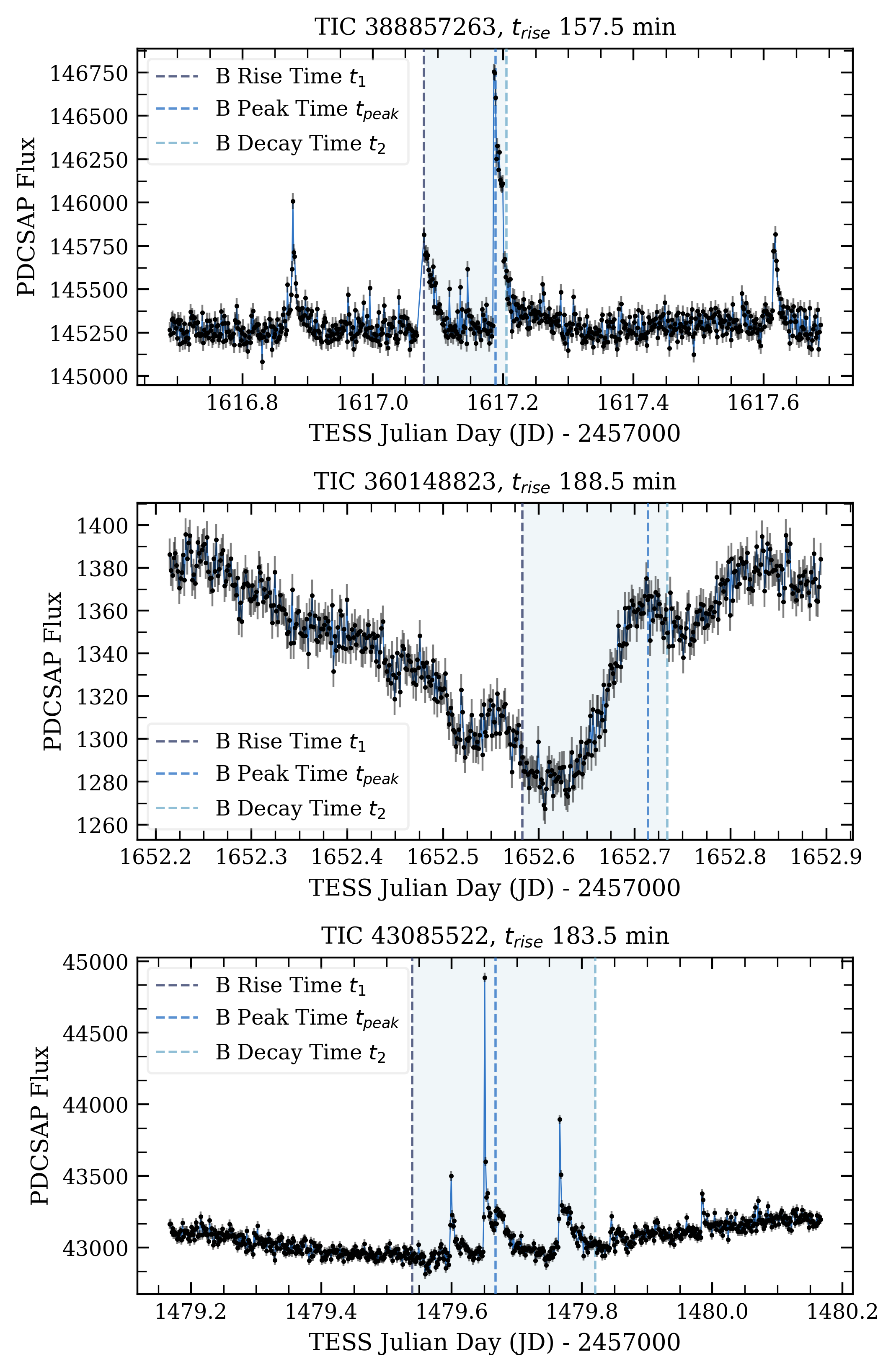}
\caption{Three examples of TESS flares from \citep{Yang_2023} with misidentified flare duration measurements compared with the true light curve structure. The three examples are recorded with excessively long flare rising times $t_{rise}$, and rise and decay ratios above 0.8, indicating a symmetric spread of the signal. The first and last flare examples are of complex, multiple flare events that have been identified as a single flare, with the very close multi-flaring peaks absorbed into a single duration measurement. The middle example shows a star with jagged and non-periodic variability that has been misidentified as a flaring event.}\label{figtesstoi}
\end{figure}

We fit two quadratic functions to each flare event. The first function measures the rising time, beginning from the first data point in the clipped light curves, and terminating at the peak flux. The second quadratic function measures the decaying time, starting at the peak flux and ending with the clipped light curve. The $t_{rise}$ modelling is constrained to a local region around the peak flux to prevent the fitting function from misidentifying other secondary flares as part of the main event. The quadratic functions work by finding the relevant root of the fitted polynomial to the time axis (t1, t2) as they intersect with the median magnitude of the light curve. The $t_{rise}$ and $t_{decay}$ durations are evaluated by subtracting the peak time from the roots and converting the durations into minutes. 
\begin{equation}
    t_{rise} = t_{peak} - t_{1}\;\;\;\;\;\;\;  t_{decay} = t_{2} - t_{peak}\, .
\end{equation}
We additionally record the ratio between $t_{rise}$ and $t_{decay}$ as $R$ to measure the relative symmetry of each flare signal, where values of $R \sim 1$ indicate that the event has equal outburst and decay durations. Most flare events will be positively skewed ($R \sim 0$), with the outburst time much quicker than the decay duration \citep{Pietras_2022}. The rise/decay algorithm can extrapolate start and end times between the finite measurements of the light curves, allowing projections forward and backward in time. Examples of the modelling process for \target\ and a TESS and Kepler flare are shown in Figure \ref{figflareexample}, which also illustrates the ability to project rise and decay times outside of the light curve array. Figure \ref{figflareloc} shows the subset of flare durations from TESS, Kepler, and AMPM plotted over the entire catalogue of TESS flares. \target\ is shown as a white diamond at $t_{rise} = 200.39$ minutes, $t_{decay} = 246.07$ minutes and $R = 0.81$, and sits well away from the median flare properties of $t_{rise} = 14.17$ minutes, $t_{decay} = 35.62$ minutes and $R = 0.33$. The rise/decay algorithm is able to project forward an end time to \target\ despite the lack of ground-based data taken during the day. However, the algorithm on \target\ has merely 90 minutes of light curve to inform the polynomial roots for $t_{decay}$, where the Tess and Kepler flares have the full duration of the flare available over several hours to days. Since the gradual tail of \target\ is missing, the $t_{decay}$ is likely overestimated, thus contributing to \target's skewed symmetry measurement of $0.81$. We indicate the broad uncertainty in the $t_{decay}$ of \target\ spans the available duration of \target's light curve from $t_{peak}$ and is shown as a white bar. 
Some Kepler flares have $t_{rise}$ and $t_{decay}$ times comparable to \target. However, flare events undergo rapid deviations in the shape of the profile, such that the Kepler 30-minute integration time has likely missed any complex fine structure \citep{Davenport_2014}. \citet{Yang_2018} estimate that flare durations measured from the 30-minute cadence Kepler light curves are overestimated by 50\% when compared to the same flare in 1-minute cadence light curves. Therefore, using long and quasi-symmetrical flares in longer cadence light curves to discredit a microlensing candidate in short cadence data ignores the intrinsic physical properties of a flare outburst. If \target\ was a flare, the minute cadence of AMPM will have detected the fast structure of the cataclysmic outburst. The complexity of long flare outbursts is best illustrated using the TESS events with $t_{rise}$ durations longer than 150 minutes and $R \geq 0.8$, with three examples shown in Figure \ref{figtesstoi}. The rise and decay times of these flares are drawn from the time measurements reported in the TESS flare database \citep{Yang_2023}. By eye-balling the light curves of these flares, as in Figure \ref{figtesstoi}, it is clear that the mechanism behind the increased $t_{rise}$ is either a misattribution of underlying stellar variability or the flare event is complex and consists of multiple consecutive outbursts \citep{Gunther_2020}. It is likely that the Kepler flares with the 30-minute sampling have absorbed some complex temporal information into a down-sampled data point, and have missed structure available present in the TESS and AMPM sampling. Given the difference in the light curve structure and event duration between \target\ and flare stars, we conclude that \target\ is not a flare event. However, high-resolution spectroscopy of the \target\ star will confirm whether it is consistent with typical M-Dwarf flaring stars, and additional rapid-cadence modelling of the star in multiple filters will help confirm or rule-out the microlensing nature of \target\ if additional, similar brightening periods are detected.

%%%%%%%%%%%%%%%%%%%%%%%% PHOEBE MODELLING %%%%%%%%%%%%%%%%%%%%%%%%%%%%%%%%%%%%
\section{The Microlensing Parameters of Phoebe}
\label{sec:parameters}
We use a Bayesian forward modelling process to determine the plausible mass range of \target. A crucial aspect of the modelling involves assumptions about the location of the lens; the dynamics of \target\ in space change depending on whether it is associated with the nearby stellar bulk of a galaxy or situated far out into the dark halo. In practical terms, we adopt appropriate spatial density profiles and velocity distributions to define the priors on the lens distance and velocity within the modelling. We apply three distinct scenarios of a lens co-located with the stellar component of the Milky Way (MW), that of the LMC, or a lens located within the overlapping dark matter halos of the MW and LMC. The three scenarios effectively constrain the mass of \target\ under the propositions that it is a planet arising from either the MW or LMC stars or a compact, non-baryonic component of the dark matter halo. The source's finite radius acts to dampen and broaden the microlensing amplification \citep{WittMao_1994, Smyth_2020}.
Therefore, we employ a realistic treatment of microlensing signal by modelling the source star with a finite radius \citep{Lee_2009}.
\subsection{The \target\ Source Star}
\label{ssec: rstar}
Using the finite source method to model microlensing relies on knowing the distance to the source and its size. We classify the source star by using colour-temperature relations to estimate the stellar radius. We note that the source is too faint to be detected by Gaia DR3, and consequently, we cannot retrieve parallax information.  However, in accordance with the very low count of foreground Galactic stars and the colour of the source, we assume it is situated within the LMC.
\newline\newline
We first cross-matched the source to SMASH DR2. Each SMASH entry is allocated an extinction value $E(B-V)$, based on the source location in the \citep{Schlegel_1998} extinction maps. For the source star $E(B-V) = 0.120$. We used the reddening correction method of DES \citep{Abbott_2018}, and using the $R_{V} = 3.1$ extinction law from \citep{Fitz_1999} to correct the magnitudes of the star.
The corrected colour and $g$ magnitude for the source are
$$
(g - r,\, g) = (0.152, 21.159) \pm (0.019, 0.012)\, .
$$
The effective temperature ($T_{\mbox{eff}}$) of the star is determined from both the polynomial relation between temperature, colour, and metallicity ($[Fe/H]$), and the SDSS g'-r' colour relations from \citep{Huang_2015}. The DECam \textit{u, g, r, i, z} filters are analogous in central wavelength and response to the SDSS filters of \textit{u', g', r', i', z'}. Therefore, we expect the SDSS empirical coefficients for $T_{\mbox{eff}}$ to function accurately for the SMASH DR2 filters. We assign the metallicity of the star to the mean LMC value $[Fe/H] = -0.42$ dex with $\sigma[Fe/H] = 0.04$ dex \citep{Choudhury_2021}. The effective temperature of the source star is $6500 \mbox{K}$.
\newline\newline
We estimate the radius ($R_{S}$) of the star from the bolometric luminosity $L_{\rm{bol}}/L_{\odot} = 4 \pi R_{S}^{2} \sigma T_{\mbox{eff}}^{4}$. An estimate of the surface gravity is necessary to construct the bolometric correction ($BC_{g}$) to the magnitude of the source. To encapsulate our uncertainty in the surface gravity of the star, we select values of log(g) between $[3.5, 4.5]$ in steps of 0.1 dex, which corresponds to a broad range in surface gravity around Main Sequence stars and solar analogues. For a star with $T_{\mbox{eff}} = 6500$ K, $A_{V} = 0.37$ mag, $[Fe/H] = -0.44$, and the array of $BC_{g}$ for the DECam g-band we interpreted for each value of surface gravity using the python package \texttt{isochrones} \citep{isochrones} in conjunction with the \texttt{MIST} synthetic photometry corrections \citep{Choi_2016} for the DECam \textit{g, r} filters. The variation in $BC_{g}$ is narrow despite the explored range in surface gravity, so we take the $BC_{g}$ as the median value of the sample, and the uncertainty as the standard deviation in the array. The luminosity of the star is then converted from,
\begin{equation}
    L/L{\odot} = 10^{-0.4(M_{\rm{bol}} - M_{\rm{bol},\odot})}\, ,
\label{eq:5}
\end{equation}
where $M_{\rm{bol}} = m_{g} + BC_{g} - 5 \log_{10}(d) + 5$, $d$ is the distance to the LMC, and $M_{\rm{bol},\odot} = 4.74$ \citep{Willmer_2018}. Finally, an estimate of the radius of the star is determined to be
$$
R_{S} = 2.3 \pm 0.1 \, R_{\odot} \, .
$$
Our aim in determining this estimate for the stellar radius is to find an appropriate value for the prior distribution on $R_{S}$. Although we quote the uncertainty in the stellar radius, we set a broad prior on the star in the MCMC sampling to encompass and marginalise over the uncertainties in the colour-temperature computation (such as the uncertainty in temperature, metallicity and surface gravity). We used a Gaussian prior on the source radius, centred on $2.3 \, R_{\odot}$, with a spread of  $1 \, R_{\odot}$. We additionally truncated the radius prior by enforcing bounds of $(0, 50] \, R_{\odot}$ to avoid excessively large or negative solutions. 
\subsection{Galactic Models}
\label{ssec:models}
We model the lensing system's density and velocity distributions to build informed priors for the microlensing parameters within the MCMC analysis. A galactic density distribution sets the prior on the distance to the lens ($D_{L}$) and source ($D_{S}$). We reproject the radial coordinates of the density models to the heliocentric frame pointing towards the central coordinates of Field 51, with the Sun's distance from the galactic centre $R_{\odot} = 8.2 \, \mbox{kpc}$ \citep{Gravity}.
The lens's distance and tangential velocity are drawn from a Gaussian prior around a Galactocentric rotational velocity component with velocity dispersion drawn from the literature. The relative tangential lens velocity is calculated by projecting to the lens plane,
\begin{equation}
    v_{\perp} = |(v_{lens} - v_{\odot}) + \frac{D_{L}}{D_{S}}(v_{\odot} - v_{source})| \, ,
\label{eq:6}
\end{equation}
where $v_{\odot}$ is the combined peculiar velocity of the Sun given by the Cartesian vector $(U, V, W) = (11.1, 12.24, 7.25) \pm (1.23, 2.05, 0.62) \, \mbox{km/s}$, with the velocity of the Local Standard of Rest (LSR) as $ V_{LSR} = 239 \pm 5 \, \mbox{km/s}$ \citep{Besla_2012}. We take the magnitude of the tangential velocity for the relative lens motion as the observer on Earth is unable to distinguish the lens' trajectory from photometry alone. Since a wide-orbiting planet is inherently bound to the stellar distribution, and FFPs are ejected away from stellar hosts at relatively slow speeds of $0.1-10\, \mbox{km/s}$ \citep{Wang_2015}, planetary microlenses generally trace the stellar density and velocity distribution \citep{DeRocco_2023}. Therefore, in the MW and LMC baryonic scenarios, we allocate the same prior distributions to both the lens and source.

\subsubsection{Dark Matter Distribution}
We use an NFW profile for the MW dark halo model \citep{Navarro_1997, Calcino_2018}. For the LMC dark halo, we use a pseudo-isothermal model with central density and scale parameters selected for the LMC surface brightness of $-17.9 M_{B}$ \citep{StaveleySmith_2003}.

\begin{equation}
   \begin{aligned}
    \varrho_{DM}(r) = \varrho_{\odot}\frac{\frac{R_{\odot}}{R_{c}}(1 + \frac{R_{\odot}}{R_{c}})^{2}}{\frac{r}{R_{c}}(1 + \frac{r}{R_{c}})^2}
     + \frac{\rho_{0,LMC}}{1 + (r/a_{LMC})^2} \, .
\label{eq:7}
\end{aligned} 
\end{equation}
\newline\newline
The NFW parameters are taken from Table 2 of \citep{Calcino_2018}, with solar DM density $\varrho_{\odot}=0.01058\,M_{\odot}\,\mbox{pc}^{-3}$, and scale radius $R_{c}=13.5\,\mbox{kpc}$. The LMC parameters are $\rho_{0,LMC}=0.01\,M_{\odot}\,\mbox{pc}^{-3}$ and $a_{LMC}=4.9\,\mbox{kpc}$ \citep{Kormendy_2016}.
\newline\newline
The velocity distribution for the joint dark matter distribution has a dispersion of $\sigma = 120 \,\mbox{km/s}$ \citep{Blaineau_2020, Battaglia_2005}. The velocity and dispersion values for the LMC dark halo are smaller than that for the Milky Way given the lower mass density of the satellite galaxy, with \citep{Kormendy_2016} measuring the typical dispersion at large radii of $M_{B} \sim -18$ spiral galaxies as $55 \mbox{km/s}$, which is comparable to the dispersion of $46 \mbox{km/s}$ in \citep{CalchiNovati_2006, Alves_2004} from a more condensed, spherical model of the LMC dark halo. However, we deliberately select the Milky Way velocity and dispersion to provide a loosely constrained prior for MCMC that more completely encapsulates the possible range of lens velocities between the observer and host star. 

\subsubsection{LMC Stellar Distribution}  \label{ssec:star}
We model the LMC stellar distribution as a double exponential disk using a reprojected coordinate system $(x', y', z', R' = \sqrt{x'^2 + y'^2})$ from the origin of LMC Field 51, that is aligned with the galaxy's inclination, $i = 25.86^{\circ}$, and position angle $\theta = 149.23^{\circ}$ \citep{Sajadian_2021, Gyuk_2000}. The centre of the LMC as $(\alpha_{0}, \delta_{0}) = (82.25^{\circ}, -69.50^{\circ})$, and the centre of LMC Field 51 is $(\alpha, \delta) = (89.97^{\circ}, -70.21^{\circ})$,
\begin{equation}
\begin{aligned}
\varrho_{LMC}(R', z') =\frac{M_{d}}{4 \pi z_{d}R_{d}^{2}}\,\mbox{exp} (\frac{-R'}{R_{d}})\,\mbox{exp}(\frac{-|z'|}{z_d})\, ,
\label{eq:8}
\end{aligned}
\end{equation}
The scale lengths are $R_{d} = 1.8  \mbox{kpc}$ and $z_{d} = 0.3 \mbox{kpc}$ %kpc
and the central mass density $M_{d} = 2.55 \times 10^{9} \, M_{\odot}$. We take the distance to the LMC as 50 \mbox{kpc}\citep{Pietrzyski_2019}. 
The LMC coordinate system transformation to $(x', y', z', R' = \sqrt{x'^2 + y'^2})$ is thoroughly described in \citep{Sajadian_2021, Weinberg_2001}.
\newline\newline
We exclude a bar component in the LMC stellar density model because Field 51 is sufficiently distant from the LMC's central bar, and disk stars dominate the stellar density in this region. There is evidence for the existence of an LMC halo population from measurements of the RR Lyrae velocity dispersion, which is elevated at $53 \pm 10$ \mbox{km/s} \citep{Alves_2004}, compared to the disk's velocity dispersion of $20.2 \pm 0.5$ \mbox{km/s} \citep{Marel_2002}. Additionally, stellar overdensity mapping from SMASH photometry suggests a low-density halo component at large radial distances, accounting for approximately 0.4\% of the total stellar mass \citep{Nidever_2019}. However, since Field 51 lies $2.75^{\circ}$ from the LMC center — well within the disk's density profile, which extends up to $15^{\circ}$ \citep{Nidever_2019} — and the disk stars dominate the stellar content at this distance, we do not model a separate LMC stellar halo. For the stellar velocity distribution, we adopt the parameters from \citep{Kallivayalil_2013}, with a galactocentric tangential velocity of $<v> = 314\,\mbox{km/s}$ and a velocity dispersion of $\sigma = 24 \,\mbox{km/s}$.

\subsubsection{MW Stellar Distribution}
The Milky Way stellar contribution is modelled with a double exponential thin and thick disk, plus a single power-law stellar halo density \citep{de_Jong_2010} as
\begin{equation}
\begin{aligned}
        \varrho(R, Z) = \varrho_{0, thin}\left (\mbox{exp} \left (\frac{R_{\odot}}{l_{1}} \right )\,\mbox{exp} \left (-\frac{R}{l_{1}} - \frac{Z}{h_{1}}\right) \right) \\
        + \varrho_{0, thick}\left (\mbox{exp}\left (\frac{R_{\odot}}{l_{2}} \right )\,\mbox{exp} \left (-\frac{R}{l_{2}} - \frac{Z}{h_{2}} \right ) \right) \\
        + \varrho_{0, halo} \left (\frac{R_{\odot}}{\sqrt{R^{2}+(\frac{Z}{q_{h}})^{2}}} \right  )^{n_{h}} \, ,
\label{eq:9}
\end{aligned}
\end{equation}
where the central densities of the thin, thick disc and halo are $\varrho_{0, thin} = 0.038 \, M_{\odot}\mbox{pc}^{-3}$, $\varrho_{0, thick} = 0.005 \, M_{\odot}\mbox{pc}^{-3}$ and $\varrho_{0, halo} = 6.61 \times 10^{-5}\, M_{\odot}\mbox{pc}^{-3}$ \citep{de_Jong_2010}. The Sun's position above the Milky Way plane is $Z_{\odot} = % 20.5 parsecs Humphreys & Larsen 1995
2.5 \,\mbox{kpc}$, and the thin and thick disc scale parameters are respectively $(l_{1}, h_{1}) = (2.6, 0.25)\, \mbox{kpc}$ \citep{Juri_2008} and $(l_{2}, h_{2}) = (4.0, 0.75) \,\mbox{kpc}$. The stellar halo parameters are $(q_{h}, n_{h}) = (0.85, -2.8)$ \citep{de_Jong_2010}. 
\newline\newline
We set the stellar velocity distribution to $ \sigma = 75\,\mbox{km/s}$. The multiple components of the galactic density have defined rotational velocities and dispersion \citep[e.g.][]{Robin_2017}, but we have no knowledge of the distinct lens location and density membership. Therefore, we construct a broad velocity prior so the MCMC walkers can explore the varied velocities of disc and halo stars.

\begin{figure}
\centering
\includegraphics[width = \columnwidth]{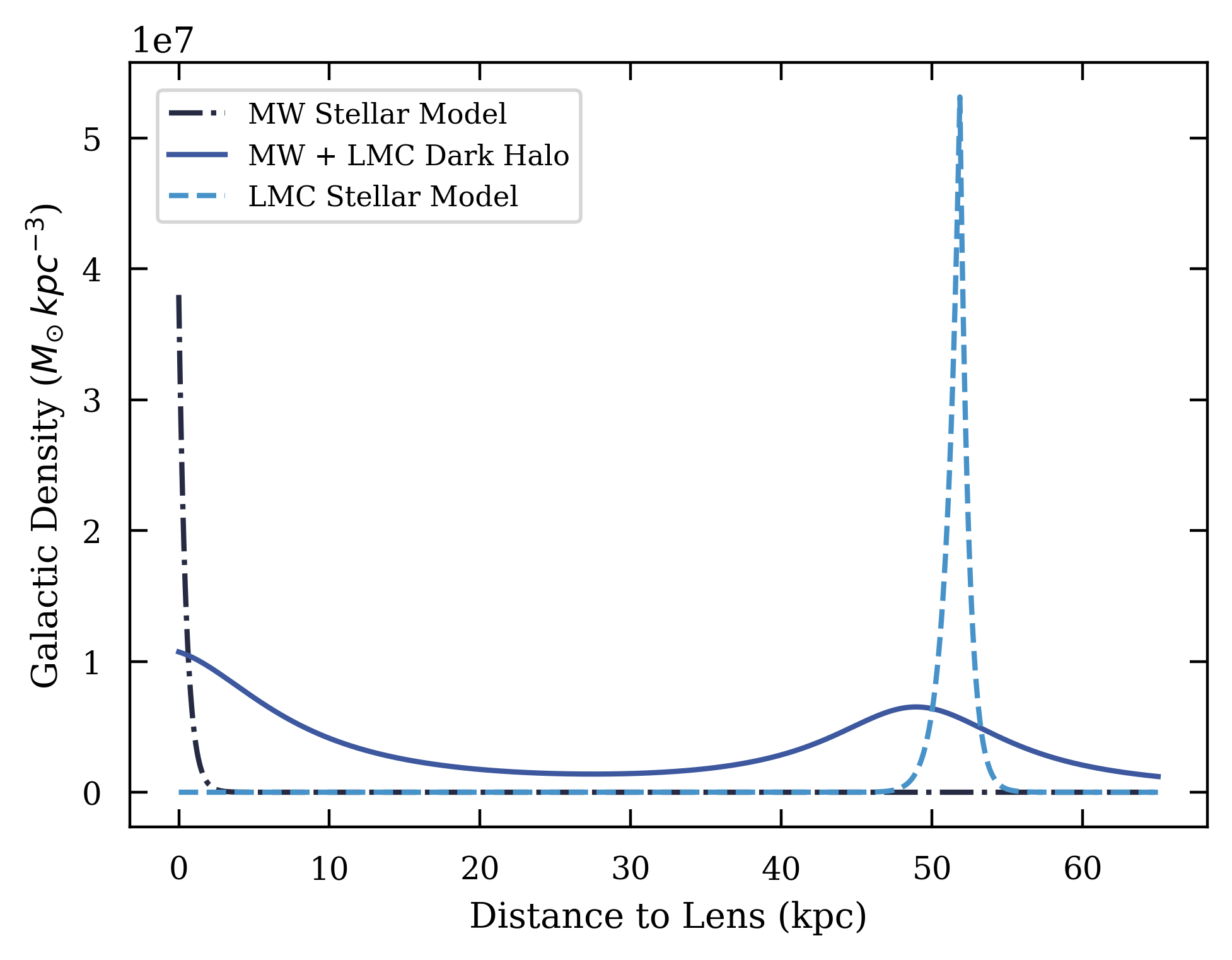}
\caption{The three density distribution functions for the MW+LMC dark matter, MW stellar and LMC stellar models, as defined in Section \ref{sec:models}. The density distribution functions are defined for a source star at the coordinates of \target\, along a dark matter-dominated sight line towards the LMC. The density function defines the priors for the location of the microlens and source along the line-of-sight.} \label{figDEN}
\end{figure}

\subsection{MCMC Microlensing Modelling}
We use MCMC sampling within the \texttt{emcee} \citep{emcee_2013} framework in conjunction with \texttt{MulensModel} \citep{Software_MulensModel} to explore a wide parameter space and $A_{FS}$ from Equation \ref{eq:AFS} amplification solutions for \target. We ran three MCMC samplers, with each sampler exploring one of the three astronomical models. Each sampler ran with 100 walkers for 200,000 steps, with a 20,000 step burn-in. Each step fed the relevant parameters into the \texttt{MulensModel} implementation of the finite source single lens model of \citep{Lee_2009}, and weighs a global $\chi^{2}$ of the parameter representation to the \texttt{Astrophot} light curve of \target. The LMC and MW baryonic and LMC and MW joint dark matter distributions define the priors for the lens and source distance and velocity, with the density distributions along the LMC line-of-sight shown in Figure \ref{figDEN}. Additionally, the priors for the distance of the lens and source are bounded by two conditions: the lens distance cannot be further towards the LMC than the source distance, and the source cannot be more than 55 \mbox{kpc}. The full description of the bounds of the priors on each parameter are summarised in Table \ref{tab:PriorTab}.  A final, separate MCMC sample ran on the \texttt{Astrophot} light curve with flat, uninformative priors to determine the four FS$-$PL event parameters, $\rho$, $t_{E}$, $t_{0}$ and $u_{0}$ as a comparison to the output from the three galactic scenario models. 

\begin{figure}
\centering
\includegraphics[width=1.\columnwidth]{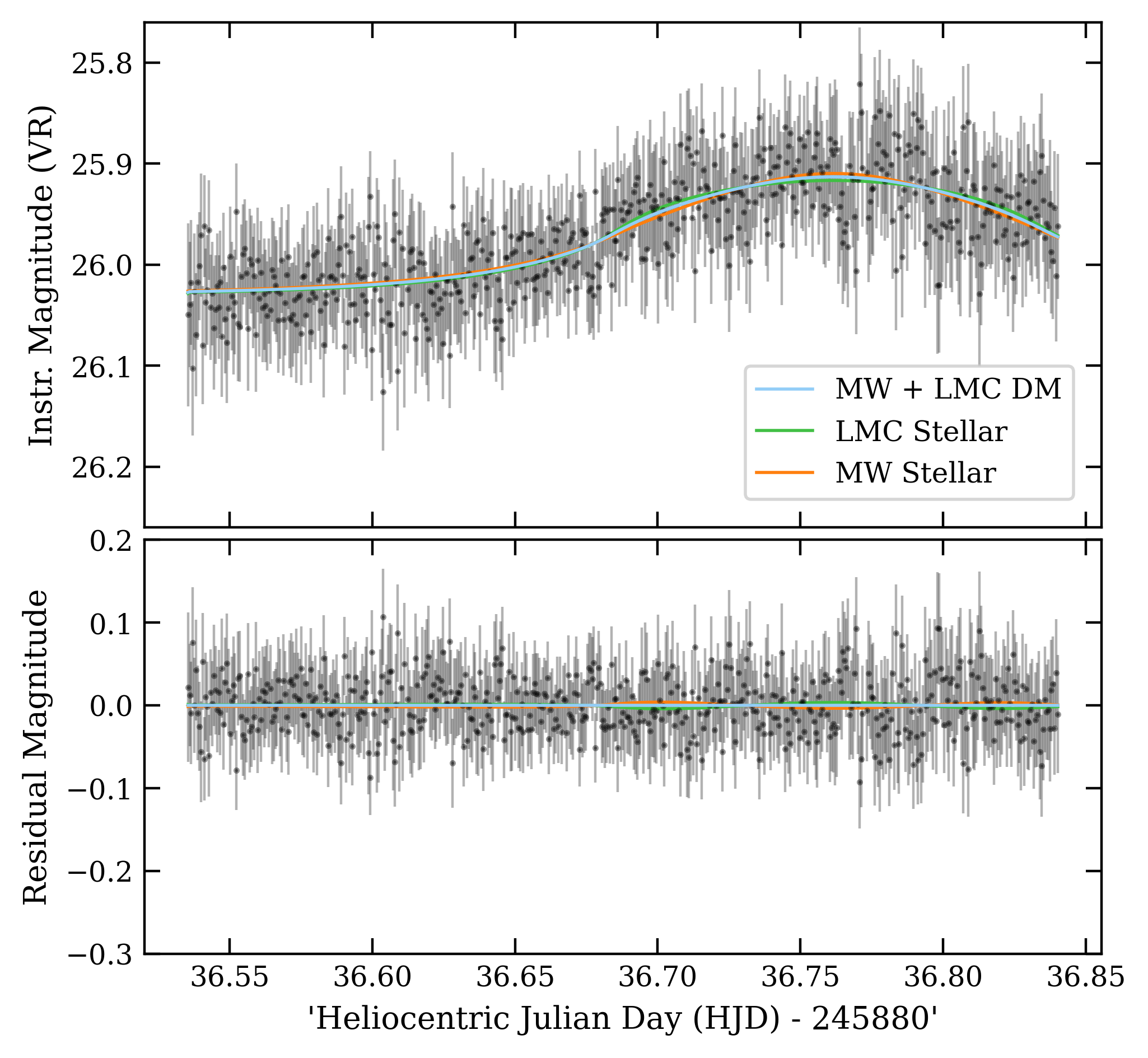}
\caption{The \texttt{Astrophot} light curve and microlensing best-fit models. The December 18th light curve with \texttt{Astrophot} \citep{Astrophot} photometry is shown with the three finite-source point-lens microlensing models generated with the lens and source star parameters with maximum likelihood in the MCMC analysis. The top panel shows the detection light curve with the MW+LMC dark matter PBH model shown in blue. As the $1\sigma$ deviation on the light curve solutions is of order $\mathcal{O}(10^{-4})$, we do not show the error margins. The MW and LMC stellar/planetary microlensing models are shown in orange and green, respectively. In the bottom panel, we show the residuals of the data to the dark matter (blue) model. We additionally show the residuals of the stellar models to the dark matter model. It is clear from both the strong similarity in the residual models that the maximum likelihood solutions for the three scenarios are a comparably good fit to the data. } \label{figMODELLC}
\end{figure}

\begin{figure}
\centering
\includegraphics[width = \columnwidth]{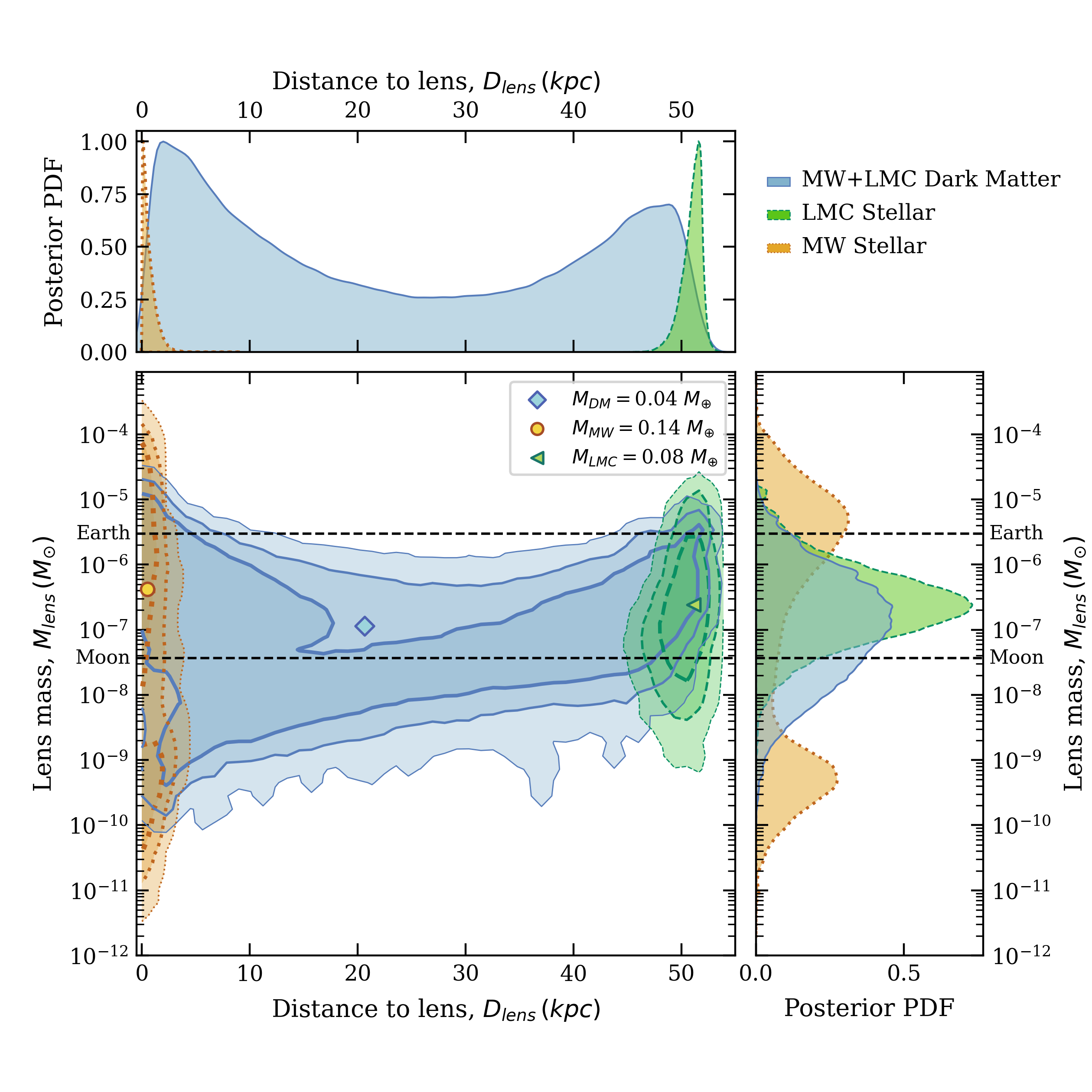}
\caption{The lens mass and distance contours from the MCMC analysis. The uppermost plot shows the Kernel Density Estimation (KDE) of the posterior distribution for the lens distance. We normalised the peaks of the distance distributions to improve the comprehensibility of the top panel. The middle image shows the contours of lens mass and distance, with the median value for lens mass and distance indicated with markers for the three galactic scenarios (the values of the lens masses are included in the legend).  The right-hand plot shows the KDE of the posterior PDF of lens mass. No relative scaling has been applied to the mass PDFs. In each galactic scenario, the median mass is always sub-terrestrial, as indicated by the dashed black lines representing the Earth ($M_{\oplus}$) and Moon masses.}\label{figCON}
\end{figure}

\renewcommand{\arraystretch}{1.5} % Default value: 1
\begin{table}[h]
\small
\centering
\begin{tabular}{ |l|l|}
\hline
\textbf{Parameter} &  \textbf{Prior Definition} \\\hline\hline
Mass (M) & Logarithmic Flat,  $0 < M \leq 100 M_{\odot}$ \\\hline
Lens Distance (DL) & See sec. \ref{sec:models}, $0 < D_{L} < D_{S}$ kpc \\\hline
Source Distance (DS) & See sec. \ref{ssec:star}, $ 0 < D_{S} \leq 55 $ kpc \\\hline
Dark Halo Velocity & V $\in N(220,120)$  km/s \\\hline
MW Stellar Velocity & V $\in N(220,75)$ km/s \\\hline
LMC Stellar Velocity & V $\in N(314,24)$ km/s \\\hline
Source Radius (RS) & $R_{S} = N(2.3, 1)\,$,  $0 < R_{S} < 450\, R_{\odot}$ \\\hline
Impact Parameter ($u_{0}$) & Flat, $0 < u_{0} < 10 $ \\\hline
Peak time in MJD ($t_{0}$) & Flat,  $58836.03577 \leq t_{0} \leq 58836.34060$ \\\hline
\end{tabular}
\caption{The definitions for the informed priors in the MCMC sampler with the FS-PL microlensing model. The function $N(v, \sigma)$ refers to the normal distribution of mean $v$ and spread $\sigma$.} \label{tab:PriorTab}
\end{table}
Figure \ref{figMODELLC} shows how each of the best-fit models from the three scenarios gives an equally good description of the data. Based on the DECam light curve alone, we cannot discriminate between the scenarios. However, we use Bayesian analysis to visualise the consequence of each scenario on the mass and distance of \target. We plot the 1, 2, and 3$\sigma$ mass-distance contours for the three galactic scenarios in Figure \ref{figCON}, with the median mass and distance indicated in panel (b). The posterior probability density function (PDF) for lens distance and mass are shown in top and right-hand panels. In the MW+LMC dark matter model, very near lens distances ($\lesssim 300$ pc from the Sun) are disfavoured. The MW and LMC stellar density models provide strongly localised posterior distance distributions due to the lower stellar density in each galaxy along the AMPM sight line. The influence of the LMC dark matter contribution with peak density at a distance of $\sim50 \mbox{kpc}$ draws the median distance solution for the MW+LMC dark matter model further into the halo of the Milky Way. We present the $16^{\mbox{th}}, 50^{\mbox{th}}, 84^{\mbox{th}}$ percentiles as the equivalent of the $-1$, $0$, and $+1 \sigma$ points of the posterior PDF for the three galactic scenarios in Table \ref{tab:results}. Each of the three scenarios results in sub-terrestrial mass scales, agreeing to within an order of magnitude for the lens mass. The microlensing parameters for each of the galactic models agree to within the uncertainties on the flat, uninformed MCMC analysis. The corner plots from each microlensing scenario are shown in Appendix \ref{appendix:Appendix}, with parameters quoted to the $16^{\mbox{th}}, 50^{\mbox{th}}, 84^{\mbox{th}}$ percentiles.

\renewcommand{\arraystretch}{1.5}

\begin{table*}
\small
\begin{tabular}{|l|lll|}
\hline
\textbf{Parameter} & \textbf{DM Lens} &  \textbf{LMC Stellar Lens} &  \textbf{MW Stellar Lens} \\\hline\hline
Lens Mass, $M_{L}$ ($M_{\oplus}$) & \multicolumn{1}{l|}{$0.026_{-0.023}^{+0.194}$} & \multicolumn{1}{l|}{$0.100_{-0.067}^{+0.192}$} &  $0.022_{-0.022}^{+1.859}$\\\hline

Lens Distance, $D_{L}$ (kpc)  & \multicolumn{1}{l|}{$18.2_{-14.1}^{+24.7}$} & \multicolumn{1}{l|}{$51.3_{-0.7}^{+0.5}$} & $0.6_{-0.4}^{+0.7}$\\\hline

Lens Velocity, $V_{L}$ (km/s) & \multicolumn{1}{l|}{$287_{-97}^{+103}$} & \multicolumn{1}{l|}{$311_{-40}^{+37}$} &  $124_{-7}^{+121}$ \\\hline

Source Distance, $D_{S}$ (kpc)  & \multicolumn{1}{l|}{$51.8_{-0.8}^{+0.7}$}  & \multicolumn{1}{l|}{$52.0_{-0.5}^{+0.8}$} &  $51.6_{-1.0}^{+0.7}$\\\hline

Source Velocity, $V_{S}$ (km/s)  &\multicolumn{1}{l|}{$313_{-25}^{+22}$} & \multicolumn{1}{l|}{$ 316_{-37}^{+35}$} &  $316_{-27}^{+26}$ \\\hline

Source Radius, $R_{S}$ ($R_{\odot}$) & \multicolumn{1}{l|}{$2.26_{-0.77}^{+0.93}$} & \multicolumn{1}{l|}{$0.99_{-0.42}^{+0.43}$} &  $2.55_{-0.87}^{+0.94}$ \\\hline

Peak Time (HJD), $t_{0}$ & \multicolumn{1}{l|}{$58836.2616_{-0.0038}^{+0.0039}$}  & \multicolumn{1}{l|}{$58836.2615_{-0.0035}^{+0.0039}$} &  $58836.2618_{-0.0036}^{+0.0033}$\\\hline

Impact Parameter, $u_{0}$ &   \multicolumn{1}{l|}{$3.06_{-1.49}^{+0.14}$}&  \multicolumn{1}{l|}{$3.15_{-0.16}^{+0.09}$} &  $1.59_{-0.05}^{+1.58}$  \\\hline

Source-Lens Radius, $\rho$ &   \multicolumn{1}{l|}{$3.49_{-3.30}^{+0.33}$}& \multicolumn{1}{l|}{$ 3.62_{-0.35}^{+0.20}$} &  $ 0.20_{-0.18}^{+3.51}$\\\hline

Einstein Time, $t_{E}$ (days) & \multicolumn{1}{l|}{$ 0.045_{-0.009}^{+0.012}$} &  \multicolumn{1}{l|}{$ 0.042_{-0.006}^{+0.008}$} &  $0.054_{-0.014}^{+0.005}$\\\hline

Einstein Radius, $\theta_{E}$ ($\mu as$) &\multicolumn{1}{l|}{${0.074}_{-0.023}^{+0.89}$}&  \multicolumn{1}{l|}{${0.026}_{-0.009}^{+0.012}$} &  ${1.145}_{-1.076}^{+10.410}$\\\hline\hline
\multicolumn{4}{|l|}{\textbf{Flat Model}: $t_{0} = 58836.2615^{+0.0040}_{-0.0033}$ \, $u_{0} = 3.575^{+0.2447}_{-1.956}$ \,  $\rho = 3.114^{+0.1135}_{-1.131}$ \, $t_{E} = 0.0434^{+0.01045}_{-0.006987}$ }\\\hline
\end{tabular}
\caption{Table of best-fit Microlens and Source parameters for \target. The microlensing parameters are quoted with $16^{\mbox{th}}, 50^{\mbox{th}} , 84^{\mbox{th}}$percentiles of the posterior probability distribution. We display the best-fit parameters for the three galactic models as described in Section \ref{ssec:models}, and the additional derived microlensing parameters $\theta_{E}, \, t_{E}, \, \rho$, as defined in Section \ref{sec:maths}. The microlensing parameters from the separate, uninformed MCMC analysis are also reported in the last row.} 
\label{tab:results}
\end{table*}

%%%%%%%%%%%%%%%%%%%%%%%% PHOEBE CONSEQUENCE %%%%%%%%%%%%%%%%%%%%%%%%%%%%%%%%%%%%
\section{The Origins of Phoebe} \label{sec:Origins}
\begin{figure}
\centering
\includegraphics[width=1.\columnwidth]{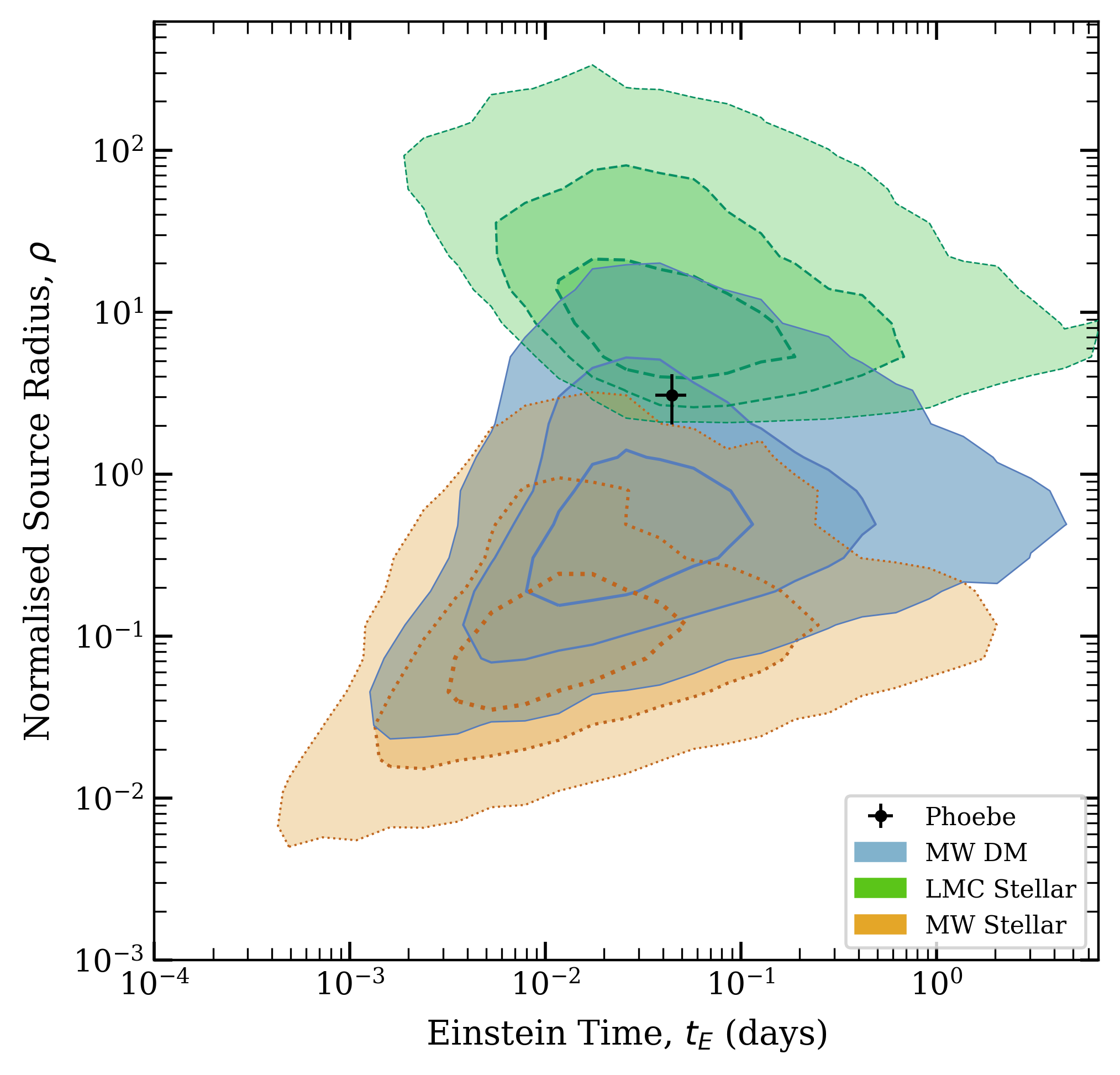}
\caption{The distribution of the microlensing event parameters of timescale ($t_{E}$) and the finite source ratio ($\rho$) for microlenses in the Milky Way stellar disk (in green), Milky Way dark matter (in blue) and Large Magellanic Cloud stellar disk (in orange). The distributions are generated for a microlens mass of $M=10^{-7}M_{\odot}$, similar to the best-fit mass of the PBH candidate, \target. The values of $t_E$ and $\rho$ for \target\ as evaluated from flat priors on the parameters are plotted with error margins in black. Naively, it seems as if the \target\ event is likely part of the blue dark matter distribution, but the event parameters also coincide with the $2\sigma$ contours for the LMC. Many more observations of microlensing events will help build robust population statistics and potentially indicate the true nature of the events. } \label{concldists}
\end{figure}

In this section, we discuss the two possible compact objects that could produce a microlensing event with the hour-long timescale of \target. It is necessary to remember that in the design of this MCMC modelling of the \target\ event, we have a global, philosophical `hyperprior' that assigns equal probability of existence to both the FFP and the PBH. The existence of one does not negate the existence of the other. In our model, the Universe is equally capable of producing both objects. Of course, the FFP may have a higher occurrence within the stellar density MW, as evidenced by the larger frequency of microlensing events within the bulge fields of MOA \citep{Sumi_2023, Nunota_2025}, KMTnet \citep{Qian_2025} and OGLE \citep{Ban_2016}. PBHs, on the other hand, should be more numerous where the dark matter density dominates in the dark halo. In this way, the population number density guides the event rate for both FFPs and PBHs, and the number density naturally varies along different lines of sight into (and outwards of) the Milky Way. 
\newline\newline
One interpretation of \target\ is a microlensing planet attached to the stellar density of the MW or LMC. A microlensing event from a foreground system of a planet orbiting a host star produces two lensing signals \citep{Han_2022}; a long-duration host event (often several days) and the shorter perturbation from the planet \citep{Wang_2021}. Due to the absence of a host event during December 14-19, we suggest that \target\ if it were associated with a stellar aspect of the galaxies is either an isolated, free-floating planet  \citep{Sumi_2023, Ryu_2021}, or a very wide-orbiting planet \citep{Wang_2021, Poleski_2021}. The vast majority of microlensing FFPs have been detected towards the MW galactic centre, as the stellar density is high and consequently, planetary events are more numerous. Two sub-terrestrial mass FFPs have been discovered towards the Galactic Bulge with Einstein timescales of $t_{E} = 0.0288^{+0.0024}_{-0.0016}$ days corresponding to a $0.3 M_{\oplus}$ lens as discovered by OGLE III \citep{Mroz_FFP12020} and $t_{E} = 0.057^{+0.016}_{-0.016}$ days with lens mass modelled between $0.37 - 0.75 M_{\oplus}$ from MOA II \citep{Koshimoto_2023}. 
It is interesting that the Einstein timescale of \target\ belonging to the MW is similar to the two sub-terrestrial events from MOA II and OGLE III, being $t_{E} = 0.054^{+0.005}_{-0.014}$. However, the AMPM sightline points well away from the Galactic Bulge. The increase in the available lens distances $D_{L}$, coupled with the different velocity distribution for disk and halo objects produces different mass profiles at the same time scales when compared to the MW bulge fields. If \target\ is in the MW, it is $565^{+661}_{-357}$ pc away from the Sun, which places it in the thick disk of the galaxy \citep{Vieira_2023} with mass $0.022^{+1.859}_{-0.022} M_{\oplus}$. If \target\ is part of the stellar density of the MW, it is among the fastest FFP detections, and is modelled with the lowest FFP mass to date, however we note that the tail of the mass distribution is skewed sufficiently to allow this event to be a terrestrial-mass FFP, as shown in the corner plots in the Appendix \ref{appendix:Appendix}. If \target\ is in the MW, this scenario results in a similar mass at $0.022 M_{\oplus}$ to the Dark Matter scenario at $0.026 M_{\oplus}$. 
\newline\newline
If \target\ is part of the LMC stellar density, it retains its FFP status, but given the altered lens velocity and distance distributions, has a completely different mass profile to the MW FFP model. 
Within the LMC, \target\ has a much higher maximum likelihood mass of $0.100 M_{\oplus}$. If in the LMC, \target\ is the first extragalactic microlensing exoplanet, and sits $700pc$ in front of the source star with mass $0.100^{+0.192}_{-0.067} M_{\oplus}$.
\newline\newline
If \target\ is instead within the dark matter halo, it is located well past the stellar disk of the MW at a distance of $18.2^{+24.7}_{-14.1}$ kpc. \target\ is a compact, self-gravitating object and virtually certain to be non-baryonic as the optical depth for baryons in the halo will be many orders of magnitude smaller than the naive estimate for the dark matter. Within the halo mass distribution, baryons make up less than 0.001\% of the mass \citep{Binney_2023} and are predominantly stars, with only a tiny mass contribution from sub-terrestrial planets. There exists a diverse range of proposed non-baryonic dark microlenses, including PBHs and axion stars \citep{Sugiyama_2023, Fujikura_2021}, however we argue that a PBH is the more natural interpretation. Moreover \target\ cannot be a stellar-remnant black hole, with typical minimum mass $\sim 5 M_{\odot}$ \citep{Ozel_2010}, as evaporation timescales cannot dissipate a stellar remnant into \target\ with mass of $M = 0.026^{+0.194}_{-0.023} M_{\oplus}$ (or $2.11$ lunar masses) within the lifetime of the universe \citep{Carr_2020}. If \target\ is a PBH, it was formed before Big Bang Nucleosynthesis, and so would be among the oldest astronomical objects yet found and offers a window into the inflationary universe. Recently Subaru-HSC has presented 12 short-duration PBH candidates with likelihood mass modelling that yields PBH masses very similar to that of \target, with the median likely mass at $10^{-7} M_{\odot}$, and the lowest mass candidate of $10^{-8} M_{\odot}$ \citep{Sugiyama_2026}.
\newline\newline
However, with only one detection from the AMPM survey, we do not have significant detection statistics to trace the velocity and distances of the underlying population of FFPs or PBHs. We cannot state the true nature of \target. Therefore, to make an estimate of \target's nature we must compare the relative likelihoods of the three possible scenarios,  MW FFP, an LMC FFP or a PBH. Using Equation \ref{eq:tau}, we calculate the optical depth of \target\ with an amplification threshold of $A_{T} = 1.1$ equal to the peak amplification of the \target\ event. We use the three galactic models as defined for the MCMC analysis in Section \ref{sec:parameters}, and generate both the PS$-$PL and FS$-$PL optical depths for a range of asteroid- to terrestrial-mass microlenses, and for the maximum likelihood solution for the distance and radius of the \target\ Star in each scenario. We also plot the maximum likelihood mass of each galactic model on the optical depth curves. Our optical depth estimation favours the dark halo by over 5 orders of magnitude compared to the combined MW and LMC stellar models. This result is purely due to the higher intervening dark matter density along our line-of-sight. However, the optical depths quoted for the MW and LMC stellar density distributions overestimate the contribution of sub-terrestrial planetary lenses. Formally, the number density in Equation \ref{eq:tau} refers to the lensing population, while we have used the source population as a proxy for planetary density. Although the number of sub-terrestrial FFPs per star is estimated to be as high as $19^{+23}_{-12}$ \citep{Sumi_2023}, the fractional difference between the masses of planetary lenses and stars is small ($\sim 10^{-7}$), which accordingly reduces the planetary density contribution in the optical depth calculation by many more orders of magnitude. Therefore, the optical depths shown here for the MW and LMC stellar models are inflated, and hence conservative, upper estimates for planetary lensing probabilities. A more accurate density contribution from sub-terrestrial planets external to the Galactic bulge may be confirmed as more microlenses are discovered in the coming decades. From a comparison of the microlensing probabilities, within the lunar mass range of \target\ it is far likelier that it is a PBH than a FFP.
\begin{figure}
\centering
\includegraphics[width=1.\columnwidth]{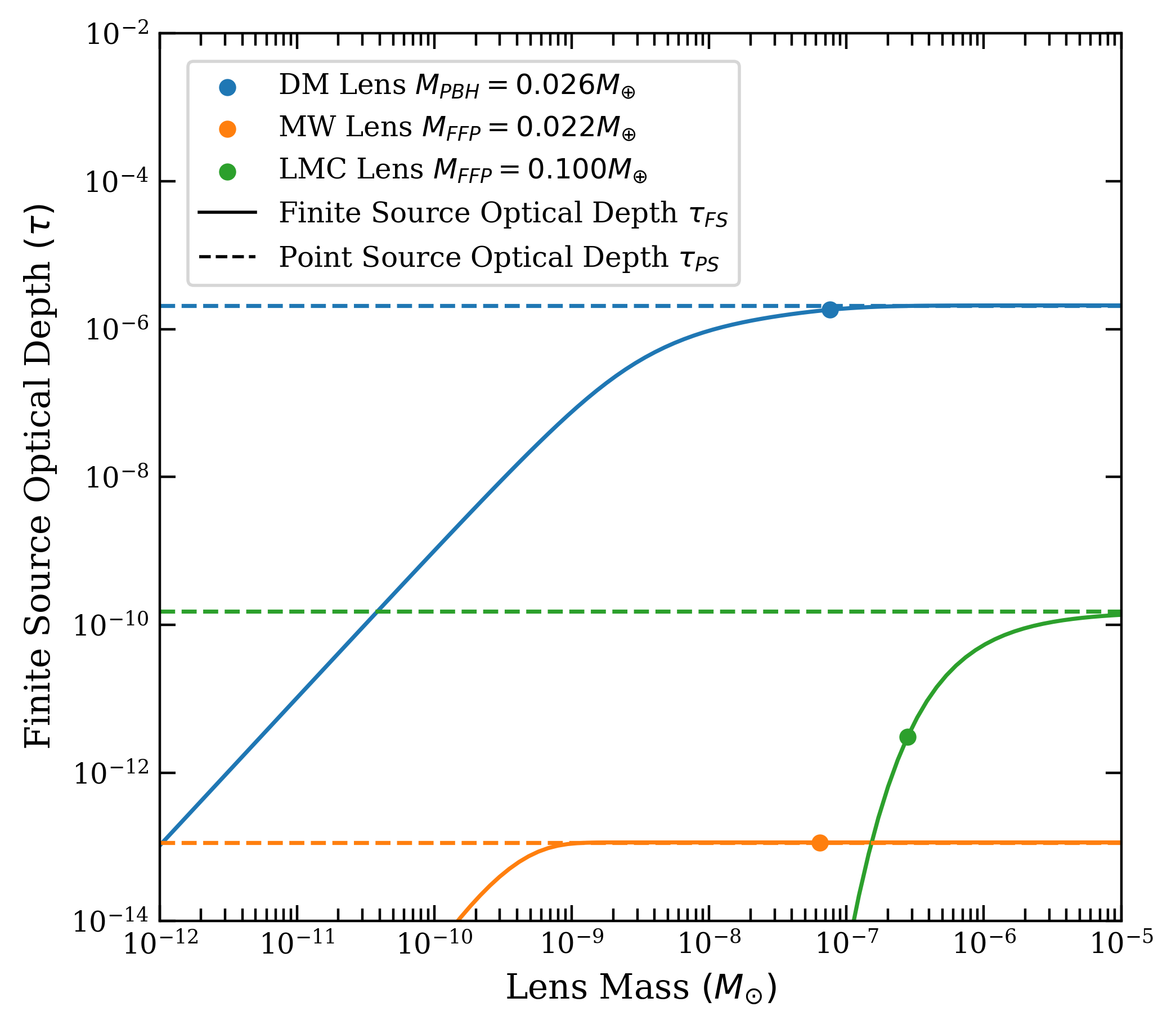}
\caption{The optical depth probability of \target\ belonging to the three microlensing scenarios of MW+LMC dark matter, MW stellar density, and LMC stellar density. The solid lines present the mass-dependent optical depth for finite source microlensing, while the dashed lines are the optical depths for a point-lens point-source approximation. The markers on each optical depth curve are the value of $\tau$ for the maximum likelihood mass solution in each galactic model, as stated in Section \ref{tab:results}.} \label{figOD}
\end{figure}
The optical depth analysis is formally a comparison of probabilities, and without multiple lunar-mass microlensing detections we cannot fix the underlying compact object population behind \target. Although the discovery of FFPs across the MW bulge and disk fields might tempt the reader to consider \target\ as another FFP, we caution that no additional confirmation of the planetary status of archival FFPs has been obtained. This is the bane of microlensing experiments; dark compact objects can be detected through lensing where other methods (radial velocity, transits) fail, but as a consequence, the true nature of the microlens remains hidden. It is entirely possible that a portion of the archival FFP events are in fact PBHs, as considered by \citep{Niikura_OGLE2019}. Until adjacent confirmation of the planetary structure of these FFPs is published, the underlying population of all microlenses must be decided through the comparison of baryonic and dark matter structures across galactic sightlines. The only way to confirm the PBH nature of the AMPM and Subaru-HSC microlensing events is to continue the current momentum of microlensing surveys and propose new experiments as the next generation of telescopes comes online. While these candidate detections are galvanising for the PBH field, the only way to unequivocally disentangle the PBH microlensing population from the planetary microlenses is to find many events across a vast range of sight lines. Many microlens detections towards the galactic bulge, the Milky Way disk (low and high in the plane) and the dark matter-dominated central and peripheral fields of the LMC and SMC will sample the event distributions of each population, i.e. \citep{DeRocco_2024, Perkins_2025}. Figure \ref{concldists} shows the microlensing time scale ($t_{E}$) and normalised source radius ($\rho$) of the three galactic distributions along the AMPM sightline: the Milky Way stellar disk and halo, the LMC  stellar halo and the joint dark matter distributions of the MW and LMC. The 1, 2 and 3$\sigma$ contours are shown on the plot for the monochromatic mass function at \target\ mass of $\sim 10^{-7} M_{\odot}$. While the $3\sigma$ contour regions overlap in the $t_{E},\rho$ parameter space, there are differences in the three distributions, which, with enough measurements of individual events, may help to determine the population of microlenses towards the LMC. 

\section{Conclusions}
In this paper, we present a full reanalysis of the single PBH candidate, \target\, from AMPM. We identified multiple avenues that might mimic a microlensing event and confirmed that the candidate is neither a repeating period star, contamination from nearby stars on the CCD detector, nor a flare star episode. We considered the three regions along the AMPM LMC sightline that \target\ may reside: the MW stellar thick disk and halo, the LMC stellar disk and the intervening joint dark matter halos of the MW and LMC. We took the three galactic distributions and designed a Bayesian MCMC modelling routine that evaluated the maximum likelihood parameters for the microlens-star system. 
\begin{itemize}\setlength{\itemsep}{3pt}
    \item[-] If in the MW, \target\ is $565^{+661}_{-357}$ pc away from the Sun, which places it in the thick disk of the galaxy with mass $0.022^{+1.859}_{-0.022} M_{\oplus}$. 
    \item[-] If in the LMC, \target\ is an extragalactic microlensing exoplanet and sits $700pc$ in front of the source star with mass $0.100^{+0.192}_{-0.067} M_{\oplus}$. 
    \item[-] If \target\ is within the density of the intervening dark matter, it is located well past the stellar disk of the MW at a distance of $18.2^{+24.7}_{-14.1}$ kpc and has a mass of $0.026^{+0.194}_{-0.023} M_{\oplus}$. 
\end{itemize}
Based on a comparison of the optical depths of the three galactic models, it is far more likely that \target\ belongs to the dark matter density and, hence, is the best candidate for a PBH.

With the detection of a PBH candidate in the direction of M31 (i.e., an even more dark matter-dominated sightline) from 39.3 hours of two-minute cadence HSC-Subaru data
\citep{Niikura_2019, Sugiyama_2026}. Our AMPM discovery of a three lunar-mass PBH supports the Subaru-HSC M31 detections of 13 PBH candidates at similar timescale. With the set of short-timescale microlensing events in the dark halo, there is enough motivation to seriously consider a population of low-mass black holes, as predicted by \cite{Carr_Hawking_1974}. To characterise the mass distribution function of the PBH population, significantly more PBH detections are required. To achieve this will require dedicated, high-cadence microlensing observations of extragalactic fields. Comparable observations of MW fields are additionally essential to trace and control the relative planet/black hole contributions. Our detection motivates the Roman and Vera C. Rubin Observatory microlensing programs to support high cadence, sit-and-stare observations to boost the sensitivity to low-mass microlenses.

\section*{Acknowledgements}

This project used data obtained with the Dark Energy Camera (DECam), which was constructed by the Dark Energy Survey (DES) collaboration. Funding for the DES Projects has been provided by the US Department of Energy, the U.S. National Science Foundation, the Ministry of Science and Education of Spain, the Science and Technology Facilities Council of the United Kingdom, the Higher Education Funding Council for England, the National Center for Supercomputing Applications at the University of Illinois at Urbana-Champaign, the Kavli Institute for Cosmological Physics at the University of Chicago, Center for Cosmology and Astro-Particle Physics at the Ohio State University, the Mitchell Institute for Fundamental Physics and Astronomy at Texas A\&M University, Financiadora de Estudos e Projetos, Fundaç\"ao Carlos Chagas Filho de Amparo à Pesquisa do Estado do Rio de Janeiro, Conselho Nacional de Desenvolvimento Científico e Tecnol\'ogico and the Ministério da Ciência, Tecnologia e Inovaç\"ao, the Deutsche Forschungsgemeinschaft and the Collaborating Institutions in the Dark Energy Survey.

The Collaborating Institutions are Argonne National Laboratory, the University of California at Santa Cruz, the University of Cambridge, Centro de Investigaciones Enérgeticas, Medioambientales y Tecnol\'ogicas–Madrid, the University of Chicago, University College London, the DES-Brazil Consortium, the University of Edinburgh, the Eidgen\"ossische Technische Hochschule (ETH) Z\"urich, Fermi National Accelerator Laboratory, the University of Illinois at Urbana-Champaign, the Institut de Ciències de l’Espai (IEEC/CSIC), the Institut de Física d’Altes Energies, Lawrence Berkeley National Laboratory, the Ludwig-Maximilians Universit\"at München and the associated Excellence Cluster Universe, the University of Michigan, NSF NOIRLab, the University of Nottingham, the Ohio State University, the OzDES Membership Consortium, the University of Pennsylvania, the University of Portsmouth, SLAC National Accelerator Laboratory, Stanford University, the University of Sussex, and Texas A\&M University.

Based on observations at NSF Cerro Tololo Inter-American Observatory, NSF NOIRLab (NOIRLab Prop. ID 2019B-0071; PI: J. Mould), which is managed by the Association of Universities for Research in Astronomy (AURA) under a cooperative agreement with the U.S. National Science Foundation.

Computation was performed on the OzSTAR national facility at Swinburne University of Technology. The OzSTAR program receives funding in part from the Astronomy National Collaborative Research Infrastructure Strategy (NCRIS) allocation provided by the Australian Government, and from the Victorian Higher Education State Investment Fund (VHESIF) provided by the Victorian Government.

This research was supported by the Australian Research Council (ARC) Centre of Excellence for Dark Matter Particle Physics (CDM; centredarkmatter.org) with Grant Number: CE200100008. RK further acknowledges the support of the Australian Government Research Training Program (RTP) Scholarship doi.org/10.82133/C42F-K220

%%%%%%%%%%%%%%%%%%%%%%%%%%%%%%%%%%%%%%%%%%%%%%%%%%
\section*{Data Availability}

The LMC Field 51 DECam data for this project are available for public access in the NOIRLab digital archive (astroarchive.noirlab.edu) under the ID 2019B-0071.
%The inclusion of a Data Availability Statement is a requirement for articles published in MNRAS. Data Availability Statements provide a standardised format for readers to understand the availability of data underlying the research results described in the article. The statement may refer to original data generated in the course of the study or to third-party data analysed in the article. The statement should describe and provide means of access, where possible, by linking to the data or providing the required accession numbers for the relevant databases or DOIs.



\bibliographystyle{mnras}
\bibliography{Ref} % if your bibtex file is called example.bib

@ARTICLE{Carr_Hawking_1974,
       author = {{Carr}, B.~J. and {Hawking}, S.~W.},
        title = "{Black holes in the early Universe}",
      journal = {\mnras},
         year = 1974,
        month = aug,
       volume = {168},
        pages = {399-416},
          doi = {10.1093/mnras/168.2.399}}

@article{Tsapras_2018,
   title={Microlensing Searches for Exoplanets},
   volume={8},
   ISSN={2076-3263},
   url={http://dx.doi.org/10.3390/geosciences8100365},
   DOI={10.3390/geosciences8100365},
   number={10},
   journal={Geosciences},
   publisher={MDPI AG},
   author={Tsapras, Yiannis},
   year={2018},
   month=Sept, pages={365} }

@article{Li_2025,
doi = {10.1088/1475-7516/2025/12/008},
url = {https://doi.org/10.1088/1475-7516/2025/12/008},
year = {2025},
month = {dec},
publisher = {IOP Publishing},
volume = {2025},
number = {12},
pages = {008},
author = {Li, Bichu and Tang, Chan-Yu and Huang, Zhuo-Ran and Liu, Lei-Hua},
title = {Microlensing of dark matter models in the Milky Way},
journal = {Journal of Cosmology and Astroparticle Physics},
}

@Inbook{Carr_2025,
author="Carr, Bernard J.
and Green, Anne M.",
editor="Byrnes, Christian
and Franciolini, Gabriele
and Harada, Tomohiro
and Pani, Paolo
and Sasaki, Misao",
title="The History of Primordial Black Holes",
bookTitle="Primordial Black Holes",
year="2025",
publisher="Springer Nature Singapore",
address="Singapore",
pages="3--33",
isbn="978-981-97-8887-3",
doi="10.1007/978-981-97-8887-3_1",
url="https://doi.org/10.1007/978-981-97-8887-3_1"
}

@article{Sugiyama_2026,
title = {Microlensing constraints on Primordial Black Hole abundance with Subaru Hyper Suprime-Cam observations of Andromeda},
author = {Sugiyama, Sunao and Takada, Masahiro and Yasuda, Naoki and Tominaga, Nozomu},
journal = {arXiv 260205840S},
year = {2026},
}

@article{Clesse_2015,
   title={Massive primordial black holes from hybrid inflation as dark matter and the seeds of galaxies},
   volume={92},
   ISSN={1550-2368},
   DOI={10.1103/physrevd.92.023524},
   number={2},
   journal={Physical Review D},
   publisher={American Physical Society (APS)},
   author={Clesse, Sébastien and García-Bellido, Juan},
   year={2015},
   month=jul }

@article{Bean_2002,
   title={Could supermassive black holes be quintessential primordial black holes?},
   volume={66},
   ISSN={1089-4918},
   DOI={10.1103/physrevd.66.063505},
   number={6},
   journal={Physical Review D},
   publisher={American Physical Society (APS)},
   author={Bean, Rachel and Magueijo, João},
   year={2002},
   month=sep }

@article{Green_2024,
    title = {Primordial black holes as a dark matter candidate - a brief overview},
    journal = {Nuclear Physics B},
    volume = {1003},
    pages = {116494},
    year = {2024},
    note = {Special Issue of Nobel Symposium 182 on Dark Matter},
    issn = {0550-3213},
    doi = {https://doi.org/10.1016/j.nuclphysb.2024.116494},
    author = {Anne M. Green}}

@article{Carr_2024,
   title={Observational evidence for primordial black holes: A positivist perspective},
   volume={1054},
   ISSN={0370-1573},
   DOI={10.1016/j.physrep.2023.11.005},
   journal={Physics Reports},
   publisher={Elsevier BV},
   author={Carr, B.J. and Clesse, S. and García-Bellido, J. and Hawkins, M.R.S. and Kühnel, F.},
   year={2024},
   month=feb, pages={1–68} }

@misc{GarciaBellido_2019,
      title={A common origin for baryons and dark matter}, 
      author={Juan García-Bellido and Bernard Carr and Sebastien Clesse},
      year={2019},
      eprint={1904.11482},
      archivePrefix={arXiv},
      primaryClass={astro-ph.CO}}

@article{Paczyski_1996,
   title={GRAVITATIONAL MICROLENSING IN THE LOCAL GROUP},
   volume={34},
   ISSN={1545-4282},
   DOI={10.1146/annurev.astro.34.1.419},
   number={1},
   journal={Annual Review of Astronomy and Astrophysics},
   publisher={Annual Reviews},
   author={Paczy\'nski, Bohdan},
   year={1996},
   month=sep, pages={419–459} }

@ARTICLE{Paczynski_1986,
       author = {{Paczy\'nski}, B.},
        title = "{Gravitational Microlensing by the Galactic Halo}",
      journal = {\apj},
         year = 1986,
        month = may,
       volume = {304},
        pages = {1},
          doi = {10.1086/164140},}

@book{Schneider_1992,
author = {Schneider, Peter and Ehlers, Jürgen and Falco, Emilio E},
address = {Berlin, Heidelberg},
copyright = {Springer-Verlag Berlin Heidelberg 1992},
edition = {1},
isbn = {1461276551},
issn = {0941-7834},
language = {eng},
organization = {SpringerLink (Online service)},
publisher = {Springer},
series = {Astronomy and Astrophysics Library},
title = {Gravitational lenses},
year = {1992},
}

@ARTICLE{WittMao_1994,
       author = {{Witt}, Hans J. and {Mao}, Shude},
        title = "{Can Lensed Stars Be Regarded as Pointlike for Microlensing by MACHOs?}",
      journal = {\apj},
         year = 1994,
        month = aug,
       volume = {430},
        pages = {505},
          doi = {10.1086/174426},
}

@ARTICLE{Griest_1991,
       author = {{Griest}, Kim},
        title = "{Galactic Microlensing as a Method of Detecting Massive Compact Halo Objects}",
      journal = {\apj},
         year = 1991,
        month = jan,
       volume = {366},
        pages = {412},
          doi = {10.1086/169575},}

@article{Griest_2011,
   title={Microlensing of Kepler Stars as a Method of Detecting Primordial Black Hole Dark Matter},
   volume={107},
   ISSN={1079-7114},
   DOI={10.1103/physrevlett.107.231101},
   number={23},
   journal={Physical Review Letters},
   publisher={American Physical Society (APS)},
   author={Griest, Kim and Lehner, Matthew J. and Cieplak, Agnieszka M. and Jain, Bhuvnesh},
   year={2011},
   month=dec }

@ARTICLE{Lee_2009,
       author = {{Lee}, C. -H. and {Riffeser}, A. and {Seitz}, S. and {Bender}, R.},
        title = "{Finite-Source Effects in Microlensing: A Precise, Easy to Implement, Fast, and Numerically Stable Formalism}",
      journal = {\apj},
         year = 2009,
        month = apr,
       volume = {695},
       number = {1},
        pages = {200-207},
          doi = {10.1088/0004-637X/695/1/200}}

@ARTICLE{NO_1984,
       author = {{Nityananda}, R. and {Ostriker}, J.~P.},
        title = "{Gravitational Lensing by Stars in a Galaxy Halo - Theory of Combined Weak and Strong Scattering}",
      journal = {Journal of Astrophysics and Astronomy},
         year = 1984,
        month = sep,
       volume = {5},
       number = {3},
        pages = {235-250},
          doi = {10.1007/BF02714541}}

@ARTICLE{VO_1983,
       author = {{Vietri}, M. and {Ostriker}, J.~P.},
        title = "{The statistics of gravitational lenses - Appaarent changes in the luminosity function of distant sources due to passage of light through a single galaxy}",
      journal = {\apj},
         year = 1983,
        month = apr,
       volume = {267},
        pages = {488-510},
          doi = {10.1086/160886}}

@ARTICLE{Cieplak_2013,
       author = {{Cieplak}, Agnieszka M. and {Griest}, Kim},
        title = "{Improved Theoretical Predictions of Microlensing Rates for the Detection of Primordial Black Hole Dark Matter}",
      journal = {\apj},
         year = 2013,
        month = apr,
       volume = {767},
       number = {2},
          eid = {145},
        pages = {145},
          doi = {10.1088/0004-637X/767/2/145},
archivePrefix = {arXiv},
       eprint = {1210.7729},
 primaryClass = {astro-ph.CO}}

@ARTICLE{Nidever_2019,
       author = {{Nidever}, David L. and {Olsen}, Knut and {Choi}, Yumi and {de Boer}, Thomas J.~L. and {Blum}, Robert D. and {Bell}, Eric F. and {Zaritsky}, Dennis and {Martin}, Nicolas F. and {Saha}, Abhijit and {Conn}, Blair C. and {Besla}, Gurtina and {van der Marel}, Roeland P. and {No{\"e}l}, Noelia E.~D. and {Monachesi}, Antonela and {Stringfellow}, Guy S. and {Massana}, Pol and {Cioni}, Maria-Rosa L. and {Gallart}, Carme and {Monelli}, Matteo and {Martinez-Delgado}, David and {Mu{\~n}oz}, Ricardo R. and {Majewski}, Steven R. and {Vivas}, A. Katherina and {Walker}, Alistair R. and {Kaleida}, Catherine and {Chu}, You-Hua},
        title = "{Exploring the Very Extended Low-surface-brightness Stellar Populations of the Large Magellanic Cloud with SMASH}",
      journal = {\apj},
         year = 2019,
        month = apr,
       volume = {874},
       number = {2},
          eid = {118},
        pages = {118},
          doi = {10.3847/1538-4357/aafaf7}}

@article{Nidever_2021,
   title={The Second Data Release of the Survey of the MAgellanic Stellar History (SMASH)},
   volume={161},
   ISSN={1538-3881},
   DOI={10.3847/1538-3881/abceb7},
   number={2},
   journal={The Astronomical Journal},
   publisher={American Astronomical Society},
   author={Nidever, David L. and Olsen, Knut and Choi, Yumi and Ruiz-Lara, Tomas and Miller, Amy E. and Johnson, L. Clifton and Bell, Cameron P. M. and Blum, Robert D. and Cioni, Maria-Rosa L. and Gallart, Carme and Majewski, Steven R. and Martin, Nicolas F. and Massana, Pol and Monachesi, Antonela and Noël, Noelia E. D. and Sakowska, Joanna D. and van der Marel, Roeland P. and Walker, Alistair R. and Zaritsky, Dennis and Bell, Eric F. and Conn, Blair C. and de Boer, Thomas J. L. and Gruendl, Robert A. and Monelli, Matteo and Muñoz, Ricardo R. and Saha, Abhijit and Vivas, A. Katherina and Bernard, Edouard and Besla, Gurtina and Carballo-Bello, Julio A. and Dorta, Antonio and Martinez-Delgado, David and Goater, Alex and Rusakov, Vadim and Stringfellow, Guy S.},
   year={2021},
   month=jan, pages={74} }

@article{Dophot_1993,
    author = "Schechter, Paul L. and Mateo, Mario and Saha, Abhijit",
    title = "{DOPHOT, a CCD photometry program: Description and tests}",
    doi = "10.1086/133316",
    journal = "Publ. Astron. Soc. Pac.",
    volume = "105",
    pages = "1342",
    year = "1993"
}

@article{VNstat1941,
 ISSN = {00034851},
 author = {J. von Neumann and R. H. Kent and H. R. Bellinson and B. I. Hart},
 journal = {The Annals of Mathematical Statistics},
 number = {2},
 pages = {153--162},
 publisher = {Institute of Mathematical Statistics},
 title = {The Mean Square Successive Difference},
 urldate = {2024-03-22},
 volume = {12},
 year = {1941}
}

@article{Kim2014,
title = "The EPOCH Project: I. Periodic variable stars in the EROS-2 LMC database {\^a}",
author = "Kim, {Dae Won} and Pavlos Protopapas and Bailer-Jones, {Coryn A.L.} and Byun, {Yong Ik} and Chang, {Seo Won} and Marquette, {Jean Baptiste} and Shin, {Min Su}",
year = "2014",
month = jun,
doi = "10.1051/0004-6361/201323252",
language = "English",
volume = "566",
journal = "Astronomy and Astrophysics",
issn = "0004-6361",
publisher = "EDP Sciences",
}

@ARTICLE{Irwin_2007,
       author = {{Irwin}, Jonathan and {Irwin}, Mike and {Aigrain}, Suzanne and {Hodgkin}, Simon and {Hebb}, Leslie and {Moraux}, Estelle},
        title = "{The Monitor project: data processing and light curve production}",
      journal = {\mnras},
         year = 2007,
        month = mar,
       volume = {375},
       number = {4},
        pages = {1449-1462},
          doi = {10.1111/j.1365-2966.2006.11408.x}}

@ARTICLE{GAIA1,
       author = {{Gaia Collaboration} and {Prusti}, T. and {de Bruijne}, J.~H.~J. and {Brown}, A.~G.~A. and {Vallenari}, A. and {Babusiaux}, C. and {Bailer-Jones}, C.~A.~L. and {Bastian}, U. and {Biermann}, M. and {Evans}, D.~W. and {Eyer}, L. and {Jansen}, F. and {Jordi}, C. and {Klioner}, S.~A. and {Lammers}, U. and {Lindegren}, L. and {Luri}, X. and {Mignard}, F. and {Milligan}, D.~J. and {Panem}, C. and {Poinsignon}, V. and {Pourbaix}, D. and {Randich}, S. and {Sarri}, G. and {Sartoretti}, P. and {Siddiqui}, H.~I. and {Soubiran}, C. and {Valette}, V. and {van Leeuwen}, F. and {Walton}, N.~A. and {Aerts}, C. and {Arenou}, F. and {Cropper}, M. and {Drimmel}, R. and {H{\o}g}, E. and {Katz}, D. and {Lattanzi}, M.~G. and {O'Mullane}, W. and {Grebel}, E.~K. and {Holland}, A.~D. and {Huc}, C. and {Passot}, X. and {Bramante}, L. and {Cacciari}, C. and {Casta{\~n}eda}, J. and {Chaoul}, L. and {Cheek}, N. and {De Angeli}, F. and {Fabricius}, C. and {Guerra}, R. and {Hern{\'a}ndez}, J. and {Jean-Antoine-Piccolo}, A. and {Masana}, E. and {Messineo}, R. and {Mowlavi}, N. and {Nienartowicz}, K. and {Ord{\'o}{\~n}ez-Blanco}, D. and {Panuzzo}, P. and {Portell}, J. and {Richards}, P.~J. and {Riello}, M. and {Seabroke}, G.~M. and {Tanga}, P. and {Th{\'e}venin}, F. and {Torra}, J. and {Els}, S.~G. and {Gracia-Abril}, G. and {Comoretto}, G. and {Garcia-Reinaldos}, M. and {Lock}, T. and {Mercier}, E. and {Altmann}, M. and {Andrae}, R. and {Astraatmadja}, T.~L. and {Bellas-Velidis}, I. and {Benson}, K. and {Berthier}, J. and {Blomme}, R. and {Busso}, G. and {Carry}, B. and {Cellino}, A. and {Clementini}, G. and {Cowell}, S. and {Creevey}, O. and {Cuypers}, J. and {Davidson}, M. and {De Ridder}, J. and {de Torres}, A. and {Delchambre}, L. and {Dell'Oro}, A. and {Ducourant}, C. and {Fr{\'e}mat}, Y. and {Garc{\'\i}a-Torres}, M. and {Gosset}, E. and {Halbwachs}, J. -L. and {Hambly}, N.~C. and {Harrison}, D.~L. and {Hauser}, M. and {Hestroffer}, D. and {Hodgkin}, S.~T. and {Huckle}, H.~E. and {Hutton}, A. and {Jasniewicz}, G. and {Jordan}, S. and {Kontizas}, M. and {Korn}, A.~J. and {Lanzafame}, A.~C. and {Manteiga}, M. and {Moitinho}, A. and {Muinonen}, K. and {Osinde}, J. and {Pancino}, E. and {Pauwels}, T. and {Petit}, J. -M. and {Recio-Blanco}, A. and {Robin}, A.~C. and {Sarro}, L.~M. and {Siopis}, C. and {Smith}, M. and {Smith}, K.~W. and {Sozzetti}, A. and {Thuillot}, W. and {van Reeven}, W. and {Viala}, Y. and {Abbas}, U. and {Abreu Aramburu}, A. and {Accart}, S. and {Aguado}, J.~J. and {Allan}, P.~M. and {Allasia}, W. and {Altavilla}, G. and {{\'A}lvarez}, M.~A. and {Alves}, J. and {Anderson}, R.~I. and {Andrei}, A.~H. and {Anglada Varela}, E. and {Antiche}, E. and {Antoja}, T. and {Ant{\'o}n}, S. and {Arcay}, B. and {Atzei}, A. and {Ayache}, L. and {Bach}, N. and {Baker}, S.~G. and {Balaguer-N{\'u}{\~n}ez}, L. and {Barache}, C. and {Barata}, C. and {Barbier}, A. and {Barblan}, F. and {Baroni}, M. and {Barrado y Navascu{\'e}s}, D. and {Barros}, M. and {Barstow}, M.~A. and {Becciani}, U. and {Bellazzini}, M. and {Bellei}, G. and {Bello Garc{\'\i}a}, A. and {Belokurov}, V. and {Bendjoya}, P. and {Berihuete}, A. and {Bianchi}, L. and {Bienaym{\'e}}, O. and {Billebaud}, F. and {Blagorodnova}, N. and {Blanco-Cuaresma}, S. and {Boch}, T. and {Bombrun}, A. and {Borrachero}, R. and {Bouquillon}, S. and {Bourda}, G. and {Bouy}, H. and {Bragaglia}, A. and {Breddels}, M.~A. and {Brouillet}, N. and {Br{\"u}semeister}, T. and {Bucciarelli}, B. and {Budnik}, F. and {Burgess}, P. and {Burgon}, R. and {Burlacu}, A. and {Busonero}, D. and {Buzzi}, R. and {Caffau}, E. and {Cambras}, J. and {Campbell}, H. and {Cancelliere}, R. and {Cantat-Gaudin}, T. and {Carlucci}, T. and {Carrasco}, J.~M. and {Castellani}, M. and {Charlot}, P. and {Charnas}, J. and {Charvet}, P. and {Chassat}, F. and {Chiavassa}, A. and {Clotet}, M. and {Cocozza}, G. and {Collins}, R.~S. and {Collins}, P. and {Costigan}, G. and {Crifo}, F. and {Cross}, N.~J.~G. and {Crosta}, M. and {Crowley}, C. and {Dafonte}, C. and {Damerdji}, Y. and {Dapergolas}, A. and {David}, P. and {David}, M. and {De Cat}, P. and {de Felice}, F. and {de Laverny}, P. and {De Luise}, F. and {De March}, R. and {de Martino}, D. and {de Souza}, R. and {Debosscher}, J. and {del Pozo}, E. and {Delbo}, M. and {Delgado}, A. and {Delgado}, H.~E. and {di Marco}, F. and {Di Matteo}, P. and {Diakite}, S. and {Distefano}, E. and {Dolding}, C. and {Dos Anjos}, S. and {Drazinos}, P. and {Dur{\'a}n}, J. and {Dzigan}, Y. and {Ecale}, E. and {Edvardsson}, B. and {Enke}, H. and {Erdmann}, M. and {Escolar}, D. and {Espina}, M. and {Evans}, N.~W. and {Eynard Bontemps}, G. and {Fabre}, C. and {Fabrizio}, M. and {Faigler}, S. and {Falc{\~a}o}, A.~J. and {Farr{\`a}s Casas}, M. and {Faye}, F. and {Federici}, L. and {Fedorets}, G. and {Fern{\'a}ndez-Hern{\'a}ndez}, J. and {Fernique}, P. and {Fienga}, A. and {Figueras}, F. and {Filippi}, F. and {Findeisen}, K. and {Fonti}, A. and {Fouesneau}, M. and {Fraile}, E. and {Fraser}, M. and {Fuchs}, J. and {Furnell}, R. and {Gai}, M. and {Galleti}, S. and {Galluccio}, L. and {Garabato}, D. and {Garc{\'\i}a-Sedano}, F. and {Gar{\'e}}, P. and {Garofalo}, A. and {Garralda}, N. and {Gavras}, P. and {Gerssen}, J. and {Geyer}, R. and {Gilmore}, G. and {Girona}, S. and {Giuffrida}, G. and {Gomes}, M. and {Gonz{\'a}lez-Marcos}, A. and {Gonz{\'a}lez-N{\'u}{\~n}ez}, J. and {Gonz{\'a}lez-Vidal}, J.~J. and {Granvik}, M. and {Guerrier}, A. and {Guillout}, P. and {Guiraud}, J. and {G{\'u}rpide}, A. and {Guti{\'e}rrez-S{\'a}nchez}, R. and {Guy}, L.~P. and {Haigron}, R. and {Hatzidimitriou}, D. and {Haywood}, M. and {Heiter}, U. and {Helmi}, A. and {Hobbs}, D. and {Hofmann}, W. and {Holl}, B. and {Holland}, G. and {Hunt}, J.~A.~S. and {Hypki}, A. and {Icardi}, V. and {Irwin}, M. and {Jevardat de Fombelle}, G. and {Jofr{\'e}}, P. and {Jonker}, P.~G. and {Jorissen}, A. and {Julbe}, F. and {Karampelas}, A. and {Kochoska}, A. and {Kohley}, R. and {Kolenberg}, K. and {Kontizas}, E. and {Koposov}, S.~E. and {Kordopatis}, G. and {Koubsky}, P. and {Kowalczyk}, A. and {Krone-Martins}, A. and {Kudryashova}, M. and {Kull}, I. and {Bachchan}, R.~K. and {Lacoste-Seris}, F. and {Lanza}, A.~F. and {Lavigne}, J. -B. and {Le Poncin-Lafitte}, C. and {Lebreton}, Y. and {Lebzelter}, T. and {Leccia}, S. and {Leclerc}, N. and {Lecoeur-Taibi}, I. and {Lemaitre}, V. and {Lenhardt}, H. and {Leroux}, F. and {Liao}, S. and {Licata}, E. and {Lindstr{\o}m}, H.~E.~P. and {Lister}, T.~A. and {Livanou}, E. and {Lobel}, A. and {L{\"o}ffler}, W. and {L{\'o}pez}, M. and {Lopez-Lozano}, A. and {Lorenz}, D. and {Loureiro}, T. and {MacDonald}, I. and {Magalh{\~a}es Fernandes}, T. and {Managau}, S. and {Mann}, R.~G. and {Mantelet}, G. and {Marchal}, O. and {Marchant}, J.~M. and {Marconi}, M. and {Marie}, J. and {Marinoni}, S. and {Marrese}, P.~M. and {Marschalk{\'o}}, G. and {Marshall}, D.~J. and {Mart{\'\i}n-Fleitas}, J.~M. and {Martino}, M. and {Mary}, N. and {Matijevi{\v{c}}}, G. and {Mazeh}, T. and {McMillan}, P.~J. and {Messina}, S. and {Mestre}, A. and {Michalik}, D. and {Millar}, N.~R. and {Miranda}, B.~M.~H. and {Molina}, D. and {Molinaro}, R. and {Molinaro}, M. and {Moln{\'a}r}, L. and {Moniez}, M. and {Montegriffo}, P. and {Monteiro}, D. and {Mor}, R. and {Mora}, A. and {Morbidelli}, R. and {Morel}, T. and {Morgenthaler}, S. and {Morley}, T. and {Morris}, D. and {Mulone}, A.~F. and {Muraveva}, T. and {Musella}, I. and {Narbonne}, J. and {Nelemans}, G. and {Nicastro}, L. and {Noval}, L. and {Ord{\'e}novic}, C. and {Ordieres-Mer{\'e}}, J. and {Osborne}, P. and {Pagani}, C. and {Pagano}, I. and {Pailler}, F. and {Palacin}, H. and {Palaversa}, L. and {Parsons}, P. and {Paulsen}, T. and {Pecoraro}, M. and {Pedrosa}, R. and {Pentik{\"a}inen}, H. and {Pereira}, J. and {Pichon}, B. and {Piersimoni}, A.~M. and {Pineau}, F. -X. and {Plachy}, E. and {Plum}, G. and {Poujoulet}, E. and {Pr{\v{s}}a}, A. and {Pulone}, L. and {Ragaini}, S. and {Rago}, S. and {Rambaux}, N. and {Ramos-Lerate}, M. and {Ranalli}, P. and {Rauw}, G. and {Read}, A. and {Regibo}, S. and {Renk}, F. and {Reyl{\'e}}, C. and {Ribeiro}, R.~A. and {Rimoldini}, L. and {Ripepi}, V. and {Riva}, A. and {Rixon}, G. and {Roelens}, M. and {Romero-G{\'o}mez}, M. and {Rowell}, N. and {Royer}, F. and {Rudolph}, A. and {Ruiz-Dern}, L. and {Sadowski}, G. and {Sagrist{\`a} Sell{\'e}s}, T. and {Sahlmann}, J. and {Salgado}, J. and {Salguero}, E. and {Sarasso}, M. and {Savietto}, H. and {Schnorhk}, A. and {Schultheis}, M. and {Sciacca}, E. and {Segol}, M. and {Segovia}, J.~C. and {Segransan}, D. and {Serpell}, E. and {Shih}, I. -C. and {Smareglia}, R. and {Smart}, R.~L. and {Smith}, C. and {Solano}, E. and {Solitro}, F. and {Sordo}, R. and {Soria Nieto}, S. and {Souchay}, J. and {Spagna}, A. and {Spoto}, F. and {Stampa}, U. and {Steele}, I.~A. and {Steidelm{\"u}ller}, H. and {Stephenson}, C.~A. and {Stoev}, H. and {Suess}, F.~F. and {S{\"u}veges}, M. and {Surdej}, J. and {Szabados}, L. and {Szegedi-Elek}, E. and {Tapiador}, D. and {Taris}, F. and {Tauran}, G. and {Taylor}, M.~B. and {Teixeira}, R. and {Terrett}, D. and {Tingley}, B. and {Trager}, S.~C. and {Turon}, C. and {Ulla}, A. and {Utrilla}, E. and {Valentini}, G. and {van Elteren}, A. and {Van Hemelryck}, E. and {van Leeuwen}, M. and {Varadi}, M. and {Vecchiato}, A. and {Veljanoski}, J. and {Via}, T. and {Vicente}, D. and {Vogt}, S. and {Voss}, H. and {Votruba}, V. and {Voutsinas}, S. and {Walmsley}, G. and {Weiler}, M. and {Weingrill}, K. and {Werner}, D. and {Wevers}, T. and {Whitehead}, G. and {Wyrzykowski}, {\L}. and {Yoldas}, A. and {{\v{Z}}erjal}, M. and {Zucker}, S. and {Zurbach}, C. and {Zwitter}, T. and {Alecu}, A. and {Allen}, M. and {Allende Prieto}, C. and {Amorim}, A. and {Anglada-Escud{\'e}}, G. and {Arsenijevic}, V. and {Azaz}, S. and {Balm}, P. and {Beck}, M. and {Bernstein}, H. -H. and {Bigot}, L. and {Bijaoui}, A. and {Blasco}, C. and {Bonfigli}, M. and {Bono}, G. and {Boudreault}, S. and {Bressan}, A. and {Brown}, S. and {Brunet}, P. -M. and {Bunclark}, P. and {Buonanno}, R. and {Butkevich}, A.~G. and {Carret}, C. and {Carrion}, C. and {Chemin}, L. and {Ch{\'e}reau}, F. and {Corcione}, L. and {Darmigny}, E. and {de Boer}, K.~S. and {de Teodoro}, P. and {de Zeeuw}, P.~T. and {Delle Luche}, C. and {Domingues}, C.~D. and {Dubath}, P. and {Fodor}, F. and {Fr{\'e}zouls}, B. and {Fries}, A. and {Fustes}, D. and {Fyfe}, D. and {Gallardo}, E. and {Gallegos}, J. and {Gardiol}, D. and {Gebran}, M. and {Gomboc}, A. and {G{\'o}mez}, A. and {Grux}, E. and {Gueguen}, A. and {Heyrovsky}, A. and {Hoar}, J. and {Iannicola}, G. and {Isasi Parache}, Y. and {Janotto}, A. -M. and {Joliet}, E. and {Jonckheere}, A. and {Keil}, R. and {Kim}, D. -W. and {Klagyivik}, P. and {Klar}, J. and {Knude}, J. and {Kochukhov}, O. and {Kolka}, I. and {Kos}, J. and {Kutka}, A. and {Lainey}, V. and {LeBouquin}, D. and {Liu}, C. and {Loreggia}, D. and {Makarov}, V.~V. and {Marseille}, M.~G. and {Martayan}, C. and {Martinez-Rubi}, O. and {Massart}, B. and {Meynadier}, F. and {Mignot}, S. and {Munari}, U. and {Nguyen}, A. -T. and {Nordlander}, T. and {Ocvirk}, P. and {O'Flaherty}, K.~S. and {Olias Sanz}, A. and {Ortiz}, P. and {Osorio}, J. and {Oszkiewicz}, D. and {Ouzounis}, A. and {Palmer}, M. and {Park}, P. and {Pasquato}, E. and {Peltzer}, C. and {Peralta}, J. and {P{\'e}turaud}, F. and {Pieniluoma}, T. and {Pigozzi}, E. and {Poels}, J. and {Prat}, G. and {Prod'homme}, T. and {Raison}, F. and {Rebordao}, J.~M. and {Risquez}, D. and {Rocca-Volmerange}, B. and {Rosen}, S. and {Ruiz-Fuertes}, M.~I. and {Russo}, F. and {Sembay}, S. and {Serraller Vizcaino}, I. and {Short}, A. and {Siebert}, A. and {Silva}, H. and {Sinachopoulos}, D. and {Slezak}, E. and {Soffel}, M. and {Sosnowska}, D. and {Strai{\v{z}}ys}, V. and {ter Linden}, M. and {Terrell}, D. and {Theil}, S. and {Tiede}, C. and {Troisi}, L. and {Tsalmantza}, P. and {Tur}, D. and {Vaccari}, M. and {Vachier}, F. and {Valles}, P. and {Van Hamme}, W. and {Veltz}, L. and {Virtanen}, J. and {Wallut}, J. -M. and {Wichmann}, R. and {Wilkinson}, M.~I. and {Ziaeepour}, H. and {Zschocke}, S.},
        title = "{The Gaia mission}",
      journal = {\aap},
         year = 2016,
        month = nov,
       volume = {595},
          eid = {A1},
        pages = {A1},
          doi = {10.1051/0004-6361/201629272}}

@ARTICLE{GAIA2,
       author = {{Gaia Collaboration} and {Vallenari}, A. and {Brown}, A.~G.~A. and {Prusti}, T. and {de Bruijne}, J.~H.~J. and {Arenou}, F. and {Babusiaux}, C. and {Biermann}, M. and {Creevey}, O.~L. and {Ducourant}, C. and {Evans}, D.~W. and {Eyer}, L. and {Guerra}, R. and {Hutton}, A. and {Jordi}, C. and {Klioner}, S.~A. and {Lammers}, U.~L. and {Lindegren}, L. and {Luri}, X. and {Mignard}, F. and {Panem}, C. and {Pourbaix}, D. and {Randich}, S. and {Sartoretti}, P. and {Soubiran}, C. and {Tanga}, P. and {Walton}, N.~A. and {Bailer-Jones}, C.~A.~L. and {Bastian}, U. and {Drimmel}, R. and {Jansen}, F. and {Katz}, D. and {Lattanzi}, M.~G. and {van Leeuwen}, F. and {Bakker}, J. and {Cacciari}, C. and {Casta{\~n}eda}, J. and {De Angeli}, F. and {Fabricius}, C. and {Fouesneau}, M. and {Fr{\'e}mat}, Y. and {Galluccio}, L. and {Guerrier}, A. and {Heiter}, U. and {Masana}, E. and {Messineo}, R. and {Mowlavi}, N. and {Nicolas}, C. and {Nienartowicz}, K. and {Pailler}, F. and {Panuzzo}, P. and {Riclet}, F. and {Roux}, W. and {Seabroke}, G.~M. and {Sordo}, R. and {Th{\'e}venin}, F. and {Gracia-Abril}, G. and {Portell}, J. and {Teyssier}, D. and {Altmann}, M. and {Andrae}, R. and {Audard}, M. and {Bellas-Velidis}, I. and {Benson}, K. and {Berthier}, J. and {Blomme}, R. and {Burgess}, P.~W. and {Busonero}, D. and {Busso}, G. and {C{\'a}novas}, H. and {Carry}, B. and {Cellino}, A. and {Cheek}, N. and {Clementini}, G. and {Damerdji}, Y. and {Davidson}, M. and {de Teodoro}, P. and {Nu{\~n}ez Campos}, M. and {Delchambre}, L. and {Dell'Oro}, A. and {Esquej}, P. and {Fern{\'a}ndez-Hern{\'a}ndez}, J. and {Fraile}, E. and {Garabato}, D. and {Garc{\'\i}a-Lario}, P. and {Gosset}, E. and {Haigron}, R. and {Halbwachs}, J. -L. and {Hambly}, N.~C. and {Harrison}, D.~L. and {Hern{\'a}ndez}, J. and {Hestroffer}, D. and {Hodgkin}, S.~T. and {Holl}, B. and {Jan{\ss}en}, K. and {Jevardat de Fombelle}, G. and {Jordan}, S. and {Krone-Martins}, A. and {Lanzafame}, A.~C. and {L{\"o}ffler}, W. and {Marchal}, O. and {Marrese}, P.~M. and {Moitinho}, A. and {Muinonen}, K. and {Osborne}, P. and {Pancino}, E. and {Pauwels}, T. and {Recio-Blanco}, A. and {Reyl{\'e}}, C. and {Riello}, M. and {Rimoldini}, L. and {Roegiers}, T. and {Rybizki}, J. and {Sarro}, L.~M. and {Siopis}, C. and {Smith}, M. and {Sozzetti}, A. and {Utrilla}, E. and {van Leeuwen}, M. and {Abbas}, U. and {{\'A}brah{\'a}m}, P. and {Abreu Aramburu}, A. and {Aerts}, C. and {Aguado}, J.~J. and {Ajaj}, M. and {Aldea-Montero}, F. and {Altavilla}, G. and {{\'A}lvarez}, M.~A. and {Alves}, J. and {Anders}, F. and {Anderson}, R.~I. and {Anglada Varela}, E. and {Antoja}, T. and {Baines}, D. and {Baker}, S.~G. and {Balaguer-N{\'u}{\~n}ez}, L. and {Balbinot}, E. and {Balog}, Z. and {Barache}, C. and {Barbato}, D. and {Barros}, M. and {Barstow}, M.~A. and {Bartolom{\'e}}, S. and {Bassilana}, J. -L. and {Bauchet}, N. and {Becciani}, U. and {Bellazzini}, M. and {Berihuete}, A. and {Bernet}, M. and {Bertone}, S. and {Bianchi}, L. and {Binnenfeld}, A. and {Blanco-Cuaresma}, S. and {Blazere}, A. and {Boch}, T. and {Bombrun}, A. and {Bossini}, D. and {Bouquillon}, S. and {Bragaglia}, A. and {Bramante}, L. and {Breedt}, E. and {Bressan}, A. and {Brouillet}, N. and {Brugaletta}, E. and {Bucciarelli}, B. and {Burlacu}, A. and {Butkevich}, A.~G. and {Buzzi}, R. and {Caffau}, E. and {Cancelliere}, R. and {Cantat-Gaudin}, T. and {Carballo}, R. and {Carlucci}, T. and {Carnerero}, M.~I. and {Carrasco}, J.~M. and {Casamiquela}, L. and {Castellani}, M. and {Castro-Ginard}, A. and {Chaoul}, L. and {Charlot}, P. and {Chemin}, L. and {Chiaramida}, V. and {Chiavassa}, A. and {Chornay}, N. and {Comoretto}, G. and {Contursi}, G. and {Cooper}, W.~J. and {Cornez}, T. and {Cowell}, S. and {Crifo}, F. and {Cropper}, M. and {Crosta}, M. and {Crowley}, C. and {Dafonte}, C. and {Dapergolas}, A. and {David}, M. and {David}, P. and {de Laverny}, P. and {De Luise}, F. and {De March}, R. and {De Ridder}, J. and {de Souza}, R. and {de Torres}, A. and {del Peloso}, E.~F. and {del Pozo}, E. and {Delbo}, M. and {Delgado}, A. and {Delisle}, J. -B. and {Demouchy}, C. and {Dharmawardena}, T.~E. and {Di Matteo}, P. and {Diakite}, S. and {Diener}, C. and {Distefano}, E. and {Dolding}, C. and {Edvardsson}, B. and {Enke}, H. and {Fabre}, C. and {Fabrizio}, M. and {Faigler}, S. and {Fedorets}, G. and {Fernique}, P. and {Fienga}, A. and {Figueras}, F. and {Fournier}, Y. and {Fouron}, C. and {Fragkoudi}, F. and {Gai}, M. and {Garcia-Gutierrez}, A. and {Garcia-Reinaldos}, M. and {Garc{\'\i}a-Torres}, M. and {Garofalo}, A. and {Gavel}, A. and {Gavras}, P. and {Gerlach}, E. and {Geyer}, R. and {Giacobbe}, P. and {Gilmore}, G. and {Girona}, S. and {Giuffrida}, G. and {Gomel}, R. and {Gomez}, A. and {Gonz{\'a}lez-N{\'u}{\~n}ez}, J. and {Gonz{\'a}lez-Santamar{\'\i}a}, I. and {Gonz{\'a}lez-Vidal}, J.~J. and {Granvik}, M. and {Guillout}, P. and {Guiraud}, J. and {Guti{\'e}rrez-S{\'a}nchez}, R. and {Guy}, L.~P. and {Hatzidimitriou}, D. and {Hauser}, M. and {Haywood}, M. and {Helmer}, A. and {Helmi}, A. and {Sarmiento}, M.~H. and {Hidalgo}, S.~L. and {Hilger}, T. and {H{\l}adczuk}, N. and {Hobbs}, D. and {Holland}, G. and {Huckle}, H.~E. and {Jardine}, K. and {Jasniewicz}, G. and {Jean-Antoine Piccolo}, A. and {Jim{\'e}nez-Arranz}, {\'O}. and {Jorissen}, A. and {Juaristi Campillo}, J. and {Julbe}, F. and {Karbevska}, L. and {Kervella}, P. and {Khanna}, S. and {Kontizas}, M. and {Kordopatis}, G. and {Korn}, A.~J. and {K{\'o}sp{\'a}l}, {\'A}. and {Kostrzewa-Rutkowska}, Z. and {Kruszy{\'n}ska}, K. and {Kun}, M. and {Laizeau}, P. and {Lambert}, S. and {Lanza}, A.~F. and {Lasne}, Y. and {Le Campion}, J. -F. and {Lebreton}, Y. and {Lebzelter}, T. and {Leccia}, S. and {Leclerc}, N. and {Lecoeur-Taibi}, I. and {Liao}, S. and {Licata}, E.~L. and {Lindstr{\o}m}, H.~E.~P. and {Lister}, T.~A. and {Livanou}, E. and {Lobel}, A. and {Lorca}, A. and {Loup}, C. and {Madrero Pardo}, P. and {Magdaleno Romeo}, A. and {Managau}, S. and {Mann}, R.~G. and {Manteiga}, M. and {Marchant}, J.~M. and {Marconi}, M. and {Marcos}, J. and {Marcos Santos}, M.~M.~S. and {Mar{\'\i}n Pina}, D. and {Marinoni}, S. and {Marocco}, F. and {Marshall}, D.~J. and {Martin Polo}, L. and {Mart{\'\i}n-Fleitas}, J.~M. and {Marton}, G. and {Mary}, N. and {Masip}, A. and {Massari}, D. and {Mastrobuono-Battisti}, A. and {Mazeh}, T. and {McMillan}, P.~J. and {Messina}, S. and {Michalik}, D. and {Millar}, N.~R. and {Mints}, A. and {Molina}, D. and {Molinaro}, R. and {Moln{\'a}r}, L. and {Monari}, G. and {Mongui{\'o}}, M. and {Montegriffo}, P. and {Montero}, A. and {Mor}, R. and {Mora}, A. and {Morbidelli}, R. and {Morel}, T. and {Morris}, D. and {Muraveva}, T. and {Murphy}, C.~P. and {Musella}, I. and {Nagy}, Z. and {Noval}, L. and {Oca{\~n}a}, F. and {Ogden}, A. and {Ordenovic}, C. and {Osinde}, J.~O. and {Pagani}, C. and {Pagano}, I. and {Palaversa}, L. and {Palicio}, P.~A. and {Pallas-Quintela}, L. and {Panahi}, A. and {Payne-Wardenaar}, S. and {Pe{\~n}alosa Esteller}, X. and {Penttil{\"a}}, A. and {Pichon}, B. and {Piersimoni}, A.~M. and {Pineau}, F. -X. and {Plachy}, E. and {Plum}, G. and {Poggio}, E. and {Pr{\v{s}}a}, A. and {Pulone}, L. and {Racero}, E. and {Ragaini}, S. and {Rainer}, M. and {Raiteri}, C.~M. and {Rambaux}, N. and {Ramos}, P. and {Ramos-Lerate}, M. and {Re Fiorentin}, P. and {Regibo}, S. and {Richards}, P.~J. and {Rios Diaz}, C. and {Ripepi}, V. and {Riva}, A. and {Rix}, H. -W. and {Rixon}, G. and {Robichon}, N. and {Robin}, A.~C. and {Robin}, C. and {Roelens}, M. and {Rogues}, H.~R.~O. and {Rohrbasser}, L. and {Romero-G{\'o}mez}, M. and {Rowell}, N. and {Royer}, F. and {Ruz Mieres}, D. and {Rybicki}, K.~A. and {Sadowski}, G. and {S{\'a}ez N{\'u}{\~n}ez}, A. and {Sagrist{\`a} Sell{\'e}s}, A. and {Sahlmann}, J. and {Salguero}, E. and {Samaras}, N. and {Sanchez Gimenez}, V. and {Sanna}, N. and {Santove{\~n}a}, R. and {Sarasso}, M. and {Schultheis}, M. and {Sciacca}, E. and {Segol}, M. and {Segovia}, J.~C. and {S{\'e}gransan}, D. and {Semeux}, D. and {Shahaf}, S. and {Siddiqui}, H.~I. and {Siebert}, A. and {Siltala}, L. and {Silvelo}, A. and {Slezak}, E. and {Slezak}, I. and {Smart}, R.~L. and {Snaith}, O.~N. and {Solano}, E. and {Solitro}, F. and {Souami}, D. and {Souchay}, J. and {Spagna}, A. and {Spina}, L. and {Spoto}, F. and {Steele}, I.~A. and {Steidelm{\"u}ller}, H. and {Stephenson}, C.~A. and {S{\"u}veges}, M. and {Surdej}, J. and {Szabados}, L. and {Szegedi-Elek}, E. and {Taris}, F. and {Taylor}, M.~B. and {Teixeira}, R. and {Tolomei}, L. and {Tonello}, N. and {Torra}, F. and {Torra}, J. and {Torralba Elipe}, G. and {Trabucchi}, M. and {Tsounis}, A.~T. and {Turon}, C. and {Ulla}, A. and {Unger}, N. and {Vaillant}, M.~V. and {van Dillen}, E. and {van Reeven}, W. and {Vanel}, O. and {Vecchiato}, A. and {Viala}, Y. and {Vicente}, D. and {Voutsinas}, S. and {Weiler}, M. and {Wevers}, T. and {Wyrzykowski}, {\L}. and {Yoldas}, A. and {Yvard}, P. and {Zhao}, H. and {Zorec}, J. and {Zucker}, S. and {Zwitter}, T.},
        title = "{Gaia Data Release 3. Summary of the content and survey properties}",
      journal = {\aap},
         year = 2023,
        month = jun,
       volume = {674},
          eid = {A1},
        pages = {A1},
          doi = {10.1051/0004-6361/202243940}}

@dataset{GAIASource,
       author = {{Gaia Collaboration}},
        title = "{VizieR Online Data Catalog: Gaia DR3 Part 1. Main source (Gaia Collaboration, 2022)}",
 howpublished = {VizieR On-line Data Catalog: I/355.  Originally published in: 2023A\&A...674A...1G; doi:10.1051/0004-63},
         year = {2022},
        month = {may},
          eid = {I/355},
          doi = {10.26093/cds/vizier.1355},
       adsurl = {https://ui.adsabs.harvard.edu/abs/2022yCat.1355....0G}
}

@article{GaiaVarInfo,
   title={GaiaData Release 3: Summary of the variability processing and analysis},
   volume={674},
   ISSN={1432-0746},
   DOI={10.1051/0004-6361/202244242},
   journal={Astronomy \& Astrophysics},
   publisher={EDP Sciences},
   author={Eyer, L. and Audard, M. and Holl, B. and Rimoldini, L. and Carnerero, M. I. and Clementini, G. and De Ridder, J. and Distefano, E. and Evans, D. W. and Gavras, P. and Gomel, R. and Lebzelter, T. and Marton, G. and Mowlavi, N. and Panahi, A. and Ripepi, V. and Wyrzykowski, Ł. and Nienartowicz, K. and Jevardat de Fombelle, G. and Lecoeur-Taibi, I. and Rohrbasser, L. and Riello, M. and García-Lario, P. and Lanzafame, A. C. and Mazeh, T. and Raiteri, C. M. and Zucker, S. and Ábrahám, P. and Aerts, C. and Aguado, J. J. and Anderson, R. I. and Bashi, D. and Binnenfeld, A. and Faigler, S. and Garofalo, A. and Karbevska, L. and Kóspál, Á and Kruszyńska, K. and Kun, M. and Lanza, A. F. and Leccia, S. and Marconi, M. and Messina, S. and Molinaro, R. and Molnár, L. and Muraveva, T. and Musella, I. and Nagy, Z. and Pagano, I. and Palaversa, L. and Plachy, E. and Prša, A. and Rybicki, K. A. and Shahaf, S. and Szabados, L. and Szegedi-Elek, E. and Trabucchi, M. and Barblan, F. and Grenon, M. and Roelens, M. and Süveges, M.},
   year={2023},
   month={jun}, pages={A13} }

@dataset{GaiaVari,
       author = {{Gaia Collaboration}},
        title = "{VizieR Online Data Catalog: Gaia DR3 Part 4. Variability (Gaia Collaboration, 2022)}",
 howpublished = {VizieR On-line Data Catalog: I/358.  Originally published in: 2023A\&A...674A..20G},
         year = {2022},
        month = {may},
          eid = {I/358},
       adsurl = {https://ui.adsabs.harvard.edu/abs/2022yCat.1358....0G}
}

@article{Griest_2014,
   title={Experiemental Limits on Primordial Black Holes Dark Matter from the First 2 Years of KEPLER Data},
   volume={786},
   ISSN={1538-4357},
   url={http://dx.doi.org/10.1088/0004-637X/786/2/158},
   DOI={10.1088/0004-637x/786/2/158},
   number={2},
   journal={The Astrophysical Journal},
   publisher={American Astronomical Society},
   author={Griest, Kim and Cieplak, Agnieszka M. and Lehner, Matthew J.},
   year={2014},
   month=apr, pages={158} }

@ARTICLE{Astrophot,
       author = {{Stone}, Connor J. and {Courteau}, St{\'e}phane and {Cuillandre}, Jean-Charles and {Hezaveh}, Yashar and {Perreault-Levasseur}, Laurence and {Arora}, Nikhil},
        title = "{AstroPhot: Fitting Everything Everywhere All at Once in Astronomical Images}",
      journal = {\mnras},
         year = 2023,
        month = aug,
          doi = {10.1093/mnras/stad2477}}

@INPROCEEDINGS{NoirlabCP,
       author = {{Valdes}, F. and {Gruendl}, R. and {DES Project}},
        title = "{The DECam Community Pipeline}",
    booktitle = {Astronomical Data Analysis Software and Systems XXIII},
         year = 2014,
       editor = {{Manset}, N. and {Forshay}, P.},
       series = {Astronomical Society of the Pacific Conference Series},
       volume = {485},
        month = may,
        pages = {379}}

@article{Robinson_2022,
   title={Understanding Accretion Variability through TESS Observations of Taurus},
   volume={935},
   ISSN={1538-4357},
   DOI={10.3847/1538-4357/ac7e51},
   number={1},
   journal={The Astrophysical Journal},
   publisher={American Astronomical Society},
   author={Robinson, Connor E. and Espaillat, Catherine C. and Rodriguez, Joseph E.},
   year={2022},
   month=aug, pages={54} }

@article{Labdon_2021,
   title={Viscous heating in the disk of the outbursting star FU Orionis},
   volume={646},
   ISSN={1432-0746},
   DOI={10.1051/0004-6361/202039370},
   journal={Astronomy \& Astrophysics},
   publisher={EDP Sciences},
   author={Labdon, Aaron and Kraus, Stefan and Davies, Claire L. and Kreplin, Alexander and Monnier, John D. and Le Bouquin, Jean-Baptiste and Anugu, Narsireddy and ten Brummelaar, Theo and Setterholm, Benjamin and Gardner, Tyler and Ennis, Jacob and Lanthermann, Cyprien and Schaefer, Gail and Laws, Anna},
   year={2021},
   month=feb, pages={A102} }

@article{Izzard_2017, 
    title={Dawes Review 6: The Impact of Companions on Stellar Evolution}, 
    volume={34}, 
    DOI={10.1017/pasa.2016.52}, 
    journal={Publications of the Astronomical Society of Australia}, 
    publisher={Cambridge University Press}, 
    author={De Marco, Orsola and Izzard, Robert G.}, 
    year={2017}, 
    pages={e001}}

@ARTICLE{Cranmer_2015,
       author = {{Cranmer}, Steven R. and {Bastien}, Fabienne A. and {Stassun}, Keivan G. and {Saar}, Steven H.},
        title = "{Stellar Granulation as the Source of High-frequency Flicker in Kepler Light Curves}",
      journal = {\apj},
         year = 2014,
        month = feb,
       volume = {781},
       number = {2},
          eid = {124},
        pages = {124},
          doi = {10.1088/0004-637X/781/2/124}}

@ARTICLE{Hillenbrand_2019,
       author = {{Hillenbrand}, Lynne A. and {Reipurth}, Bo and {Connelley}, Michael and {Cutri}, Roc M. and {Isaacson}, Howard},
        title = "{Gaia 19ajj: A Young Star Brightening Due to Enhanced Accretion and Reduced Extinction}",
      journal = {\aj},
         year = 2019,
        month = dec,
       volume = {158},
       number = {6},
          eid = {240},
        pages = {240},
          doi = {10.3847/1538-3881/ab4e16}}

@article{Hakala_2019,
   title={TESS observations of the asynchronous polar CD Ind: mapping the changing accretion geometry},
   volume={486},
   ISSN={1365-2966},
   DOI={10.1093/mnras/stz992},
   number={2},
   journal={Monthly Notices of the Royal Astronomical Society},
   publisher={Oxford University Press (OUP)},
   author={Hakala, Pasi and Ramsay, Gavin and Potter, Stephen B and Beardmore, Andrew and Buckley, David A H and Wynn, Graham},
   year={2019},
   month=apr, pages={2549–2556} }

@MISC{lightkurve,
   author = {{Lightkurve Collaboration} and {Cardoso}, J.~V.~d.~M. and
             {Hedges}, C. and {Gully-Santiago}, M. and {Saunders}, N. and
             {Cody}, A.~M. and {Barclay}, T. and {Hall}, O. and
             {Sagear}, S. and {Turtelboom}, E. and {Zhang}, J. and
             {Tzanidakis}, A. and {Mighell}, K. and {Coughlin}, J. and
             {Bell}, K. and {Berta-Thompson}, Z. and {Williams}, P. and
             {Dotson}, J. and {Barentsen}, G.},
    title = "{Lightkurve: Kepler and TESS time series analysis in Python}",
    howpublished = {Astrophysics Source Code Library},
    year = 2018,
    month = dec}

@article{Yan_2021,
    author = {Yan, Y and He, H and Li, C and Esamdin, A and Tan, B L and Zhang, L Y and Wang, H},
    title = "{Characteristic time of stellar flares on Sun-like stars}",
    journal = {Monthly Notices of the Royal Astronomical Society: Letters},
    volume = {505},
    number = {1},
    pages = {L79-L83},
    year = {2021},
    month = {06},
    issn = {1745-3925},
    doi = {10.1093/mnrasl/slab055}}

@article{Webb_2021,
	doi = {10.1093/mnras/stab1798},
	year = {2021},
	month = {jun},
	volume = {506},
	number = {2},
	pages = {2089--2103},
	author = {S Webb and C Flynn and J Cooke and J Zhang and A Mahabal and T M C Abbott and R Allen and I Andreoni and S A Bird and S Goode and M Lochner and T Pritchard},
	title = {The Deeper, Wider, Faster programme: exploring stellar flare activity with deep, fast cadenced {DECam} imaging via machine learning},
	journal = {Monthly Notices of the Royal Astronomical Society}}

@article{Jackman_2021,
	doi = {10.1093/mnras/stab979},
	year = 2021,
	month = {apr},
	publisher = {Oxford University Press ({OUP})},
	volume = {504},
	number = {3},
	pages = {3246--3264},
	author = {James A G Jackman and Peter J Wheatley and Jack S Acton and David R Anderson and Daniel Bayliss and Joshua T Briegal and Matthew R Burleigh and Sarah L Casewell and Boris T Gänsicke and Samuel Gill and Edward Gillen and Michael R Goad and Maximilian N Günther and Beth A Henderson and Simon T Hodgkin and James S Jenkins and Chloe Pugh and Didier Queloz and Liam Raynard and Rosanna H Tilbrook and Christopher A Watson and Richard G West},
	title = {Stellar flares detected with the Next Generation Transit Survey},
	journal = {Monthly Notices of the Royal Astronomical Society}}

@article{Kowalski_2024,
  title={Stellar flares},
  author={Kowalski, A.F.},
  journal={Living Reviews in Solar Physics},
  volume={21},
  number={1},
  year={2024},
  pages={1},
  doi={10.1007/s41116-024-00039-4}}

@article{Yang_2024,
    doi = {10.3847/2041-8213/ad6c3c},
    year = {2024},
    month = {aug},
    publisher = {The American Astronomical Society},
    volume = {972},
    number = {1},
    pages = {L12},
    author = {Hongjing Yang and Weicheng Zang and Tianjun Gan and Renkun Kuang and Andrew Gould and Shude Mao},
    title = {How Rare Are TESS Free-floating Planets?},
    journal = {The Astrophysical Journal Letters}}

@article{Mroz_TESS2024,
   title={TESS Free-floating Planet Candidate Is Likely a Stellar Flare},
   volume={73},
   ISSN={00015237},
   DOI={10.32023/0001-5237/73.4.1},
   number={4},
   journal={Acta Astronomica},
   publisher={Copernicus Foundation for Polish Astronomy},
   author={Mr\'oz, P.},
   year={2024},
   month={May},
   pages={259–264} }

@article{Pietras_2022,
   title={Statistical Analysis of Stellar Flares from the First Three Years of TESS Observations},
   volume={935},
   ISSN={1538-4357},
   DOI={10.3847/1538-4357/ac8352},
   number={2},
   journal={The Astrophysical Journal},
   publisher={American Astronomical Society},
   author={Pietras, M. and Falewicz, R. and Siarkowski, M. and Bicz, K. and Preś, P.},
   year={2022},
   month=aug, pages={143} }

@ARTICLE{Yang_2023,
       author = {{Yang}, Zilu and {Zhang}, Liyun and {Meng}, Gang and {Han}, Xianming L. and {Misra}, Prabhakar and {Yang}, Jiawei and {Pi}, Qingfeng},
        title = "{Properties of flare events based on light curves from the TESS survey}",
      journal = {\aap},
         year = 2023,
        month = jan,
       volume = {669},
          eid = {A15},
        pages = {A15},
          doi = {10.1051/0004-6361/202142710}}

@article{Gunther_2020,
   title={Stellar Flares from the First TESS Data Release: Exploring a New Sample of M Dwarfs},
   volume={159},
   ISSN={1538-3881},
   DOI={10.3847/1538-3881/ab5d3a},
   number={2},
   journal={The Astronomical Journal},
   publisher={American Astronomical Society},
   author={Günther, Maximilian N. and Zhan, Zhuchang and Seager, Sara and Rimmer, Paul B. and Ranjan, Sukrit and Stassun, Keivan G. and Oelkers, Ryan J. and Daylan, Tansu and Newton, Elisabeth and Kristiansen, Martti H. and Olah, Katalin and Gillen, Edward and Rappaport, Saul and Ricker, George R. and Vanderspek, Roland K. and Latham, David W. and Winn, Joshua N. and Jenkins, Jon M. and Glidden, Ana and Fausnaugh, Michael and Levine, Alan M. and Dittmann, Jason A. and Quinn, Samuel N. and Krishnamurthy, Akshata and Ting, Eric B.},
   year={2020},
   month=jan, pages={60} }

@article{Green26,
year={2026},
journal={arXiv2602.15974},
author={    Green, A.} 

}

@ARTICLE{Davenport_2014,
       author = {{Davenport}, James R.~A. and {Hawley}, Suzanne L. and {Hebb}, Leslie and {Wisniewski}, John P. and {Kowalski}, Adam F. and {Johnson}, Emily C. and {Malatesta}, Michael and {Peraza}, Jesus and {Keil}, Marcus and {Silverberg}, Steven M. and {Jansen}, Tiffany C. and {Scheffler}, Matthew S. and {Berdis}, Jodi R. and {Larsen}, Daniel M. and {Hilton}, Eric J.},
        title = "{Kepler Flares. II. The Temporal Morphology of White-light Flares on GJ 1243}",
      journal = {\apj},
         year = 2014,
        month = dec,
       volume = {797},
       number = {2},
          eid = {122},
        pages = {122},
          doi = {10.1088/0004-637X/797/2/122}}

@misc{Kunimoto_2024,
      title={Searching for Free-Floating Planets with TESS: I. Discovery of a First Terrestrial-Mass Candidate}, 
      author={Michelle Kunimoto and William DeRocco and Nolan Smyth and Steve Bryson},
      year={2024},
      eprint={2404.11666},
      archivePrefix={arXiv},
      primaryClass={astro-ph.EP}}

@article{Davenport_2016,
    doi = {10.3847/0004-637X/829/1/23},
    year = {2016},
    month = {sep},
    publisher = {The American Astronomical Society},
    volume = {829},
    number = {1},
    pages = {23},
    author = {James R. A. Davenport},
    title = {THE KEPLER CATALOG OF STELLAR FLARES},
    journal = {The Astrophysical Journal}}

@ARTICLE{Yang_2018,
       author = {{Yang}, Huiqin and {Liu}, Jifeng and {Qiao}, Erlin and {Zhang}, Haotong and {Gao}, Qing and {Cui}, Kaiming and {Han}, Henggeng},
        title = "{Do Long-cadence Data of the Kepler Spacecraft Capture Basic Properties of Flares?}",
      journal = {\apj},
         year = 2018,
        month = jun,
       volume = {859},
       number = {2},
          eid = {87},
        pages = {87},
          doi = {10.3847/1538-4357/aabd31}}

@article{Smyth_2020,
   title={Updated constraints on asteroid-mass primordial black holes as dark matter},
   volume={101},
   ISSN={2470-0029},
   DOI={10.1103/physrevd.101.063005},
   number={6},
   journal={Physical Review D},
   publisher={American Physical Society (APS)},
   author={Smyth, Nolan and Profumo, Stefano and English, Samuel and Jeltema, Tesla and McKinnon, Kevin and Guhathakurta, Puragra},
   year={2020},
   month=mar }

@article{Schlegel_1998,
   title={Maps of Dust Infrared Emission for Use in Estimation of Reddening and Cosmic Microwave Background Radiation Foregrounds},
   volume={500},
   ISSN={1538-4357},
   DOI={10.1086/305772},
   number={2},
   journal={The Astrophysical Journal},
   publisher={American Astronomical Society},
   author={Schlegel, David J. and Finkbeiner, Douglas P. and Davis, Marc},
   year={1998},
   month=jun, pages={525–553} }

@article{Abbott_2018,
   title={The Dark Energy Survey: Data Release 1},
   volume={239},
   ISSN={1538-4365},
   DOI={10.3847/1538-4365/aae9f0},
   number={2},
   journal={The Astrophysical Journal Supplement Series},
   publisher={American Astronomical Society},
   author={Abbott, T. M. C. and Abdalla, F. B. and Allam, S. and Amara, A. and Annis, J. and Asorey, J. and Avila, S. and Ballester, O. and Banerji, M. and Barkhouse, W. and Baruah, L. and Baumer, M. and Bechtol, K. and Becker, M. R. and Benoit-Lévy, A. and Bernstein, G. M. and Bertin, E. and Blazek, J. and Bocquet, S. and Brooks, D. and Brout, D. and Buckley-Geer, E. and Burke, D. L. and Busti, V. and Campisano, R. and Cardiel-Sas, L. and Rosell, A. Carnero and Kind, M. Carrasco and Carretero, J. and Castander, F. J. and Cawthon, R. and Chang, C. and Chen, X. and Conselice, C. and Costa, G. and Crocce, M. and Cunha, C. E. and D’Andrea, C. B. and Costa, L. N. da and Das, R. and Daues, G. and Davis, T. M. and Davis, C. and Vicente, J. De and DePoy, D. L. and DeRose, J. and Desai, S. and Diehl, H. T. and Dietrich, J. P. and Dodelson, S. and Doel, P. and Drlica-Wagner, A. and Eifler, T. F. and Elliott, A. E. and Evrard, A. E. and Farahi, A. and Neto, A. Fausti and Fernandez, E. and Finley, D. A. and Flaugher, B. and Foley, R. J. and Fosalba, P. and Friedel, D. N. and Frieman, J. and García-Bellido, J. and Gaztanaga, E. and Gerdes, D. W. and Giannantonio, T. and Gill, M. S. S. and Glazebrook, K. and Goldstein, D. A. and Gower, M. and Gruen, D. and Gruendl, R. A. and Gschwend, J. and Gupta, R. R. and Gutierrez, G. and Hamilton, S. and Hartley, W. G. and Hinton, S. R. and Hislop, J. M. and Hollowood, D. and Honscheid, K. and Hoyle, B. and Huterer, D. and Jain, B. and James, D. J. and Jeltema, T. and Johnson, M. W. G. and Johnson, M. D. and Kacprzak, T. and Kent, S. and Khullar, G. and Klein, M. and Kovacs, A. and Koziol, A. M. G. and Krause, E. and Kremin, A. and Kron, R. and Kuehn, K. and Kuhlmann, S. and Kuropatkin, N. and Lahav, O. and Lasker, J. and Li, T. S. and Li, R. T. and Liddle, A. R. and Lima, M. and Lin, H. and López-Reyes, P. and MacCrann, N. and Maia, M. A. G. and Maloney, J. D. and Manera, M. and March, M. and Marriner, J. and Marshall, J. L. and Martini, P. and McClintock, T. and McKay, T. and McMahon, R. G. and Melchior, P. and Menanteau, F. and Miller, C. J. and Miquel, R. and Mohr, J. J. and Morganson, E. and Mould, J. and Neilsen, E. and Nichol, R. C. and Nogueira, F. and Nord, B. and Nugent, P. and Nunes, L. and Ogando, R. L. C. and Old, L. and Pace, A. B. and Palmese, A. and Paz-Chinchón, F. and Peiris, H. V. and Percival, W. J. and Petravick, D. and Plazas, A. A. and Poh, J. and Pond, C. and Porredon, A. and Pujol, A. and Refregier, A. and Reil, K. and Ricker, P. M. and Rollins, R. P. and Romer, A. K. and Roodman, A. and Rooney, P. and Ross, A. J. and Rykoff, E. S. and Sako, M. and Sanchez, M. L. and Sanchez, E. and Santiago, B. and Saro, A. and Scarpine, V. and Scolnic, D. and Serrano, S. and Sevilla-Noarbe, I. and Sheldon, E. and Shipp, N. and Silveira, M. L. and Smith, M. and Smith, R. C. and Smith, J. A. and Soares-Santos, M. and Sobreira, F. and Song, J. and Stebbins, A. and Suchyta, E. and Sullivan, M. and Swanson, M. E. C. and Tarle, G. and Thaler, J. and Thomas, D. and Thomas, R. C. and Troxel, M. A. and Tucker, D. L. and Vikram, V. and Vivas, A. K. and Walker, A. R. and Wechsler, R. H. and Weller, J. and Wester, W. and Wolf, R. C. and Wu, H. and Yanny, B. and Zenteno, A. and Zhang, Y. and Zuntz, J. and Juneau, S. and Fitzpatrick, M. and Nikutta, R. and Nidever, D. and Olsen, K. and Scott, A.},
   year={2018},
   month=nov, pages={18} }

@ARTICLE{Fitz_1999,
       author = {{Fitzpatrick}, Edward L.},
        title = "{Correcting for the Effects of Interstellar Extinction}",
      journal = {\pasp},
         year = 1999,
        month = jan,
       volume = {111},
       number = {755},
        pages = {63-75},
          doi = {10.1086/316293}}

@ARTICLE{Huang_2015,
       author = {{Huang}, Y. and {Liu}, X. -W. and {Yuan}, H. -B. and {Xiang}, M. -S. and {Chen}, B. -Q. and {Zhang}, H. -W.},
        title = "{Empirical metallicity-dependent calibrations of effective temperature against colours for dwarfs and giants based on interferometric data}",
      journal = {\mnras},
         year = 2015,
        month = dec,
       volume = {454},
       number = {3},
        pages = {2863-2889},
          doi = {10.1093/mnras/stv1991}}

@ARTICLE{Choudhury_2021,
       author = {{Choudhury}, Samyaday and {de Grijs}, Richard and {Bekki}, Kenji and {Cioni}, Maria-Rosa L. and {Ivanov}, Valentin D. and {van Loon}, Jacco Th and {Miller}, Amy E. and {Niederhofer}, Florian and {Oliveira}, Joana M. and {Ripepi}, Vincenzo and {Sun}, Ning-Chen and {Subramanian}, Smitha},
        title = "{The VMC survey - XLIV: mapping metallicity trends in the large magellanic cloud using near-infrared passbands}",
      journal = {\mnras},
         year = 2021,
        month = nov,
       volume = {507},
       number = {4},
        pages = {4752-4763},
          doi = {10.1093/mnras/stab2446}}

@MISC{isochrones,
       author = {{Morton}, Timothy D.},
        title = "{isochrones: Stellar model grid package}",
 howpublished = {Astrophysics Source Code Library, record ascl:1503.010},
         year = 2015,
        month = mar,
          eid = {ascl:1503.010},
        pages = {ascl:1503.010},}

@article{Choi_2016,
   title={MESA ISOCHRONES AND STELLAR TRACKS (MIST). I. SOLAR-SCALED MODELS},
   volume={823},
   ISSN={1538-4357},
   DOI={10.3847/0004-637x/823/2/102},
   number={2},
   journal={The Astrophysical Journal},
   publisher={American Astronomical Society},
   author={Choi, Jieun and Dotter, Aaron and Conroy, Charlie and Cantiello, Matteo and Paxton, Bill and Johnson, Benjamin D.},
   year={2016},
   month=may, pages={102} }

@article{Willmer_2018,
   title={The Absolute Magnitude of the Sun in Several Filters},
   volume={236},
   ISSN={1538-4365},
   DOI={10.3847/1538-4365/aabfdf},
   number={2},
   journal={The Astrophysical Journal Supplement Series},
   publisher={American Astronomical Society},
   author={Willmer, Christopher N. A.},
   year={2018},
   month=jun, pages={47} }

@article{Gravity,
    author = {Abuter, R. and Amorim, A. and Bauböck, M. and Berger, J. P. and Bonnet, H. and Brandner, W. and Clénet, Y. and Coudé du Foresto, V. and de Zeeuw, P. T. and Dexter, J. and Duvert, G. and Eckart, A. and Eisenhauer, F. and Förster Schreiber, N. M. and Garcia, P. and Gao, F. and Gendron, E. and Genzel, R. and Gerhard, O. and Gillessen, S. and Habibi, M. and Haubois, X. and Henning, T. and Hippler, S. and Horrobin, M. and Jiménez-Rosales, A. and Jocou, L. and Kervella, P. and Lacour, S. and Lapeyrère, V. and Le Bouquin, J.-B. and Léna, P. and Ott, T. and Paumard, T. and Perraut, K. and Perrin, G. and Pfuhl, O. and Rabien, S. and Rodriguez Coira, G. and Rousset, G. and Scheithauer, S. and Sternberg, A. and Straub, O. and Straubmeier, C. and Sturm, E. and Tacconi, L. J. and Vincent, F. and von Fellenberg, S. and Waisberg, I. and Widmann, F. and Wieprecht, E. and Wiezorrek, E. and Woillez, J. and Yazici, S.},
    collaboration = "Gravity",
    title = "{A geometric distance measurement to the Galactic center black hole with 0.3\% uncertainty}",
    primaryClass = "astro-ph.GA",
    doi = "10.1051/0004-6361/201935656",
    journal = "Astron. Astrophys.",
    volume = "625",
    year = "2019"}

@article{Besla_2012,
   title={The origin of the microlensing events observed towards the LMC and the stellar counterpart of the Magellanic stream},
   volume={428},
   ISSN={0035-8711},
   DOI={10.1093/mnras/sts192},
   number={3},
   journal={Monthly Notices of the Royal Astronomical Society},
   publisher={Oxford University Press (OUP)},
   author={Besla, Gurtina and Hernquist, Lars and Loeb, Abraham},
   year={2012},
   month=nov, pages={2342–2365} }

@article{Wang_2015,
   title={Close encounters involving free-floating planets in star clusters},
   volume={449},
   ISSN={1365-2966},
   DOI={10.1093/mnras/stv542},
   number={4},
   journal={Monthly Notices of the Royal Astronomical Society},
   publisher={Oxford University Press (OUP)},
   author={Wang, Long and Kouwenhoven, M. B. N. and Zheng, Xiaochen and Church, Ross P. and Davies, Melvyn B.},
   year={2015},
   month=apr, pages={3543–3558} }

@article{DeRocco_2023,
   title={Constraints on sub-terrestrial free-floating planets from Subaru microlensing observations},
   volume={527},
   ISSN={1365-2966},
   DOI={10.1093/mnras/stad3824},
   number={3},
   journal={Monthly Notices of the Royal Astronomical Society},
   publisher={Oxford University Press (OUP)},
   author={DeRocco, William and Smyth, Nolan and Profumo, Stefano},
   year={2023},
   month=nov, pages={8921–8930} }

@article{Calcino_2018,
   title={Updating the MACHO fraction of the Milky Way dark halowith improved mass models},
   volume={479},
   ISSN={1365-2966},
   DOI={10.1093/mnras/sty1368},
   number={3},
   journal={Monthly Notices of the Royal Astronomical Society},
   publisher={Oxford University Press (OUP)},
   author={Calcino, Josh and García-Bellido, Juan and Davis, Tamara M},
   year={2018},
   month=may, pages={2889–2905} }

@article{Navarro_1997,
   title={A Universal Density Profile from Hierarchical Clustering},
   volume={490},
   ISSN={1538-4357},
   url={http://dx.doi.org/10.1086/304888},
   DOI={10.1086/304888},
   number={2},
   journal={The Astrophysical Journal},
   publisher={American Astronomical Society},
   author={Navarro, Julio F. and Frenk, Carlos S. and White, Simon D. M.},
   year={1997},
   month=dec, pages={493–508} }

@ARTICLE{StaveleySmith_2003,
       author = {{Staveley-Smith}, L. and {Kim}, S. and {Calabretta}, M.~R. and {Haynes}, R.~F. and {Kesteven}, M.~J.},
        title = "{A new look at the large-scale HI structure of the Large Magellanic Cloud}",
      journal = {\mnras},
         year = 2003,
        month = feb,
       volume = {339},
       number = {1},
        pages = {87-104},
          doi = {10.1046/j.1365-8711.2003.06146.x}}

@article{Kormendy_2016,
   title={SCALING LAWS FOR DARK MATTER HALOS IN LATE-TYPE AND DWARF SPHEROIDAL GALAXIES},
   volume={817},
   ISSN={1538-4357},
   DOI={10.3847/0004-637x/817/2/84},
   number={2},
   journal={The Astrophysical Journal},
   publisher={American Astronomical Society},
   author={Kormendy, John and Freeman, K. C.},
   year={2016},
   month=jan, pages={84} }

@article{Blaineau_2020,
   title={Parallax in microlensing toward the Magellanic Clouds: Effect on detection efficiency and detectability},
   volume={636},
   ISSN={1432-0746},
   DOI={10.1051/0004-6361/202038005},
   journal={Astronomy \& Astrophysics},
   publisher={EDP Sciences},
   author={Blaineau, T. and Moniez, M.},
   year={2020},
   month=apr, pages={L9} }

@article{Battaglia_2005,
   title={The radial velocity dispersion profile of the Galactic halo: constraining the density profile of the dark halo of the Milky Way},
   volume={364},
   ISSN={1365-2966},
   DOI={10.1111/j.1365-2966.2005.09367.x},
   number={2},
   journal={Monthly Notices of the Royal Astronomical Society},
   publisher={Oxford University Press (OUP)},
   author={Battaglia, Giuseppina and Helmi, Amina and Morrison, Heather and Harding, Paul and Olszewski, Edward W. and Mateo, Mario and Freeman, Kenneth C. and Norris, John and Shectman, Stephen A.},
   year={2005},
   month=dec, pages={433–442} }

@article{Alcock_2000,
   title={The MACHO Project: Microlensing Results from 5.7 Years of Large Magellanic Cloud Observations},
   volume={550},
   ISSN={1538-4357},
   DOI={10.1086/309512},
   number={1},
   journal={The Astrophysical Journal},
   publisher={American Astronomical Society},
   author={Alcock, C. and Allsman, R. A. and Alves, D. R. and Axelrod, T. S. and Becker, A. C. and Bennett, D. P. and Cook, K. H. and Dalal, N. and Drake, A. J. and Freeman, K. C. and Geha, M. and Griest, K. and Lehner, M. J. and Marshall, S. L. and Minniti, D. and Nelson, C. A. and Peterson, B. A. and Popowski, P. and Pratt, M. R. and Quinn, P. J. and Stubbs, C. W. and Sutherland, W. and Tomaney, A. B. and Vandehei, T. and Welch, D.},
   year={2001},
   month=oct, pages={169} }

@ARTICLE{CalchiNovati_2006,
       author = {{Calchi Novati}, S. and {de Luca}, F. and {Jetzer}, Ph. and {Scarpetta}, G.},
        title = "{Microlensing towards the Large Magellanic Cloud: a study of the LMC halo contribution}",
      journal = {\aap},
         year = 2006,
        month = nov,
       volume = {459},
       number = {2},
        pages = {407-414},
          doi = {10.1051/0004-6361:20065149}}

@article{Alves_2004,
   title={The Stellar Halo in the Large Magellanic Cloud: Mass, Luminosity, and Microlensing Predictions},
   volume={601},
   ISSN={1538-4357},
   DOI={10.1086/382130},
   number={2},
   journal={The Astrophysical Journal},
   publisher={American Astronomical Society},
   author={Alves, David R.},
   year={2004},
   month=jan, pages={L151–L154} }

@article{Sajadian_2021,
   title={On the detection of free-floating planets through microlensing towards the Magellanic Clouds},
   volume={506},
   ISSN={1365-2966},
   DOI={10.1093/mnras/stab1907},
   number={3},
   journal={Monthly Notices of the Royal Astronomical Society},
   publisher={Oxford University Press (OUP)},
   author={Sajadian, Sedighe},
   year={2021},
   month=jul, pages={3615–3628} }

@article{Gyuk_2000,
   title={Self‐lensing Models of the Large Magellanic Cloud},
   volume={535},
   ISSN={1538-4357},
   DOI={10.1086/308834},
   number={1},
   journal={The Astrophysical Journal},
   publisher={American Astronomical Society},
   author={Gyuk, G. and Dalal, N. and Griest, K.},
   year={2000},
   month=may, pages={90–103} }

@ARTICLE{Pietrzyski_2019,
       author = {{Pietrzy{\'n}ski}, G. and {Graczyk}, D. and {Gallenne}, A. and {Gieren}, W. and {Thompson}, I.~B. and {Pilecki}, B. and {Karczmarek}, P. and {G{\'o}rski}, M. and {Suchomska}, K. and {Taormina}, M. and {Zgirski}, B. and {Wielg{\'o}rski}, P. and {Ko{\l}aczkowski}, Z. and {Konorski}, P. and {Villanova}, S. and {Nardetto}, N. and {Kervella}, P. and {Bresolin}, F. and {Kudritzki}, R.~P. and {Storm}, J. and {Smolec}, R. and {Narloch}, W.},
        title = "{A distance to the Large Magellanic Cloud that is precise to one per cent}",
      journal = {\nat},
         year = 2019,
        month = mar,
       volume = {567},
       number = {7747},
        pages = {200-203},
          doi = {10.1038/s41586-019-0999-4}}

@article{Weinberg_2001,
   title={Structure of the Large Magellanic Cloud from 2MASS},
   volume={548},
   ISSN={1538-4357},
   DOI={10.1086/319001},
   number={2},
   journal={The Astrophysical Journal},
   publisher={American Astronomical Society},
   author={Weinberg, Martin D. and Nikolaev, Sergei},
   year={2001},
   month=feb, pages={712–726} }

@ARTICLE{Marel_2002,
       author = {{van der Marel}, Roeland P. and {Alves}, David R. and {Hardy}, Eduardo and {Suntzeff}, Nicholas B.},
        title = "{New Understanding of Large Magellanic Cloud Structure, Dynamics, and Orbit from Carbon Star Kinematics}",
      journal = {\aj},
         year = 2002,
        month = nov,
       volume = {124},
       number = {5},
        pages = {2639-2663},
          doi = {10.1086/343775}}

@article{Kallivayalil_2013,
   title={THIRD-EPOCH MAGELLANIC CLOUD PROPER MOTIONS. I.HUBBLE SPACE TELESCOPE/WFC3 DATA AND ORBIT IMPLICATIONS},
   volume={764},
   ISSN={1538-4357},
   DOI={10.1088/0004-637x/764/2/161},
   number={2},
   journal={The Astrophysical Journal},
   publisher={American Astronomical Society},
   author={Kallivayalil, Nitya and van der Marel, Roeland P. and Besla, Gurtina and Anderson, Jay and Alcock, Charles},
   year={2013},
   month=feb, pages={161} }

@article{de_Jong_2010,
   title={MAPPING THE STELLAR STRUCTURE OF THE MILKY WAY THICK DISK AND HALO USING SEGUE PHOTOMETRY},
   volume={714},
   ISSN={1538-4357},
   DOI={10.1088/0004-637x/714/1/663},
   number={1},
   journal={The Astrophysical Journal},
   publisher={American Astronomical Society},
   author={de Jong, Jelte T. A. and Yanny, Brian and Rix, Hans-Walter and Dolphin, Andrew E. and Martin, Nicolas F. and Beers, Timothy C.},
   year={2010},
   month=apr, pages={663–674} }

@article{Juri_2008,
   title={The Milky Way Tomography with SDSS. I. Stellar Number Density Distribution},
   volume={673},
   ISSN={1538-4357},
   DOI={10.1086/523619},
   number={2},
   journal={The Astrophysical Journal},
   publisher={American Astronomical Society},
   author={Jurić, Mario and Ivezić, Željko and Brooks, Alyson and Lupton, Robert H. and Schlegel, David and Finkbeiner, Douglas and Padmanabhan, Nikhil and Bond, Nicholas and Sesar, Branimir and Rockosi, Constance M. and Knapp, Gillian R. and Gunn, James E. and Sumi, Takahiro and Schneider, Donald P. and Barentine, J. C. and Brewington, Howard J. and Brinkmann, J. and Fukugita, Masataka and Harvanek, Michael and Kleinman, S. J. and Krzesinski, Jurek and Long, Dan and Neilsen, Jr., Eric H. and Nitta, Atsuko and Snedden, Stephanie A. and York, Donald G.},
   year={2008},
   month=feb, pages={864–914} }

@article{Robin_2017,
   title={Kinematics of the local disk from the RAVE survey and the Gaia first data release},
   volume={605},
   ISSN={1432-0746},
   DOI={10.1051/0004-6361/201630217},
   journal={Astronomy \& Astrophysics},
   publisher={EDP Sciences},
   author={Robin, Annie C. and Bienaymé, Olivier and Fernández-Trincado, José G. and Reylé, Céline},
   year={2017},
   month=aug, pages={A1} }

@article{emcee_2013,
   title={<tt>emcee</tt>: The MCMC Hammer},
   volume={125},
   ISSN={1538-3873},
   DOI={10.1086/670067},
   number={925},
   journal={Publications of the Astronomical Society of the Pacific},
   publisher={IOP Publishing},
   author={Foreman-Mackey, Daniel and Hogg, David W. and Lang, Dustin and Goodman, Jonathan},
   year={2013},
   month=mar, pages={306–312} }

@misc{Sumi_2023,
      title={Free-Floating planet Mass Function from MOA-II 9-year survey towards the Galactic Bulge}, 
      author={Takahiro Sumi and Naoki koshimoto and David P. Bennett and Nicholas J. Rattenbury and Fumio Abe and Richard Barry and Aparna Bhattacharya and Ian A. Bond and Hirosane Fujii and Akihiko Fukui and Ryusei Hamada and Yuki Hirao and Stela Ishitani Silva and Yoshitaka Itow and Rintaro Kirikawa and Iona Kondo and Yutaka Matsubara and Shota Miyazaki and Yasushi Muraki and Greg Olmschenk and Clement Ranc and Yuki Satoh and Daisuke Suzuki and Mio Tomoyoshi and Paul . J. Tristram and Aikaterini Vandorou and Hibiki Yama and Kansuke Yamashita},
      year={2023},
      eprint={2303.08280},
      archivePrefix={arXiv},
      primaryClass={astro-ph.EP}}

@article{Nunota_2025,
    doi = {10.3847/1538-4357/ada352},
    url = {https://dx.doi.org/10.3847/1538-4357/ada352},
    year = {2025},
    month = {jan},
    publisher = {The American Astronomical Society},
    volume = {979},
    number = {2},
    pages = {123},
    author = {Nunota, Kansuke and Sumi, Takahiro and Koshimoto, Naoki and Rattenbury, Nicholas J. and Abe, Fumio and Barry, Richard and Bennett, David P. and Bhattacharya, Aparna and Fukui, Akihiko and Hamada, Ryusei and Hamada, Shunya and Hamasaki, Naoto and Hirao, Yuki and Ishitani Silva, Stela and Itow, Yoshitaka and Matsubara, Yutaka and Miyazaki, Shota and Muraki, Yasushi and Nagai, Tsutsumi and Olmschenk, Greg and Ranc, Clement and Satoh, Yuki K. and Suzuki, Daisuke and Tristram, Paul. J. and Vandorou, Aikaterini and Yama, Hibiki and (MOA collaboration)},
    title = {The Microlensing Event Rate and Optical Depth from MOA-II 9 Yr Survey Toward the Galactic Bulge},
    journal = {The Astrophysical Journal}
    }

@misc{Qian_2025,
      title={Systematic Search for FFPs in KMTNet Full-Frame Images. I. Photometry Pipeline}, 
      author={Qiyue Qian and Hongjing Yang and Weicheng Zang and Yoon-Hyun Ryu and Shude Mao and Renkun Kuang and Jiyuan Zhang and Michael D. Albrow and Sun-Ju Chung and Andrew Gould and Cheongho Han and Kyu-Ha Hwang and Youn Kil Jung and In-Gu Shin and Yossi Shvartzvald and Jennifer C. Yee and Sang-Mok Cha and Dong-Jin Kim and Hyoun-Woo Kim and Seung-Lee Kim and Chung-Uk Lee and Dong-Joo Lee and Yongseok Lee and Byeong-Gon Park and Richard W. Pogge},
      year={2025},
      eprint={2503.24097},
      archivePrefix={arXiv},
      primaryClass={astro-ph.EP},
      url={https://arxiv.org/abs/2503.24097}, 
}

@article{Ban_2016,
   title={The microlensing rate and distribution of free-floating planets towards the Galactic bulge},
   volume={595},
   ISSN={1432-0746},
   url={http://dx.doi.org/10.1051/0004-6361/201629166},
   DOI={10.1051/0004-6361/201629166},
   journal={Astronomy & Astrophysics},
   publisher={EDP Sciences},
   author={Ban, M. and Kerins, E. and Robin, A. C.},
   year={2016},
   month=oct, pages={A53} }

@article{Han_2022,
   title={KMT-2021-BLG-1898: Planetary microlensing event involved with binary source stars},
   volume={663},
   ISSN={1432-0746},
   DOI={10.1051/0004-6361/202243102},
   journal={Astronomy \& Astrophysics},
   publisher={EDP Sciences},
   author={Han, Cheongho and Gould, Andrew and Kim, Doeon and Jung, Youn Kil and Albrow, Michael D. and Chung, Sun-Ju and Hwang, Kyu-Ha and Lee, Chung-Uk and Ryu, Yoon-Hyun and Shin, In-Gu and Shvartzvald, Yossi and Yee, Jennifer C. and Zang, Weicheng and Cha, Sang-Mok and Kim, Dong-Jin and Kim, Seung-Lee and Lee, Dong-Joo and Lee, Yongseok and Park, Byeong-Gon and Pogge, Richard W.},
   year={2022},
   month=jul, pages={A145} }

@ARTICLE{Wang_2021,
       author = {{Wang}, Hanyue and {Zang}, Weicheng and {Zhu}, Wei and {Hwang}, Kyu-Ha and {Udalski}, Andrzej and {Gould}, Andrew and {Han}, Cheongho and {Albrow}, Michael D. and {Chung}, Sun-Ju and {Jung}, Youn Kil and {Kim}, Doeon and {Ryu}, Yoon-Hyun and {Shin}, In-Gu and {Shvartzvald}, Yossi and {Yee}, Jennifer C. and {Cha}, Sang-Mok and {Kim}, Dong-Jin and {Kim}, Hyoun-Woo and {Kim}, Seung-Lee and {Lee}, Chung-Uk and {Lee}, Dong-Joo and {Lee}, Yongseok and {Park}, Byeong-Gon and {Pogge}, Richard W. and {Poleski}, Radoslaw and {Mr{\'o}z}, Przemek and {Skowron}, Jan and {Szyma{\'n}ski}, Micha{\l} K. and {Soszy{\'n}ski}, Igor and {Pietrukowicz}, Pawe{\l} and {Koz{\l}owski}, Szymon and {Ulaczyk}, Krzysztof and {Rybicki}, Krzysztof A. and {Iwanek}, Patryk and {Wrona}, Marcin and {Gromadzki}, Mariusz and {Yang}, Hongjing and {Mao}, Shude and {Zhang}, Xiangyu},
        title = "{Systematic Korea Microlensing Telescope Network planetary anomaly search - III. One wide-orbit planet and two stellar binaries}",
      journal = {\mnras},
         year = 2022,
        month = feb,
       volume = {510},
       number = {2},
        pages = {1778-1790},
          doi = {10.1093/mnras/stab3581}}

@article{Ryu_2021,
   title={KMT-2017-BLG-2820 and the Nature of the Free-floating Planet Population},
   volume={161},
   ISSN={1538-3881},
   DOI={10.3847/1538-3881/abd55f},
   number={3},
   journal={The Astronomical Journal},
   publisher={American Astronomical Society},
   author={Ryu, Yoon-Hyun and Mróz, Przemek and Gould, Andrew and Hwang, Kyu-Ha and Kim, Hyoun-Woo and Yee, Jennifer C. and Albrow, Michael D. and Chung, Sun-Ju and Jung, Youn Kil and Shin, In-Gu and Shvartzvald, Yossi and Zang, Weicheng and Cha, Sang-Mok and Kim, Dong-Jin and Kim, Seung-Lee and Lee, Chung-Uk and Lee, Dong-Joo and Lee, Yongseok and Park, Byeong-Gon and Han, Cheongho and Pogge, Richard W. and Udalski, Andrzej and Poleski, Radek and Skowron, Jan and Szymański, Michał K. and Soszyński, Igor and Pietrukowicz, Paweł and Kozłowski, Szymon and Ulaczyk, Krzysztof and Rybicki, Krzysztof A. and Iwanek, Patryk},
   year={2021},
   month=feb, pages={126}}

@article{Poleski_2021, place={PL},
   title={Wide-Orbit Exoplanets are Common. Analysis of Nearly 20 Years of OGLE Microlensing Survey Data},
   volume={71},
   ISSN={00015237},
   DOI={10.32023/0001-5237/71.1.1},
   number={1},
   journal={Acta Astronomica},
   publisher={Copernicus Foundation for Polish Astronomy},
   author={Poleski, R. and Skowron, J. and Mróz, P. and Udalski, A. and Szymański, M.K. and Pietrukowicz, P. and Ulaczyk, K. and Rybicki, K. and Iwanek, P. and Wrona, M. and Gromadzki, M.},
   year={2021},
   month={Mar},
   pages={1–23} }

@article{Mroz_FFP12020,
   title={A Terrestrial-mass Rogue Planet Candidate Detected in the Shortest-timescale Microlensing Event},
   volume={903},
   ISSN={2041-8213},
   DOI={10.3847/2041-8213/abbfad},
   number={1},
   journal={The Astrophysical Journal Letters},
   publisher={American Astronomical Society},
   author={Mróz, Przemek and Poleski, Radosław and Gould, Andrew and Udalski, Andrzej and Sumi, Takahiro and Szymański, Michał K. and Soszyński, Igor and Pietrukowicz, Paweł and Kozłowski, Szymon and Skowron, Jan and Ulaczyk, Krzysztof and Albrow, Michael D. and Chung, Sun-Ju and Han, Cheongho and Hwang, Kyu-Ha and Jung, Youn Kil and Kim, Hyoun-Woo and Ryu, Yoon-Hyun and Shin, In-Gu and Shvartzvald, Yossi and Yee, Jennifer C. and Zang, Weicheng and Cha, Sang-Mok and Kim, Dong-Jin and Kim, Seung-Lee and Lee, Chung-Uk and Lee, Dong-Joo and Lee, Yongseok and Park, Byeong-Gon and Pogge, Richard W.},
   year={2020},
   month=oct, pages={L11} }

@misc{Koshimoto_2023,
      title={Terrestrial and Neptune mass free-floating planet candidates from the MOA-II 9-year Galactic Bulge survey}, 
      author={Naoki Koshimoto and Takahiro Sumi and David P. Bennett and Valerio Bozza and Przemek Mróz and Andrzej Udalski and Nicholas J. Rattenbury and Fumio Abe and Richard Barry and Aparna Bhattacharya and Ian A. Bond and Hirosane Fujii and Akihiko Fukui and Ryusei Hamada and Yuki Hirao and Stela Ishitani Silva and Yoshitaka Itow and Rintaro Kirikawa and Iona Kondo and Yutaka Matsubara and Shota Miyazaki and Yasushi Muraki and Greg Olmschenk and Clément Ranc and Yuki Satoh and Daisuke Suzuki and Mio Tomoyoshi and Paul J. Tristram and Aikaterini Vandorou and Hibiki Yama and Kansuke Yamashita},
      year={2023},
      eprint={2303.08279},
      archivePrefix={arXiv},
      primaryClass={astro-ph.EP}}

@Article{Vieira_2023,
    AUTHOR = {Vieira, Katherine and Korchagin, Vladimir and Carraro, Giovanni and Lutsenko, Artem},
    TITLE = {Vertical Structure of the Milky Way Disk with Gaia DR3},
    JOURNAL = {Galaxies},
    VOLUME = {11},
    YEAR = {2023},
    NUMBER = {3},
    ARTICLE-NUMBER = {77},
    ISSN = {2075-4434},
    DOI = {10.3390/galaxies11030077}}

@ARTICLE{Binney_2023,
       author = {{Binney}, James and {Vasiliev}, Eugene},
        title = "{Self-consistent models of our Galaxy}",
      journal = {\mnras},
         year = 2023,
        month = apr,
       volume = {520},
       number = {2},
        pages = {1832-1847},
          doi = {10.1093/mnras/stad094}}

@article{Sugiyama_2023,
   title={Possible evidence of axion stars in HSC and OGLE microlensing events},
   volume={840},
   ISSN={0370-2693},
   DOI={10.1016/j.physletb.2023.137891},
   journal={Physics Letters B},
   publisher={Elsevier BV},
   author={Sugiyama, Sunao and Takada, Masahiro and Kusenko, Alexander},
   year={2023},
   month=may, pages={137891} }

@article{Fujikura_2021,
   title={Microlensing constraints on axion stars including finite lens and source size effects},
   volume={104},
   ISSN={2470-0029},
   DOI={10.1103/physrevd.104.123012},
   number={12},
   journal={Physical Review D},
   publisher={American Physical Society (APS)},
   author={Fujikura, Kohei and Hertzberg, Mark P. and Schiappacasse, Enrico D. and Yamaguchi, Masahide},
   year={2021},}

@article{Ozel_2010,
   title={THE BLACK HOLE MASS DISTRIBUTION IN THE GALAXY},
   volume={725},
   ISSN={1538-4357},
   DOI={10.1088/0004-637x/725/2/1918},
   number={2},
   journal={The Astrophysical Journal},
   publisher={American Astronomical Society},
   author={Özel, Feryal and Psaltis, Dimitrios and Narayan, Ramesh and McClintock, Jeffrey E.},
   year={2010},
   month=dec, pages={1918–1927} }

@article{Carr_2020,
   title={Primordial Black Holes as Dark Matter: Recent Developments},
   volume={70},
   ISSN={1545-4134},
   DOI={10.1146/annurev-nucl-050520-125911},
   number={1},
   journal={Annual Review of Nuclear and Particle Science},
   publisher={Annual Reviews},
   author={Carr, Bernard and Kühnel, Florian},
   year={2020},
   month=oct, pages={355–394} }

@article{Niikura_OGLE2019,
   title={Constraints on Earth-mass primordial black holes from OGLE 5-year microlensing events},
   volume={99},
   ISSN={2470-0029},
   url={http://dx.doi.org/10.1103/PhysRevD.99.083503},
   DOI={10.1103/physrevd.99.083503},
   number={8},
   journal={Physical Review D},
   publisher={American Physical Society (APS)},
   author={Niikura, Hiroko and Takada, Masahiro and Yokoyama, Shuichiro and Sumi, Takahiro and Masaki, Shogo},
   year={2019},
   month=apr }

@article{Niikura_2019,
   title={Microlensing constraints on primordial black holes with Subaru/HSC Andromeda observations},
   volume={3},
   ISSN={2397-3366},
   DOI={10.1038/s41550-019-0723-1},
   number={6},
   journal={Nature Astronomy},
   publisher={Springer Science and Business Media LLC},
   author={Niikura, Hiroko and Takada, Masahiro and Yasuda, Naoki and Lupton, Robert H. and Sumi, Takahiro and More, Surhud and Kurita, Toshiki and Sugiyama, Sunao and More, Anupreeta and Oguri, Masamune and Chiba, Masashi},
   year={2019},
   month=apr, pages={524–534} }

@article{DeRocco_2024,
   title={Revealing terrestrial-mass primordial black holes with the Nancy Grace Roman Space Telescope},
   volume={109},
   ISSN={2470-0029},
   url={http://dx.doi.org/10.1103/PhysRevD.109.023013},
   DOI={10.1103/physrevd.109.023013},
   number={2},
   journal={Physical Review D},
   publisher={American Physical Society (APS)},
   author={DeRocco, William and Frangipane, Evan and Hamer, Nick and Profumo, Stefano and Smyth, Nolan},
   year={2024},
   month=jan }

@misc{Perkins_2025,
      title={Hints of an Anomalous Lens Population towards the Galactic Bulge}, 
      author={Scott E. Perkins and Peter McGill and William A. Dawson and Ming-Feng Ho and Natasha S. Abrams and Simeon Bird and Jessica R. Lu},
      year={2025},
      eprint={2503.22037},
      archivePrefix={arXiv},
      primaryClass={astro-ph.GA},
      url={https://arxiv.org/abs/2503.22037}, 
}

@software{Software_MulensModel,author = {Poleski, Radek and Yee, Jennifer},license = {MIT},title = {{MulensModel}},url = {https://github.com/rpoleski/MulensModel}}

\appendix \section{Appendix}\label{appendix:Appendix}

\begin{figure}
\includegraphics[width=1.\columnwidth]{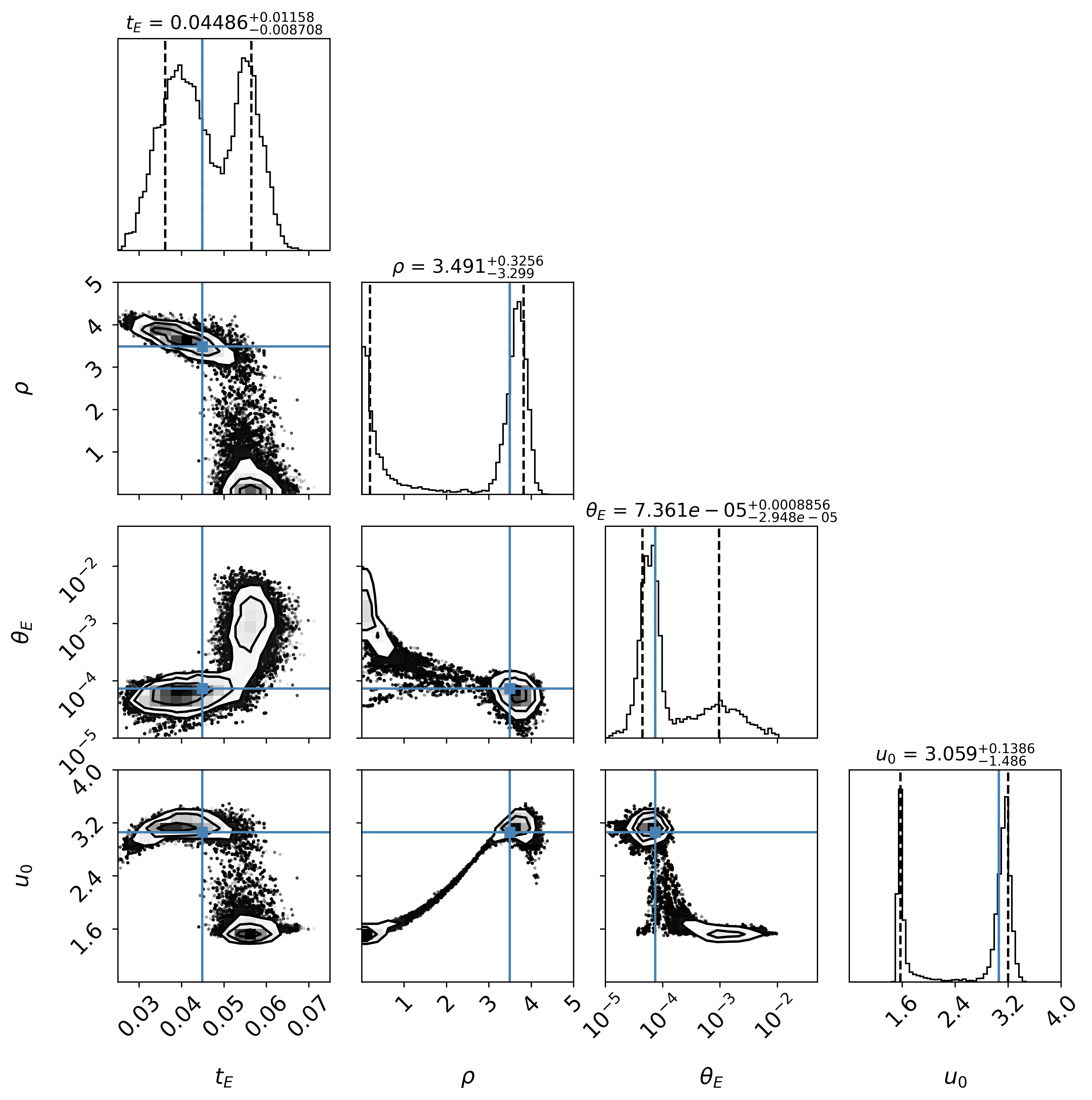}
\caption{\textbf{Dark Matter PBH microlens light curve parameters model}.} \label{figappDML}
\end{figure}

\begin{figure}
\includegraphics[width=1.\columnwidth]{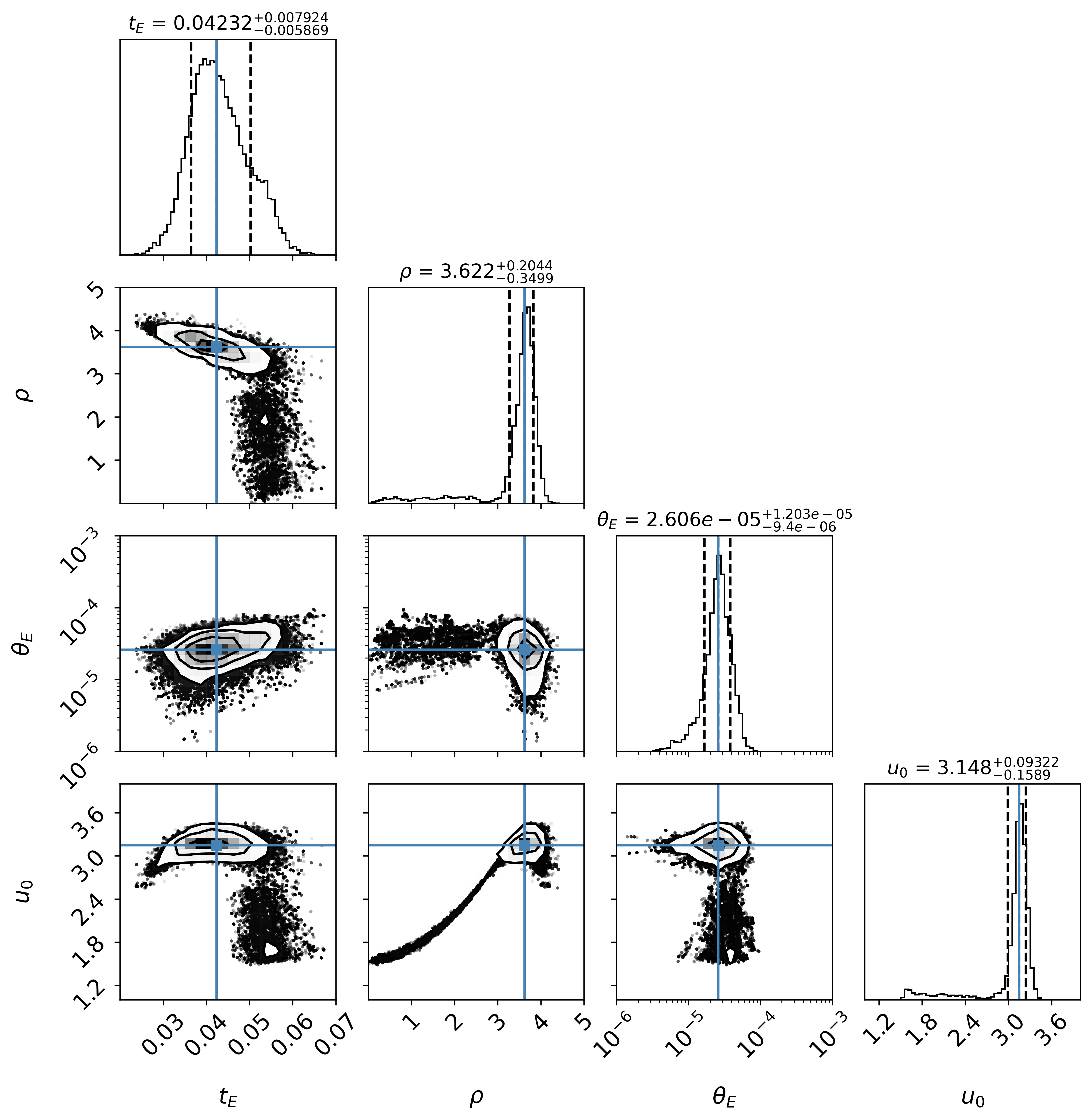}
\caption{\textbf{LMC planetary microlens light curve parameters model}.} \label{figappLMCL}
\end{figure}

\begin{figure}
\includegraphics[width=1.\columnwidth]{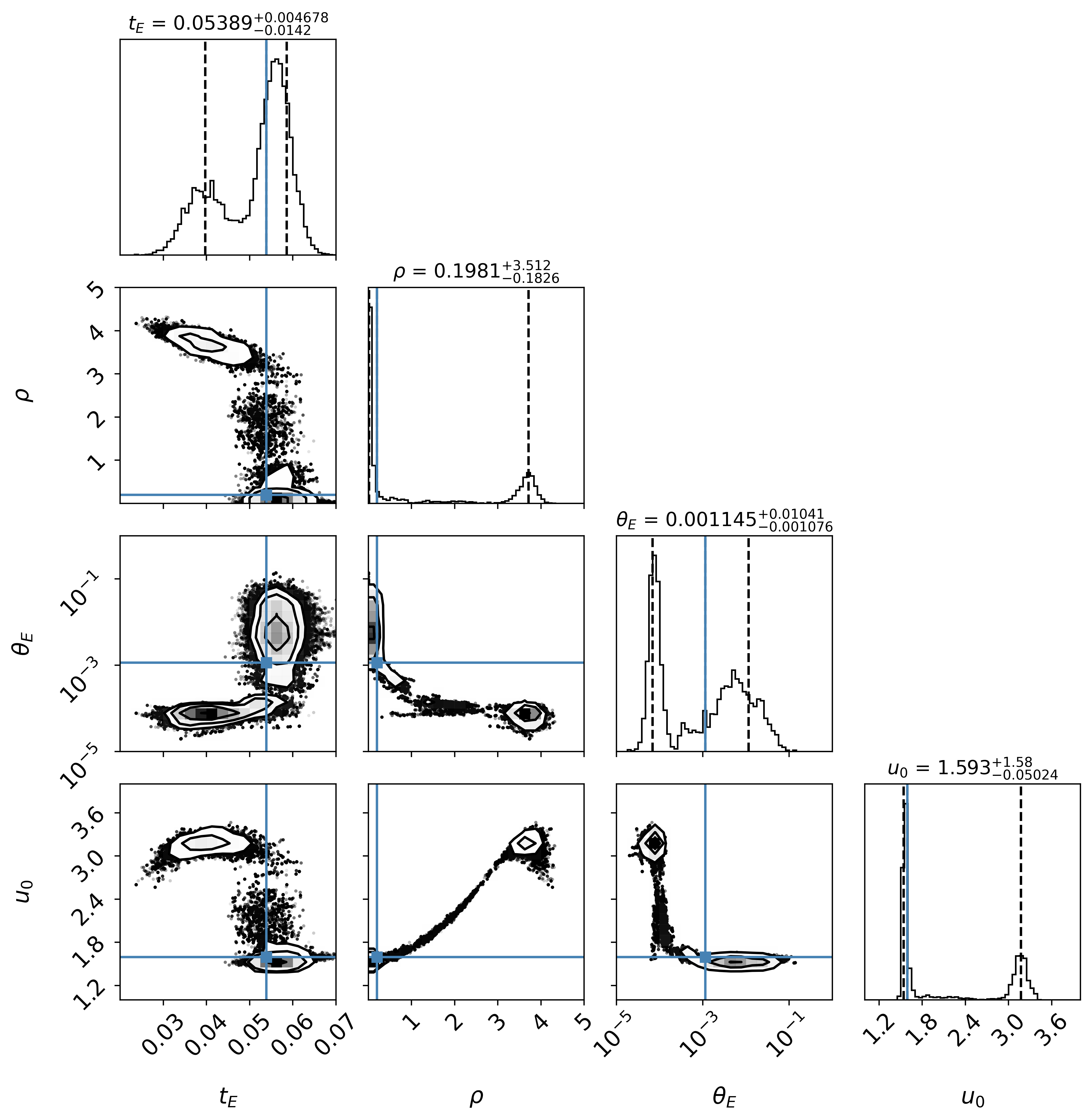}
\caption{\textbf{MW planetary microlens light curve parameters model}.} \label{figappMWL}
\end{figure}

\begin{figure*}
\includegraphics[scale = 0.25]{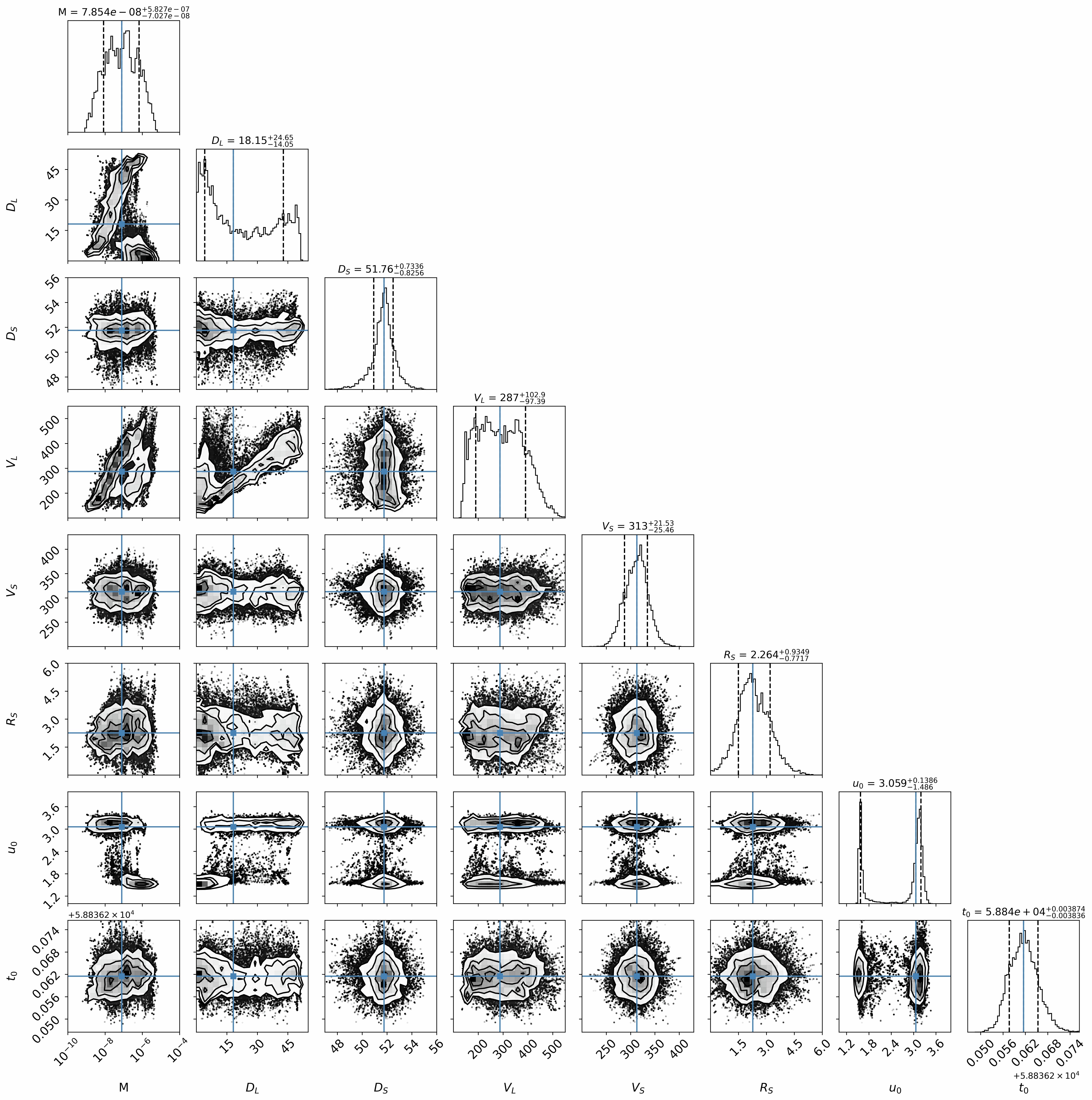}
\caption{\textbf{Dark Matter PBH microlens galactic parameters model}.} \label{figappDML}
\end{figure*}

\begin{figure*}
\includegraphics[scale = 0.25]{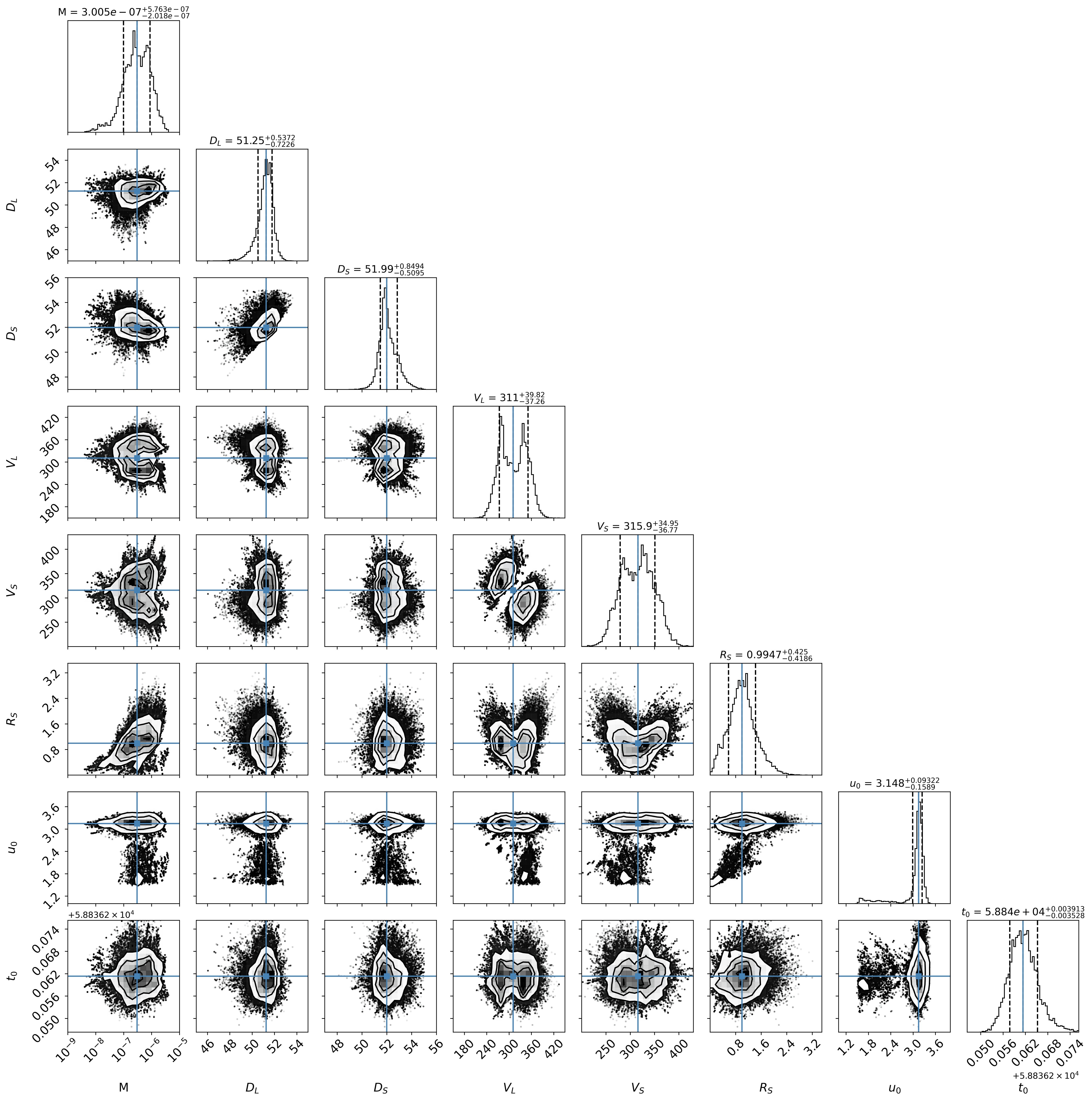}
\caption{\textbf{LMC planetary microlens galactic parameters model}.} \label{figappLMCL}
\end{figure*}

\begin{figure*}
\includegraphics[scale = 0.25]{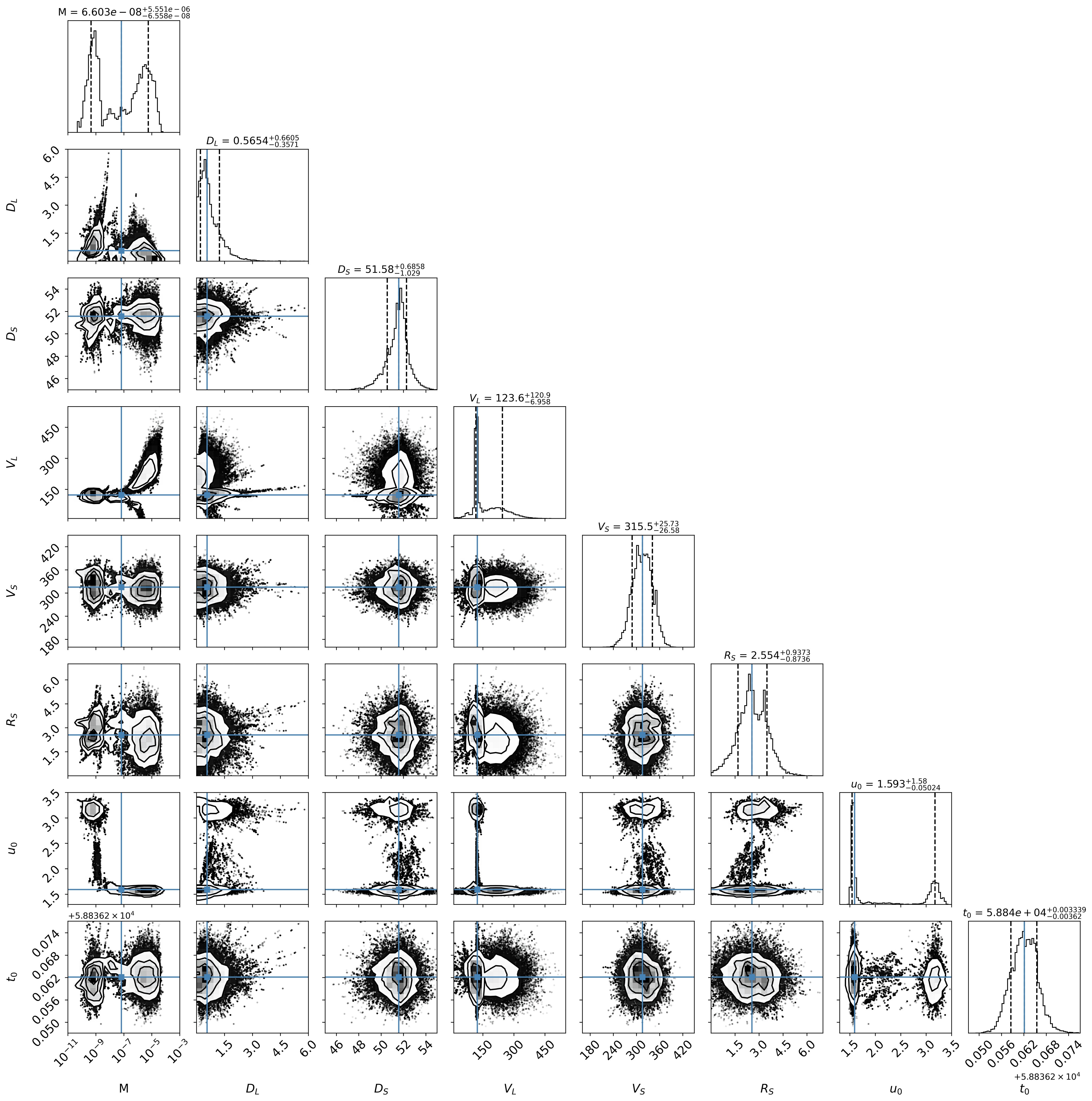}
\caption{\textbf{MW planetary microlens galactic parameters model}.} \label{figappMWL}
\end{figure*}

% Don't change these lines
\bsp	% typesetting comment
\label{lastpage}
\end{document}